\documentclass[journal]{IEEEtran}
\usepackage{times}
\usepackage{epsfig}
\usepackage{graphicx}
\usepackage{svg}
\usepackage{amsmath}
\usepackage{amssymb,pifont}
\usepackage{overpic}
\usepackage{multirow}
\usepackage{booktabs}
\usepackage{bm}
\usepackage{colortbl}

\usepackage{cite}
\usepackage{makecell}
\usepackage{boldline}
\setcellgapes{3pt}
\usepackage{arydshln}
\usepackage{tabu}
\usepackage{subfigure}
\usepackage{caption}
\usepackage{balance}
\usepackage{xcolor}
\usepackage[breaklinks=true,bookmarks=false]{hyperref}
\UseRawInputEncoding
\usepackage{stfloats}
\usepackage{algorithm}  
\usepackage{algorithmic}  

\usepackage{makecell} 
\usepackage{cleveref}
\crefformat{section}{\S#2#1#3} 
\crefformat{subsection}{\S#2#1#3}
\crefformat{subsubsection}{\S#2#1#3}
\usepackage{overpic}

\usepackage{color}

\begin{document}
\title{NLHD: A Pixel-Level Non-Local  Retinex Model for Low-Light Image Enhancement}

\author{
  Hao Hou, 
  Yingkun Hou, \textit{Senior Member, IEEE}, 
  Yuxuan Shi,
  Benzheng Wei,
  Jun Xu, \textit{Member, IEEE}
\thanks{
This work is partly supported by Natural Science Foundation of Shandong Province (Nos. ZR2020MF038, ZR2020KF013), Natural Science Foundation of China (Nos. 62002176, 61872225), the Fundamental Research Funds for the Central Universities, Nankai University (No. 63211099). Y Hou is the corresponding author (email: ykhou@tsu.edu.cn).
}
\thanks{
H Hou is with College of Intelligence and Information Engineering, Shandong University of Traditional Chinese Medicine, Ji'nan, 250355,  China.
\par Y Hou is with School of Information Science and Technology, Taishan University, Tai'an, 271000, China.
\par Y Shi and J Xu are with School of Statistics and Data Science, Nankai University, Tianjin, China.
\par B Wei and H Hou are with the Center for Medical Artificial Intelligence, Shandong University of Traditional Chinese Medicine, Ji'nan, 250355, China.
}
}


\maketitle

\begin{abstract}
Retinex model has been successfully applied to low-light image enhancement in many existing methods. More appropriate decomposition of a low-light image to illumination and reflectance components can help achieve better image enhancement. In this paper, we propose a new pixel-level non-local Haar transform based illumination and reflectance decomposition method (NLHD). The unique low-frequency coefficient of Haar transform on each similar pixel group is used to reconstruct the illumination component, and the rest of all high-frequency coefficients are employed to reconstruct the reflectance component. The complete similarity of pixels in a matched similar pixel group and the simple separable Haar transform help to obtain more appropriate image decomposition; thus, the image is hardly sharpened in the image brightness enhancement procedure. The exponential transform and logarithmic transform are respectively implemented on the illumination component. Then a minimum fusion strategy on the results of these two transforms is utilized to achieve more natural illumination component enhancement. It can alleviate the mosaic artifacts produced in the darker regions by the exponential transform with a gamma value less than 1 and reduce information loss caused by excessive enhancement of the brighter regions due to the logarithmic transform. Finally, the Retinex model is applied to the enhanced illumination and reflectance to achieve image enhancement. We also develop a local noise level estimation based noise suppression method and a non-local saturation reduction based color deviation correction method. These two methods can respectively attenuate noise or color deviation usually presented in the enhanced results of the extremely dark low-light images. Experiments on benchmark datasets show that the proposed method can achieve better low-light image enhancement results on subjective and objective evaluations than most existing methods.
\end{abstract}

\begin{IEEEkeywords}
Pixel-level self-similarity; Haar transform; frequency decomposition; Retinex model; low-light image enhancement.
\end{IEEEkeywords}

%
\IEEEpeerreviewmaketitle

\section{Introduction}
\label{sec:intro}
\begin{figure}[t!]
\centering
\subfigure{
\begin{minipage}[t]{0.23\textwidth}
\centering
\raisebox{-0.15cm}{\includegraphics[width=1\textwidth]{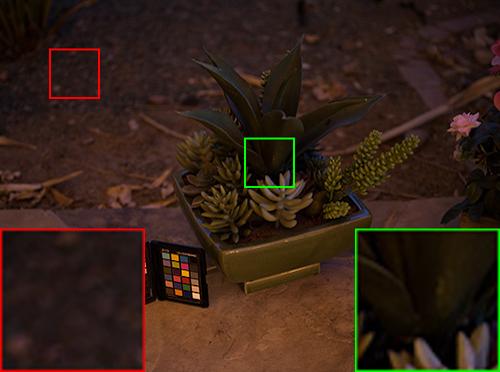}}
{\footnotesize (a) Original}
\end{minipage}
\begin{minipage}[t]{0.23\textwidth}
\centering
\raisebox{-0.15cm}{\includegraphics[width=1\textwidth]{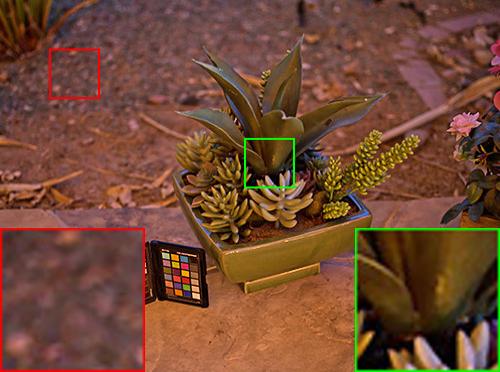}}
{\footnotesize (b) STAR~\cite{star_tip2020}}
\end{minipage}
}\vspace{-3mm}
\subfigure{
\begin{minipage}[t]{0.23\textwidth}
\centering
\raisebox{-0.15cm}{\includegraphics[width=1\textwidth]{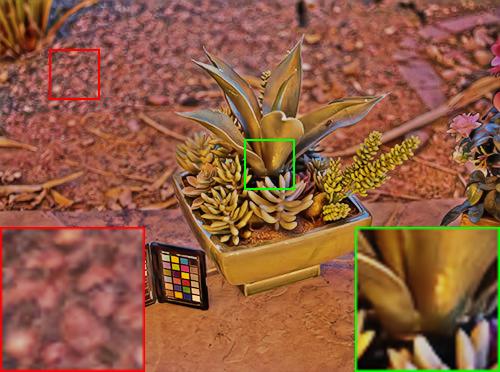}}
{\footnotesize (c) KinD++~\cite{2021KinD++}}
\end{minipage}
\begin{minipage}[t]{0.23\textwidth}
\centering
\raisebox{-0.15cm}{\includegraphics[width=1\textwidth]{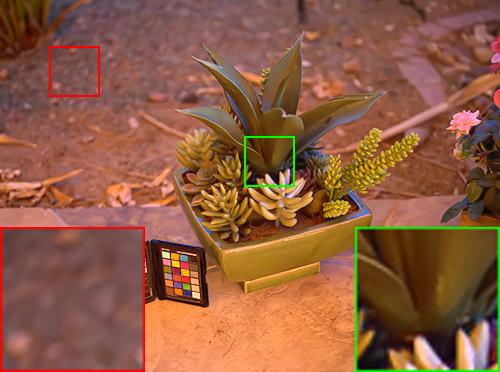}}
{\footnotesize (d) NLHD (Ours)}
\end{minipage}
}\vspace{-3mm}
\caption{Comparison of the low-light image enhancement results among the two latest methods and our method. The highlighted parts in the red and green boxes are used to compare the sharpening phenomenon. One can see that our method hardly introduces a sharpening phenomenon under the premise of effective brightness enhancement. The original low-light image \textbf{(a)} and the brightness enhanced images by STAR~\cite{star_tip2020} \textbf{(b)}, KinD++~\cite{2021KinD++} \textbf{(c)}, and our NLHD \textbf{(d)}, respectively.\label{fig1}}\vspace{-3mm}
\end{figure}

\IEEEPARstart{T}{he} images captured under low-light environments contain unavoidable quality degradation such as poor visibility and intense noise~\cite{Wei_2020_CVPR}, limiting the accuracy of subsequent applications like object recognition~\cite{buchsbaum1980spatial} and detection~\cite{YOLO,li2018dsfd,glenn_jocher_2021_4679653}. Intuitively, the visibility of the images captured in poor-light environments could be rescued by some camera work of good flash, high ISO, and the long exposure time, etc. However, these techniques produce other image degradation such as color deviation (by flash), intense noise (by high ISO), or blurring (by long exposure time), etc. To this end, many low-light image enhancement methods have been developed to achieve brightness enhancement on the captured low-light images.  

Generally, a better low-light image enhancement method can effectively improve the visibility of the dark regions in a low-light image, which is usually achieved by increasing the brightness of the dark regions. In order to obtain satisfactory perceptual quality, the naturalness of the enhanced images is essential~\cite{npe_tip2013}, especially, the image should not be over-sharpened in the procedure of brightness enhancement. Traditional low-light image enhancement methods are mainly formulated under the Retinex framework~\cite{ssr1997,msrcr1997,npe_tip2013,pmiesire_tip2015,lime_tip2017,mf2016,jiep2017,li2018structure,star_tip2020}. The early Retinex decomposition is achieved by smoothing techniques~\cite{ssr1997,msrcr1997}, based on the assumption that the illumination component is piece-wisely smooth~\cite{ssr1997}. However, the over-smoothed illumination component must lead to a large amount of structure and texture information getting into the reflectance component, which further causes the image to sharpen within the brightness enhancement procedure. Since the early Retinex model based method regards illumination component removal as a default choice and does not limit the range of the reflectance component, it can not effectively maintain the naturalness of non-uniform illumination image~\cite{npe_tip2013}. 

To simultaneously estimate the illumination and reflectance components, in recent years, some new methods employ more reasonable image priors~\cite{pmiesire_tip2015,jiep2017,star_tip2020} or variational model~\cite{wvm2016} for low-light image enhancement, with the help of local derivative decomposition techniques. These methods decompose the low-light image into the illumination and reflectance components and then manipulate the illumination component to achieve the image enhancement.
The recently proposed Structure and Texture Aware Retinex Model (STAR)~\cite{star_tip2020} decomposes the image through local derivative to alleviate the over smoothing of illumination component, which can reduce the sharpening phenomenon of the enhanced images to a certain extent, an enhanced image by STAR can be seen from Fig.~\ref{fig1}(b).
Most of deep learning based low-light image enhancement methods~\cite{SICE2018,drde_bmvc2018,mbllen_bmvc2018,2021KinD++} are still based on the Retinex model ones. The illumination component is also obtained by local filtering~\cite{SICE2018} or convolution operation~\cite{drde_bmvc2018,mbllen_bmvc2018,2021KinD++}. Some of these methods still over smooth the illumination component. The enhanced images by the latest Kindling the Darkness++ (KinD++)~\cite{2021KinD++} method also produces an intense sharpening phenomenon. Please refer to the highlighted parts by the red and green boxes, respectively, in Fig.~\ref{fig1}(c). 

In summary, most existing methods can only represent the local singularity because the decomposition operation is restricted to local image blocks or patches. This kind of local filtering method is challenging to control the smoothness of illumination components effectively and generally generates excessive smoothness illumination components. Compared with the above local methods, the non-local filtering method~\cite{NLM2005,bm3d2007,comments_BM3D,nlh_tip2020} can effectively control the smoothness, which has an excellent performance in image denoising in recent years. The application of the non-local idea to Retinex decomposition may also be a meaningful attempt.

This paper proposes a new pixel-level non-local Haar decomposition (NLHD) based Retinex model and further develops a new low-light image enhancement method. The NLHD method makes full use of the pixel-level non-local self-similarity in the images. The block-matching and row-matching operations are utilized to get 2D similar pixel groups; the Haar transform on these similar pixel groups can appropriately decompose the image into illumination and reflectance components to more effectively alleviate image sharpening in the image brightness enhancement procedure; an enhanced image by the proposed NLHD method can be seen from Fig.~\ref{fig1}(d).  Furthermore, a minimum fusion strategy is explored to fully use the advantages of logarithmic transformation and exponential transformation simultaneously to alleviate the mosaic artifacts in the darker regions and reduce over-exposure in the brighter regions.

The extremely dark low-light images or extremely dark regions in a low-light image generally accompany intense noise, such as the LOL dataset~\cite{drde_bmvc2018}. This paper proposes a local noise level estimation method and a simplified NLH version to suppress noise effectively. Noticing that the enhanced images of extremely dark low-light images usually suffer a color deviation problem, we also propose a new non-local saturation reduction based color deviation correction method to achieve better subjective visual quality and objective evaluation performance of the enhanced extremely dark low-light images.

The main contributions of this paper are summarized in three aspects:
\begin{itemize}
\vspace{-0mm}
    \item We take full advantage of the NLH frequency decomposition ability to achieve more appropriate illumination and reflectance components in the Retinex model. 
    \vspace{-0mm}
    \item We propose to enhance illumination components by exponential transform and logarithmic transform synchronously, and a novel minimum fusion strategy to achieve much better low-light image enhancement results than most existing methods.
    \item A local noise level estimation method and a simplified NLH version are utilized to effectively suppress noise in the  low-light image enhancement procedure. A new non-local saturation reduction based color deviation correction method help achieve better subjective visual quality and objective evaluation performance of the enhanced extremely dark low-light images.
    \vspace{-0mm}
\end{itemize}

The rest of the paper is organized as follows: \S\ref{Related Work} summarizes the related works, and \S\ref{Proposed Method} elaborates the proposed NLHD method. \S\ref{noise and color} presents the proposed noise suppression method and color deviation correction method. Then, \S\ref{Experiments} presents experimental details and reports comparison results on Retinex decomposition and low-light image enhancement. As a practical application, we conduct face detection experiments on enhanced low-light images of Dark Face dataset~\cite{ug2_dark_face} to verify the effectiveness of the proposed method in \S\ref{face detection}. We also present the ablation study of the proposed method in \S\ref{Ablation Study}. Finally, we make a conclusion remark in \S\ref{Conclusion}.

\section{Related Work}
\label{Related Work}

\subsection{Retinex Model}

Simplified Retinex theory assumes that we can decompose an observed image into an illumination component and a reflectance component. As a seminal work, Single Scale Retinex (SSR)~\cite{ssr1997} and Multi-Scale Retinex (MSR)~\cite{msrcr1997} implement Gaussian filtering on the images, take the low-frequency results as the illumination component, and remove the filtered results from the original images to obtain reflectance component. However, these methods are prone to introduce color distortion and halo artifacts in regions with large brightness deviations. In addition, the over-smoothed illumination component results in excessive structural and texture information getting into the reflectance component so that the enhancement of the reflectance component produces an intense sharpening phenomenon.  

In recent years, the Retinex model is still widely applied to the low-light image enhancement task. Multi-scale derived images Fusion (MF)~\cite{mf2016} adjusts the illumination component by fusing multiple derivatives of its initial estimation. Low-Rank Regularized Retinex Model (LR3M)~\cite{LR3M} suppresses the noise in the reflectance component by adding a low-rank constraint into the decomposition model but misses enough details from the illumination component. A Weighted Variational Model (WVM)~\cite{wvm2016} was proposed to estimate the reflectance and illumination simultaneously. The Joint intrinsic-extrinsic Prior model (JieP)~\cite{jiep2017} enhances the illumination component through exponential transformation and obtains ideal low-light image enhancement results. Low-Light Image Enhancement via Illumination Map Estimation (LIME)~\cite{lime_tip2017} estimates the brightness component of each pixel by finding the maximum value in the RGB channels. The structure prior is added to the initial illumination component to refine it into the final illumination component. Finally, the enhanced image is generated according to the illumination component. The Structure and Texture Aware Retinex model (STAR)~\cite{star_tip2020} well estimates both the illumination and reflectance components for image enhancement.  Because the above methods preserve part of the structure and texture information in the illumination component, the sharpening phenomenon has been alleviated. 

Unlike previous Retinex-based methods, our Non-local Haar Decomposition (NLHD) method makes full use of the pixel-level non-local self-similarity prior to decompose the low-light images; it can achieve more appropriate illumination and reflectance components. Because the illumination component preserves much more structure and texture information, the enhanced image has almost no sharpening phenomenon. 
\subsection{Deep Low-light Image Enhancement}
Deep learning has achieved great success in many fields. Many deep learning based low-light image enhancement methods have also been developed in recent years. 

Retinex Decomposition based Generative Adversarial Network (RDGAN)~\cite{rdgan_icme2019} integrates the Retinex model into a deep learning framework, splitting the Generative Adversarial Network (GAN)~\cite{GAN2014} into a Retinex decomposition network module and a fusion reinforcement network module. Deep Retinex Decomposition for Low-Light Enhancement Network (Retinex-Net)~\cite{issrlle_acmmm2020} combines Retinex model with semantic information perception to enhance segmented image regions. Low-light Net (LLNet)~\cite{LLNet2017} achieves the low-light image enhancement based on a deep auto-encoder network. Multi-Branch Low-Light image Enhancement Network  (MBLLEN)~\cite{mbllen_bmvc2018} extracts image features at different levels, enhances the image through multiple sub-networks, and finally generates the output image through multi-branch fusion. Kindling the Darkness (KinD)~\cite{kind_acmmm2019} added an adjustment network to adjust the illumination, which alleviates the problem of overexposure. Deep Light Enhancement Generative Adversarial Network (EnlightenGAN)~\cite{enlightengan_tip2021} uses the information extracted from the input image itself to regularize the unpaired training and proposes a global-local discrimination structure, a self regularized perceptual loss fusion and attention mechanism; four non-reference loss functions are designed to satisfy the non-reference training. Because most of the above deep learning based methods combine the Retinex model, the Retinex decomposition is still achieved by convolution operation without considering non-local prior; the over-smoothed illumination component still causes sharpening.
\subsection{Non-local Self-Similarity}
Generally, there are many similar image blocks in an image, which may not come from the same local region. This property is usually called non-local self-similarity. Non-local self-similarity was initially proposed by the Non-local Means (NLM)~\cite{NLM2005} to solve the image denoising problem. NLM estimates each pixel by computing a weighted average of some similar pixels obtained by implementing image block-matching, so NLM is a spatial domain method. BM3D~\cite{bm3d2007,comments_BM3D} is a transform domain non-local method that combines the non-local idea with discrete wavelet transform (DWT) and discrete cosine transform (DCT). BM3D is a patch-level non-local method, i.e., the signal in a single image patch is still local, so it is not an utterly non-local method. In order to better apply the non-local idea to the image denoising problem, NLH~\cite{nlh_tip2020} proposed to further search similar pixels in a similar image patch group and can achieve better image denoising results. 

In the NLH method, a similar image block group is obtained by block-matching as the same as NLM and BM3D, then scan each image block to a column vector, stack all the vectors to get a 2D matrix, finally implement row-matching to obtain similar pixel groups. A separable Haar transform is implemented on each similar pixel group to achieve the signal decomposition.

In this paper, we apply the NLH idea to decompose the image into illumination and reflectance components and further achieve low-light image enhancement. The proposed image decomposition method makes full use of the pixel-level non-local self-similarity in the images, so the enhanced images by our method look much more natural than the enhanced images by most existing local filtering based Retinex decomposition methods, especially the enhanced images by our method have almost no sharpening phenomenon.  

\begin{figure*}[t!] 
\flushleft
\includegraphics[width=1\textwidth]{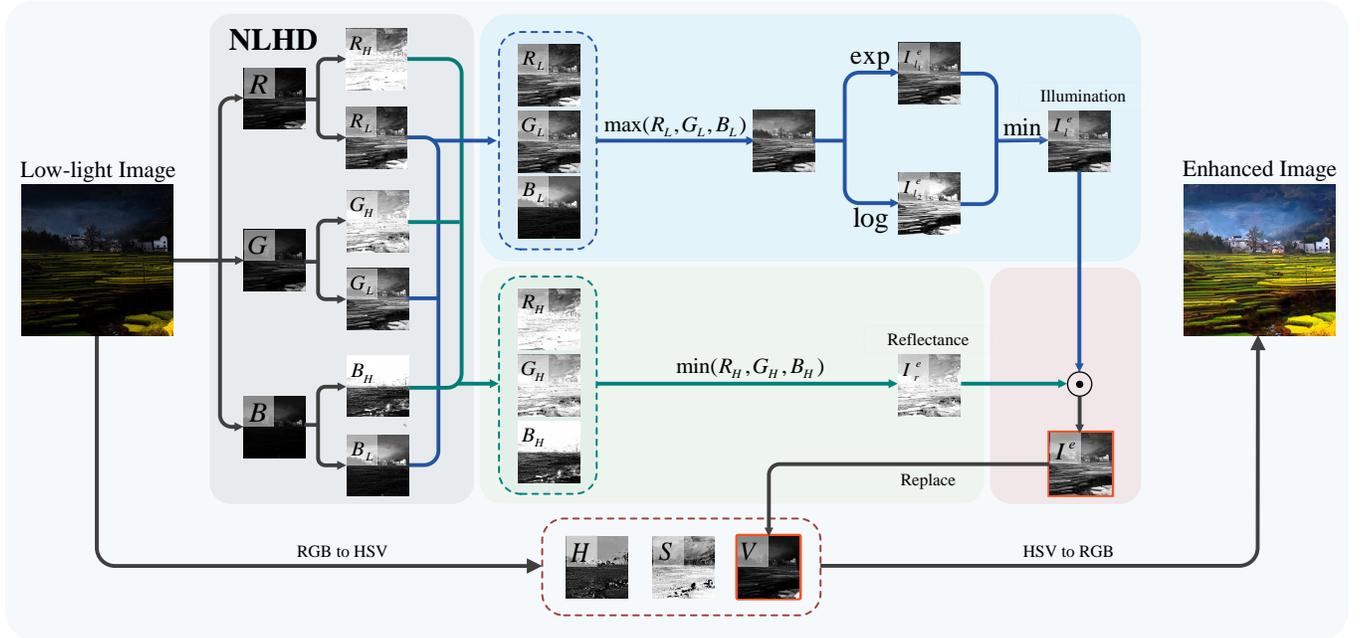}
\caption{Illustration of the non-local Haar decomposition based Retinex model for low-light image enhancement. In order to completely show the whole frame, we scale down the size of each channel image. Where $\max$ and $\min$ are the maximum and minimum values fusion operation, respectively. $\exp$ and $\log$ are exponential transformation and logarithmic transformation,  respectively. $\odot$ means element-wise multiplication operation. The specific details of the NLHD can be seen in Fig.~\ref{Decomposition}.}
\label{Frame}
\end{figure*}

\begin{figure*}[ht] 
\flushleft
\includegraphics[width=1\textwidth]{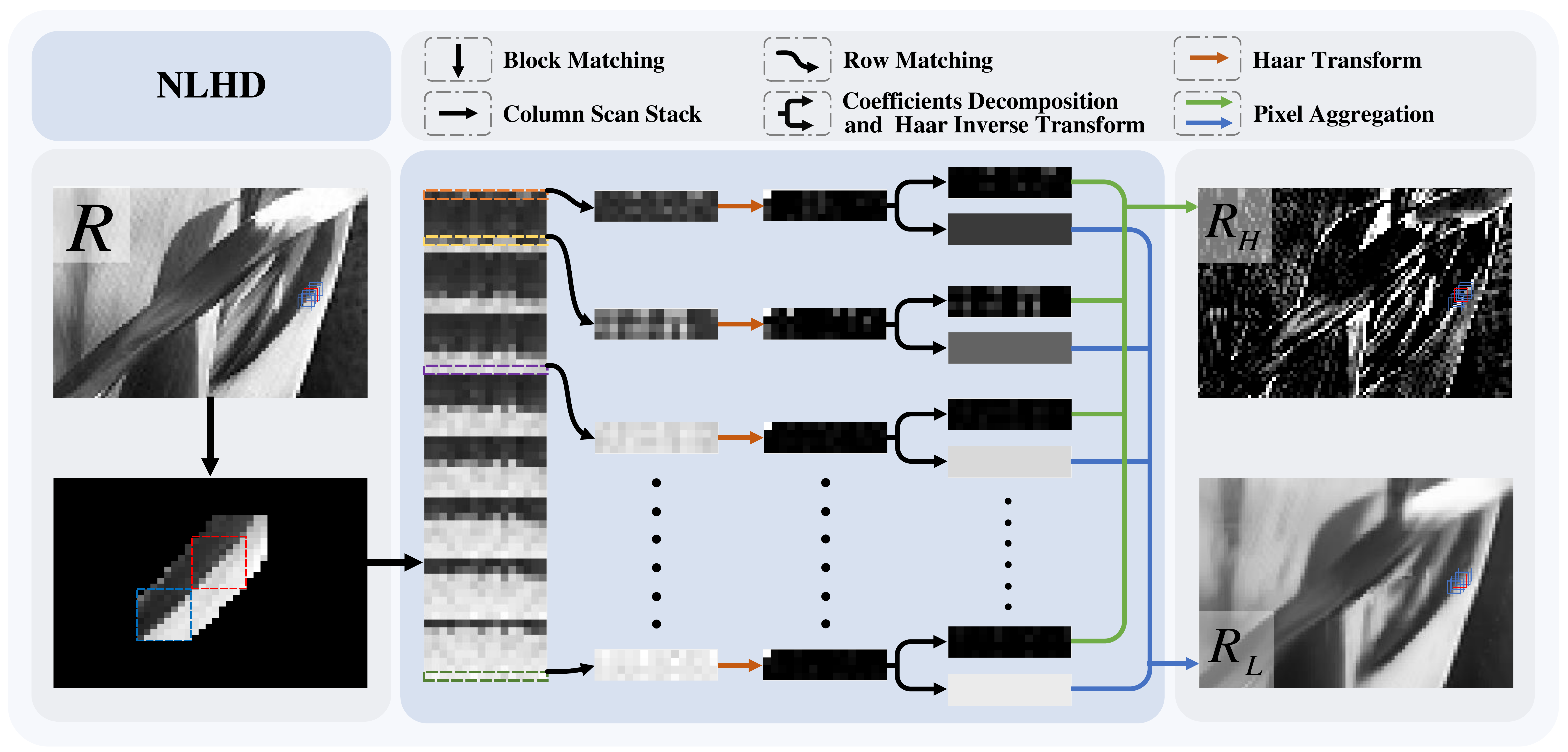}
\caption{Illustration of the non-local Haar decomposition in Fig.~\ref{Frame}. We only give the R channel situation here, G channel and B channel are similar to R channel. In order to improve the visual effect of $R_{H}$, we increased its brightness to a certain extent.}
\label{Decomposition}
\end{figure*}
\section{Proposed Method}\label{Proposed Method}
The Retinex model has been proved to be an effective low-light image enhancement method, more appropriate illumination and reflectance decomposition is the key of the Retinex model. In order to better apply the Retinex model to achieve more effective low-light image enhancement, a new pixel-level non-local Haar transform based illumination and reflectance decomposition (NLHD) method is proposed in this section.

The proposed method includes the following four steps: (1) The similar pixel groups are obtained by image block-matching and row-matching; (2) The Haar transform is implemented on each similar pixel group, and the illumination component is obtained by low-frequency reconstruction of the transform coefficients, the high-frequency coefficients reconstruction obtains the reflectance component; (3) The illumination components are enhanced by exponential transform and logarithmic transform respectively, and then use a minimum fusion strategy on both of the enhanced illumination components to get the final enhanced illumination components; (4) The Retinex model is used to achieve the image enhancement by combining the enhanced illumination and reflectance components. Fig.~\ref{Frame} shows the flow chart of the proposed low-light image enhancement method, and the details of NLHD can be seen in Fig.~\ref{Decomposition}.

\subsection{Background on Retinex Model}\label{Background on Retinex Model}
Retinex model is developed based on the human visual cognition system. The model decomposes the image into two components: illumination and reflectance, i.e.,
\begin{equation}\label{dot}
O = I \odot R
\end{equation}
where $O$ is an image, $I\in \mathbb{R}^{m\times n}$ is the illumination component of $O$, representing the incident light image, namely the brightness of the light on the object surface in the image, which determines the dynamic range that the pixel value can reach in the image; $R\in \mathbb{R}^{m\times n}$ is the reflectance property image of the objects, representing the internal attribute of the image, namely the structure of the objects in the image. The basic idea of the Retinex model is to remove or reduce the influence of incident light on the image by some methods to preserve the object's natural attribute as much as possible. Therefore, estimating or decomposing the illumination component $I$ and reflectance component $R$ is the key problem of the Retinex model.
\par
In essence, the illumination component $I$ represents the brightness of the object surface in the image, which comes mainly from the smoother regions of the image; However, the reflectance component $R$ is generally distributed in the small structural parts such as texture in the image. Therefore, the Retinex decomposition should be attributed to better decomposing the image's smooth regions and texture regions. Based on this prior knowledge, we propose a new Retinex decomposition method based on the pixel-level non-local Haar transform in this paper. We present the details of the proposed method in~\S\ref{pixel group} and~\S\ref{HTID}.

\subsection{Similar Pixel Group}\label{pixel group}
Given a low-light color image $I\in \mathbb{R}^{h\times w\times c}$ in RGB color space. We implement block-matching and row-matching in RGB space to obtain similar pixel groups. We firstly compute the mean value of a reference image block in each R, G, B channel, according to the maximum mean value to decide the block-matching channel $C_{m}$. According to a certain sliding step size, a reference image block $B_{r}$ with the size of $\sqrt{n}\times \sqrt{n}$ is selected in $C_{m}$ to implement block-matching by Euclidean distance to get a similar block group with $N_{2}$ blocks, then extract the same position's blocks in other two channels, so we can get three similar block groups. Each image block with the size of $\sqrt{n}\times \sqrt{n}$ is stretched into a column vector $V_{l}\in \mathbb{R}^{\sqrt{n}\times \sqrt{n}}(l = 1,...,N_{2})$ in each channel respectively. All $V_{l}$ are stacked   into a matrix $M_{b}$ with $\sqrt{n}\times \sqrt{n}$ rows and $N_{2}$ columns, we finally get three $M_{b}$ all together in three color channels respectively.
In order to better mine the self-similarity in the images, we further implement row-matching in each $M_{b}$. Each row $R_{r}$ is used as the reference row to calculate the Euclidean distance with all other rows to find the most similar $N_{3}-1$ rows, put together $R_{r}$ with the most similar $N_{3}-1$ rows to construct a dimension $N_{3}\times N_{2}$ similar pixel matrix $M_{s}$. Specifically, for the i-th row as the reference row, the Euclidean distance between the i-th row and all other rows is calculated as follows,

\begin{equation} \label{1}
d_{ij} = \left \| M_{b}^{i} - M_{b}^{j}\right \|_{2}
\end{equation} 

\subsection{Haar Transform and Image Decomposition}\label{HTID}
In order to effectively decompose the image into illumination and reflectance components in the Retinex model, we implement separable Haar transform on the similar pixel group $M_{s}$, i.e., the vertical and horizontal separable lifting transformation. The transform can be represented as the following,
\begin{equation} \label{2}
    C_{h} = H_{l}\times M_{s}\times H_{r}
\end{equation}
where $C_{h}$ is the transform spectral matrix of the Haar transform, $H_{l}$ and $H_{r}$ is the Haar transform matrix.
Due to the characteristics of separable lifting Haar transform, $C_{h}(1,1)$ is the weighted average of all the pixels in the $M_{s}$, which we define as the low-frequency coefficient, we only use $C_{h}(1,1)$ to reconstruct the ideal illumination component $I_{l}$ by the inverse Haar transform. Instead, use the rest of $C_{h}-C_{h}(1,1)$ coefficients to reconstruct the reflectance component $I_{r}$ by the inverse Haar transform. The specific detail of NLHD is shown in Fig.~\ref{Decomposition}.

\subsection{Illumination and Reflectance Enhancement}\label{component enhancement}
Because the illumination component and the reflectance component have their respective characteristics, we respectively implement different enhancement operations on illumination component $I_{l}$ and reflectance component $I_{r}$. Three reflectance components $I_{r_{r}}$, $I_{r_{g}}$ and $I_{r_{b}}$ can be initially obtained by $R$, $G$ and $B$ channels respectively, one can use the minimum value fusion of these reflectance components as the final reflectance component $I_{r}$, but the following method can achieve the best image enhancement results in our experiments,
\begin{equation} \label{3}
I_{r} = \min(I_{r_{r}},|I_{r_{g}}|,|I_{r_{b}}|)
\end{equation}
where $| \cdot |$ is the absolute value sign, and we finally get the enhanced reflectance component $I_{r}^{e}$ by simply adding 1.0 to $I_{r}$, i.e., $I_{r}^{e} = 1.0+I_{r}$.

As for illumination component $I_{l}$, we also initially obtained three illumination components $I_{l_{r}}$, $I_{l_{g}}$  and $I_{l_{b}}$, we firstly get bright illumination component $I_{l}$ by maximum value fusing  among the above three illumination components as the following, 
\begin{equation} \label{4}
I_{l} = \max(I_{l_{r}},I_{l_{g}},I_{l_{b}})
\end{equation}
\par We use different index ${\gamma_{1}}$ and ${\gamma_{2}}$ to achieve $I_{l}$ enhancement respectively, where ${\gamma_{1}}$ and ${\gamma_{2}}$ are both adaptively calculated as follows,
\begin{equation} \label{5}
{\gamma_{1}} = \min\bigg(\frac{||{I_{l}(I_{l}<\alpha_{1})}||_{0}}{||{I_{l}}||_{0}},\beta_{1}\bigg)
\end{equation}
we get ${\gamma_{2}}$ by the two following situations:
if~$\frac{||{I_{l}(I_{l}<\theta_{1})||_{0}}}{||{I_{l}(I_{l}<\theta_{2})||_{0}}}>\theta$
\begin{equation} \label{6}
{\gamma_{2}} = \min\bigg(\frac{||{I_{l}(I_{l}<\alpha_{2})}||_{0}}{||{I_{l}}||_{0}},\beta_{2}\bigg)
\end{equation}
else
\begin{equation} \label{7}
{\gamma_{2}} = \min\bigg(\frac{||{I_{l}(I_{l}<\alpha_{3})}||_{0}}{||{I_{l}}||_{0}},\beta_{2}\bigg)
\end{equation}
where $||\cdot||_{0}$ means the number of non-zero elements. In this stage, we set $\alpha_{1} = 0.35$, $\alpha_{2} = 0.005$, $\alpha_{3} = 0.05$, $\beta_{1} = 0.45$, $\beta_{2} = 0.61$, $\theta = 0.9$, $\theta_{1} = 0.05$, $\theta_{2} = 0.15$.
Two different enhanced $I_{l_{1}}^{e}$ and $I_{l_{2}}^{e}$ can be obtained by the following formulas,
\begin{equation} \label{8}
I_{l_{1}}^{e} = (I_{l})^{\gamma_{1}}
\end{equation}
\begin{equation} \label{9}
I_{l_{2}}^{e} = \frac{1}{\gamma_{2}}\log_{2}(1+I_{l})+I_{l}
\end{equation}
Here, if we only use exponential transformation to enhance the illumination component, the extremely dark regions in the low-light images are generally over-enhanced, some mosaic artifacts are usually introduced, the noise is amplified as well. If we only use logarithmic transformation to enhance the illumination components, the relatively bright regions in the low-light images usually present over-exposure phenomenon. In order to solve these problems, we propose to enhance the illumination component as the following,
\begin{equation} \label{10}
I_{l}^{e} = \min(I_{l_{1}}^{e},I_{l_{2}}^{e})
\end{equation}
\par Because real-world low-light images present various illumination levels, better-enhanced results of some extremely dark images could not be achieved just by the above procedure. We use a further algorithm~\ref{algorithm1} to achieve balanced enhancement results on various initial illumination levels.

\begin{algorithm}[t]
	\caption{Illumination Component Enhancement}
	\label{algorithm1}
	\begin{algorithmic}[1]
		\STATE \textbf{Input:} low-light image $\bm{O}$, parameters $\gamma_1,\gamma_2$,\\ 
\quad\quad\quad mean value $m_O$ of $\bm{O}$, \\
 \quad\quad\quad  maximum iteration number $K$: \\
 \quad\quad\quad       $K = (1.0-m_O)\times 30$;
\\
\STATE \textbf{Initialization}:  Decomposed $I_l^0$ by NLHD;\\
\STATE
\quad
Compute $\gamma_1$ by Eqn. (\ref{5});\\
\STATE
\quad
Compute $\gamma_2$ by Eqn. (\ref{6}-\ref{7});\\
		\FOR{$k = 0,\cdots ,K-1$}
		\STATE Update $I_l^{k+1}$ by  Eqn. (\ref{8}-\ref{10});
		\IF {mean($I_l^{k+1}) > 0.6 $ or min($I_l^{k+1}) > 0.1$}
		\STATE Stop;
		\ELSE \item {$I_l^{k+1}$ = $I_l^{k}$ + $0.15 \times I_l^{0}$};
		\ENDIF 
		\ENDFOR
		\STATE \textbf{Output}: Enhanced illumination component $I_l^e$ = $I_l^{k+1}$.
	\end{algorithmic}
\end{algorithm}

\par Finally, the illumination component and the reflectance component are applied to the Retinex model, that is, 
\begin{equation} \label{11}
{I^{e} = I_{l}^{e}\odot I_{r}^{e}}
\end{equation}
where $I^{e}$ is the initial enhancement image, $\odot$ is the element-wise multiplication operation. The V channel in HSV color space is replaced by $I^{e}$, then HSV color space is converted to RGB color space to achieve the final RGB enhancement image.
\section{Noise Suppression and Color Deviation Correction }\label{noise and color}
The extremely dark low-light images or extremely dark regions in a low-light image generally accompany intense noise, such as the LOL dataset~\cite{drde_bmvc2018}, the noise is usually further amplified in the process of enhancement, it seriously influences the subjective visual quality and objective evaluation of the enhancement results. Therefore, how to effectively suppress the noise in the low-light images is also a significant problem. In this paper, we propose a local noise level estimation method and a simplified NLH version to effectively suppress noise for achieving better low-light image enhancement.
\par The enhanced images of extremely dark low-light images usually suffer color deviation problems as well. In this section, we also propose a new non-local saturation reduction based color deviation correction method to achieve better subjective visual quality and objective evaluation performance of the enhanced images.
\subsection{Noise Suppression}
The noise distribution in the low-light images is usually uneven between brighter regions and darker regions. The darker the regions, the noise is the stronger. So the traditional image denoising method is not entirely suitable for the low-light image noise suppression problem. The pixel-level non-local Haar transform (NLH) was initially proposed to solve the image denoising problem. However, NLH estimates noise level on the whole image, so it is unsuitable for low-light image noise estimation. 
\par In this paper, we further develop a new local noise estimation method and combine it with the NLH method to achieve better noise suppression in the low-light image enhancement procedure. The proposed method has a little bit different from the NLH method; we use the minimum row-matching distances to estimate local noise level, further use the mean value of the similar pixel group, the local noise level, and the standard deviation of minimum row-matching distances to decide the hard-thresholding parameter value. The local noise standard deviation $\sigma$ is computed as the following,

\begin{equation} \label{12}
 \sigma = \sqrt{\frac{1}{N_1^2}\sum\limits_{n=1}^{N_1^2}d_{n}^2}
\end{equation}
where $d_{n}$ is the minimum row distance in each row-matching procedure referring to Eq.\ref{1}, $N_1$ is the size of image block. 
\par We adaptively select the hard-thresholding parameter according to the similar pixel group mean value $m_{s}$. The smaller the mean value, the larger the hard-thresholding parameter. Besides, we also use minimum row-matching distance standard deviation to adaptively decide hard-thresholding parameter; the larger the distance standard deviation value, the smaller hard-thresholding parameter to better preserve the edge details in images, such as textures and contours.
The local minimum row-matching distance standard deviation can be calculated as the following, 

\begin{equation} \label{13}
 \sigma_d  = \sqrt{\sum\limits_{n=1}^{N_1^2}\Bigg(d_{n}-\frac{1}{N_1^2}\sum\limits_{n=1}^{N_1^2}d_{n}\Bigg)^2}
\end{equation}

\par We use parameters $N_1 = 6$, $N_{step} = 5$, $N_s = 13$, $N_2 = 16$, $N_3 = 4$ in the noise suppression procedure, where $N_1$ is the image block size, $N_{step}$ is the sliding step size of reference image block, $N_s$ is the block-matching searching neighbor radius,  $N_2$ is the number of similar image blocks, and $N_3$ is the number of the rows in each similar pixel group. The hard-thresholding parameter $Thr$ is computed as the following,
\begin{equation} \label{14}
Thr = \frac{\sigma}{m_s(k\sigma_d)}
\end{equation}
where $m_s$ is the mean value of the similar pixel group, the parameter $k$ adjusts $\sigma_d$ value such that it is close to the value of $m_s$, $k=1000$ in our experiments. The role of this formula is that we allocate larger $Thr$ in the darker regions in the images; this corresponds to the fact that the darker the regions, the stronger the noise. Besides, the larger $\sigma_d$ means the richer image details, so we should allocate the smaller $Thr$ to better preserve image details. On the contrary, we should use larger $Thr$ to improve the denoising ability. Meanwhile, we also use structural hard-thresholding similar to the original NLH method. We only implement two steps iteration here and remove the Wiener filtering stage in the original NLH method.
Because we use a group of low complexity parameters and the simplified procedure, the running speed of this algorithm is relatively much faster than the original NLH denoising method. 

\subsection{Color Deviation Correction}
 In this subsection, we propose a non-local saturation reduction method to attenuate the color deviation problem. CIELab color space can be effectively used to measure the color deviation degree; we transform denoised RGB low-light image to CIELab color space, then compute the mean value of $a$ channel and $b$ channel, the global color deviation degree $D_c$ can be measured as the following, 
 \begin{equation} \label{15}
D_{c}=\sqrt{M_a^2+M_b^2}
\end{equation}
where $M_a$ and $M_b$ are the mean values of channel $a$ and channel $b$ respectively. In channel $a$, the positive values represent the red color; the negative values refer to the green color. Similarly, the positive values in channel $b$ represent the yellow color; the negative values refer to the blue color. The zero value of $D_c$ means that the color is well-balanced; the larger $D_c$ value means the stronger color deviation in the image.
\par We propose a pixel-level non-local color deviation correction method in this paper, input the $S$ channel of HSV space to illumination decomposition procedure, synchronously using the block-matching and row-matching results to get similar pixel group $G_{S}$ in $S$ channel, using the formula $\hat{G}_{S}=G_{S}^\gamma$ to reduce the saturation, where $\gamma > 1$, the larger $\gamma$ value, the stronger degree of the saturation reduction. The $\gamma$ is computed as the following,
\begin{equation} \label{16}
    \gamma = min(1 +\frac{ k \times D_{c}}{min_m+std_m},\alpha) 
\end{equation}
where $min_m = min(m_R,m_G,m_B)$ is the minimum value of similar pixel group mean values $m_R$, $m_G$ and $m_B$ for three channels in RGB color space, $std_m = std(m_R,m_G,m_B)$ is the standard deviation of these three mean values, $k$ is an empirical parameter used to control the saturation reduction, we take its value as 0.013 in our experiments. $\alpha$ is an upper limitation to void the over reduction of the saturation; we take its value as 4.5 in our experiments. Because the darker images or the darker regions in the images usually result in the stronger color deviation, and the smaller standard deviation among the three color channels means the stronger color deviation as well, so we use the larger $\gamma$ value to reduce the saturation in these two situations. All the block-matching and row-matching parameters are the same ones as the illumination decomposition procedure.
\section{Experiments}
\label{Experiments}
In this section, we qualitatively and quantitatively evaluate the Retinex decomposition performance of the proposed NLHD method. We also conduct ablation experiments on some essential parameters of the proposed method to further explore and gain deep insight into the proposed NLHD model. All codes were run on a MacBook Pro laptop with an Intel Core i9-9900K CPU and 32GB memory except for KinD++~\cite{2021KinD++}, the code of KinD++ was run on a computer equipped with Tesla-V100 GPU.
\par \textbf{Datasets.}
In order to verify the effectiveness of the proposed NLHD Retinex decomposition method, we conduct extensive comparison experiments on low-light image enhancement problem. We use 35 low-light images collected in~\cite{jiep2017,pmiesire_tip2015,wvm2016,lime_tip2017} in~\cite{star_tip2020} that is called \textbf{35images} and 200 random extracted images from Dark Face dataset~\cite{ug2_dark_face} that is called \textbf{200darkface} here respectively. We also use the \textbf{LOL} dataset~\cite{drde_bmvc2018} including 500 extremely dark low-light images and their corresponding high-light version to further verify the effectiveness of the proposed noise suppression and color deviation correction method.

\textbf{Objective metrics.}\label{Metrics}
We use different objective metrics to evaluate the performance of the low-light image enhancement. The comparison methods are evaluated according to the following two common indicators, the method of measuring natural image quality without reference image quality assessment (NIQE)~\cite{niqe} and the lightness order error (LOE)~\cite{npe_tip2013} between the original image $I$ and its enhanced version $I^{e}$. The smaller the NIQE value is, the higher the image quality. The smaller the LOE value is, the better the lightness order is preserved. Because there is no ground truth in 35images and 200darkface datasets, we can not compute the peak signal-to-noise ratio (PSNR) and structural similarity index (SSIM)~\cite{SSIM}. As for the LOL dataset, we compare SSIM, PSNR, and $\Delta$E on this dataset to achieve a more objective evaluation because there is ground truth in this dataset.   

\subsection{NLHD Based Retinex Decomposition}\label{Retinex decomposition}
The proposed NLHD method can decompose images to illumination and reflectance components in Retinex model, the low frequency coefficients are used to reconstruct the illumination component, the high frequency coefficients are used to reconstruct the reflectance component. We have both 5 parameters in each reconstruction, take patch size $\sqrt{n}=11$, number of similar patches $m=16$, number of rows in row matching $N_{3}=16$, sliding step size of reference patches $N_{step}=10$ and the window size $N_{s}=23$ for searching similar patches to obtain reflectance component in each RGB channels. $\sqrt{n}=6$, $m=8$, $N_{3}=2$, $N_{step}=6$ and $N_{s}=13$ are used to obtain the illumination component in each RGB channels. We use Eq.\ref{3} and Eq.\ref{4} to achieve reflectance component $I_r$ and illumination component $I_l$ respectively.

\begin{figure*} [htb]
\vspace{-5mm}
\centering
\subfigure{
\begin{minipage}[t]{0.24\textwidth}
\centering
\raisebox{-0.15cm}{\includegraphics[width=1\textwidth]{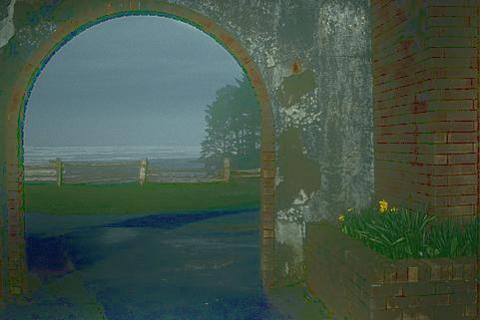}}
{\footnotesize (a) Illumination by SIRE~\cite{pmiesire_tip2015}}
\end{minipage}
\begin{minipage}[t]{0.24\textwidth}
\centering
\raisebox{-0.15cm}{\includegraphics[width=1\textwidth]{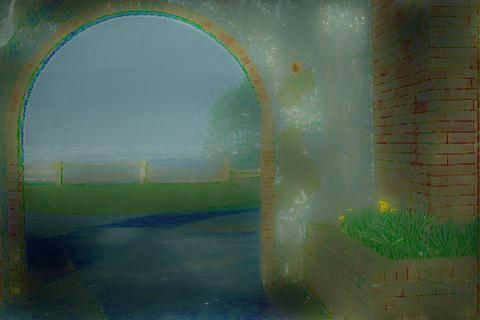}}
{\footnotesize (b) Illumination by WVM~\cite{wvm2016}}
\end{minipage}
\begin{minipage}[t]{0.24\textwidth}
\centering
\raisebox{-0.15cm}{\includegraphics[width=1\textwidth]{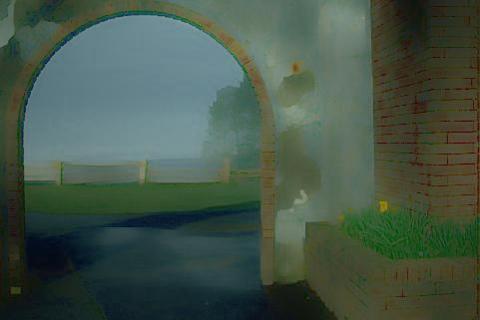}}
{\footnotesize (c) Illumination by JieP~\cite{jiep2017}}
\end{minipage}
\begin{minipage}[t]{0.24\textwidth}
\centering
\raisebox{-0.15cm}{\includegraphics[width=1\textwidth]{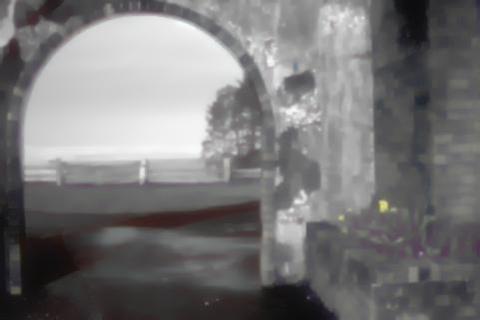}}
{\footnotesize (d) Illumination by RRM~\cite{li2018structure}}
\end{minipage}
}\vspace{-3mm}
\subfigure{
\begin{minipage}[t]{0.24\textwidth}
\centering
\raisebox{-0.15cm}{\includegraphics[width=1\textwidth]{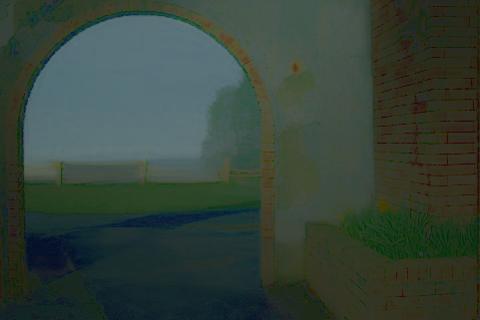}}
{\footnotesize (e) Illumination by LIME~\cite{lime_tip2017} }
\end{minipage}
\begin{minipage}[t]{0.24\textwidth}
\centering
\raisebox{-0.15cm}{\includegraphics[width=1\textwidth]{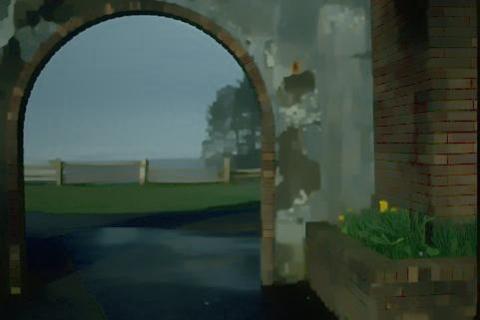}}
{\footnotesize (f) Illumination by STAR~\cite{star_tip2020} }
\end{minipage}
\begin{minipage}[t]{0.24\textwidth}
\centering
\raisebox{-0.15cm}{\includegraphics[width=1\textwidth]{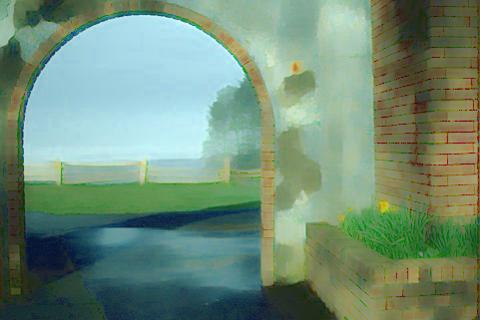}}
{\footnotesize (g) Illumination by KinD++~\cite{2021KinD++} }
\end{minipage}
\begin{minipage}[t]{0.24\textwidth}
\centering
\raisebox{-0.15cm}{\includegraphics[width=1\textwidth]{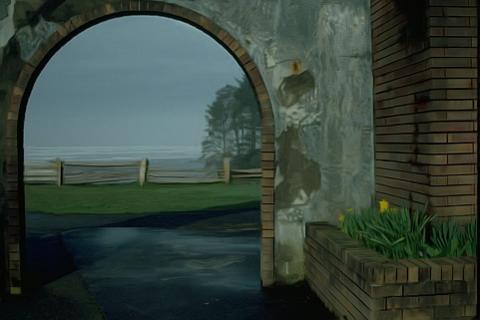}}
{\footnotesize (h) Illumination by our NLHD }
\end{minipage}
}\vspace{-3mm}
\subfigure{
\begin{minipage}[t]{0.24\textwidth}
\centering
\raisebox{-0.15cm}{\includegraphics[width=1\textwidth]{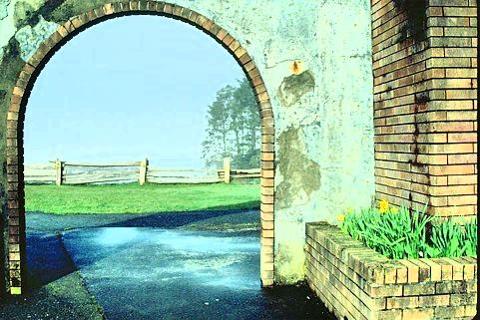}}
{\footnotesize (i) Reflectance by SIRE~\cite{pmiesire_tip2015} }
\end{minipage}
\begin{minipage}[t]{0.24\textwidth}
\centering
\raisebox{-0.15cm}{\includegraphics[width=1\textwidth]{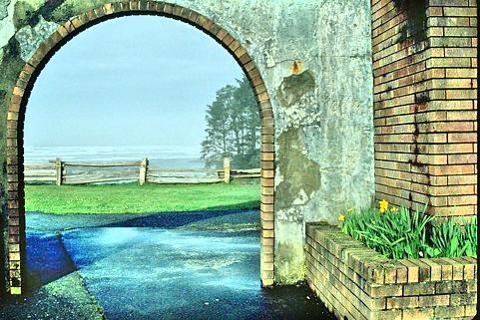}}
{\footnotesize (j) Reflectance by WVM~\cite{wvm2016} }
\end{minipage}
\begin{minipage}[t]{0.24\textwidth}
\centering
\raisebox{-0.15cm}{\includegraphics[width=1\textwidth]{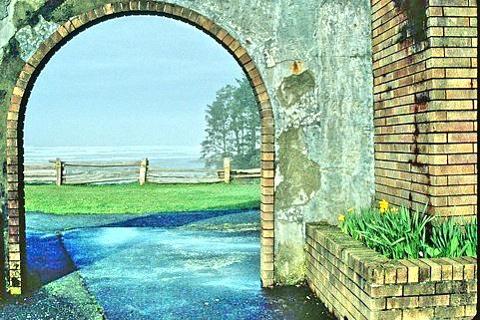}}
{\footnotesize (k) Reflectance by JieP~\cite{jiep2017} }
\end{minipage}
\begin{minipage}[t]{0.24\textwidth}
\centering
\raisebox{-0.15cm}{\includegraphics[width=1\textwidth]{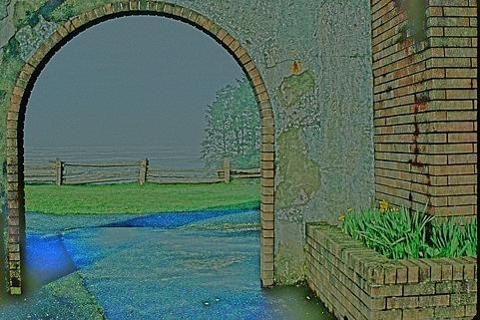}}
{\footnotesize (l) Reflectance by RRM~\cite{li2018structure} }
\end{minipage}
}\vspace{-3mm}
\subfigure{
\begin{minipage}[t]{0.24\textwidth}
\centering
\raisebox{-0.15cm}{\includegraphics[width=1\textwidth]{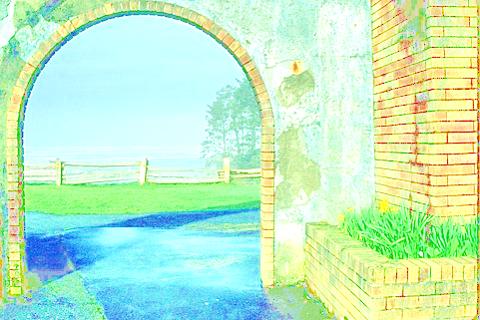}}
{\footnotesize (m) Reflectance by LIME~\cite{lime_tip2017} }
\end{minipage}
\begin{minipage}[t]{0.24\textwidth}
\centering
\raisebox{-0.15cm}{\includegraphics[width=1\textwidth]{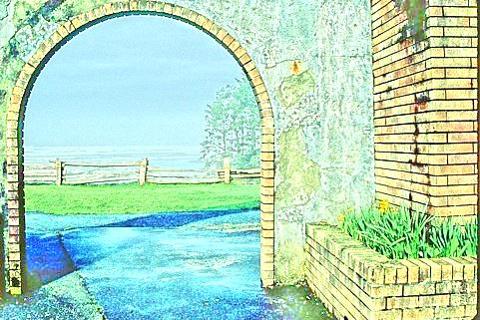}}
{\footnotesize (n) Reflectance by STAR~\cite{star_tip2020} }
\end{minipage}
\begin{minipage}[t]{0.24\textwidth}
\centering
\raisebox{-0.15cm}{\includegraphics[width=1\textwidth]{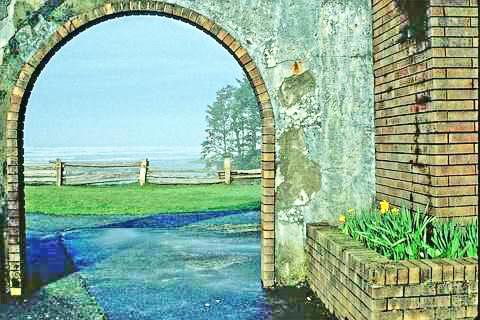}}
{\footnotesize (o) Reflectance by KinD++~\cite{2021KinD++} }
\end{minipage}
\begin{minipage}[t]{0.24\textwidth}
\centering
\raisebox{-0.15cm}{\includegraphics[width=1\textwidth]{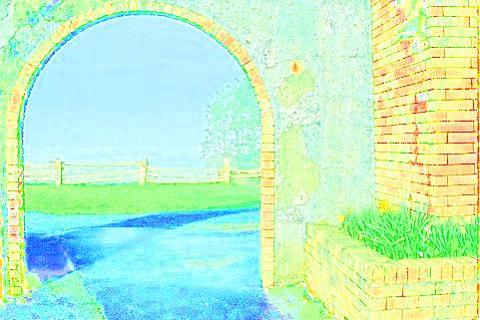}}
{\footnotesize (p) Reflectance by our NLHD }
\end{minipage}
}
\caption{ Comparisons of illumination and reflectance components obtained by various Retinex decomposition methods on the image ``2'' in the 35images dataset.}
\label{landr}
 \vspace{-2mm}
\end{figure*}

The general Retinex algorithms assume that the illumination component is usually smooth, but that is not necessarily the case; one can see from the recent Retinex decomposition methods~\cite{jiep2017,star_tip2020} that the illumination component preserves much larger structure information, but better low-light image enhancement results can be achieved. Generally, there is no ground truth of illumination and reflectance~\cite{star_tip2020}, so the quantitative evaluation of the existing Retinex decomposition method has always been a difficult problem. In order to evaluate the effectiveness of the proposed NLHD method, we qualitatively compare the illumination and reflectance with some of the most advanced Retinex models, including CRM~\cite{Ying_2017_ICCV}, SIRE~\cite{pmiesire_tip2015}, Weighted Variational Model (WVM)~\cite{wvm2016}, Joint intrinsic-extrinsic Prior (JieP) model~\cite{jiep2017}, Robust Retinex Model (RRM)~\cite{li2018structure}, NPE~\cite{npe_tip2013}, LIME~\cite{lime_tip2017}, STAR~\cite{star_tip2020}, and the latest deep learning based method KinD++~\cite{2021KinD++}. Similar to these methods, we replace the decomposed illumination and reflectance component to the V channel of HSV color space and return the decomposed components to RGB space. We compare the Retinex decomposition results of the above methods in Fig.~\ref{landr}. We can see that the proposed NLHD method can effectively achieve the smoothing effect and preserve main structure information in the illumination component; the reflectance component can better extract the tiny details of the low-light images. 

\subsection{Low-light Image Enhancement}\label{low-light image enhancement}
\begin{figure*}[ht!]
 \vspace{-1mm}
\centering
\begin{minipage}[t]{0.156\textwidth}
\vspace{12.5mm}
\subfigure{\raisebox{-0.15cm}{\includegraphics[width=1\textwidth]{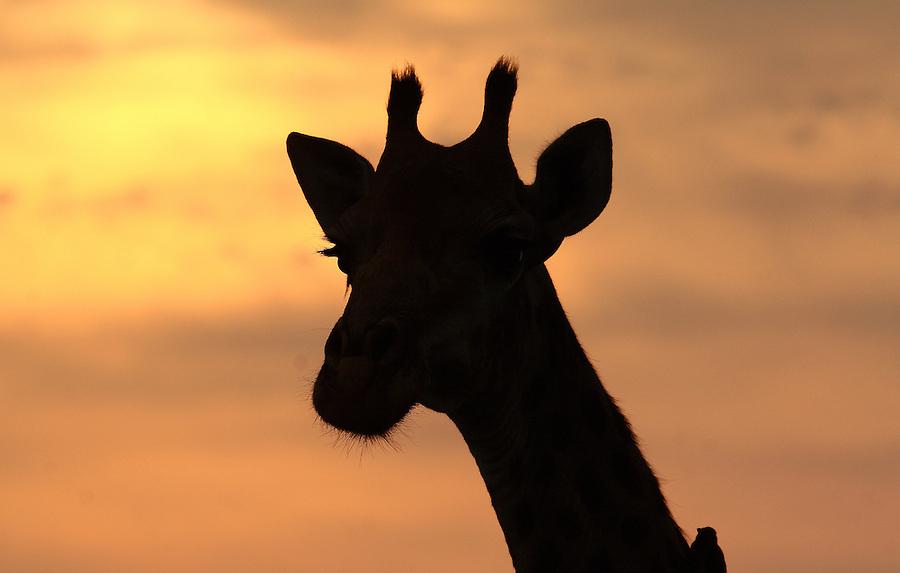}}}
\centering{\footnotesize (a) Original }
\end{minipage}
\begin{minipage}[t]{0.156\textwidth}
\subfigure{\raisebox{-0.15cm}{\includegraphics[width=1\textwidth]{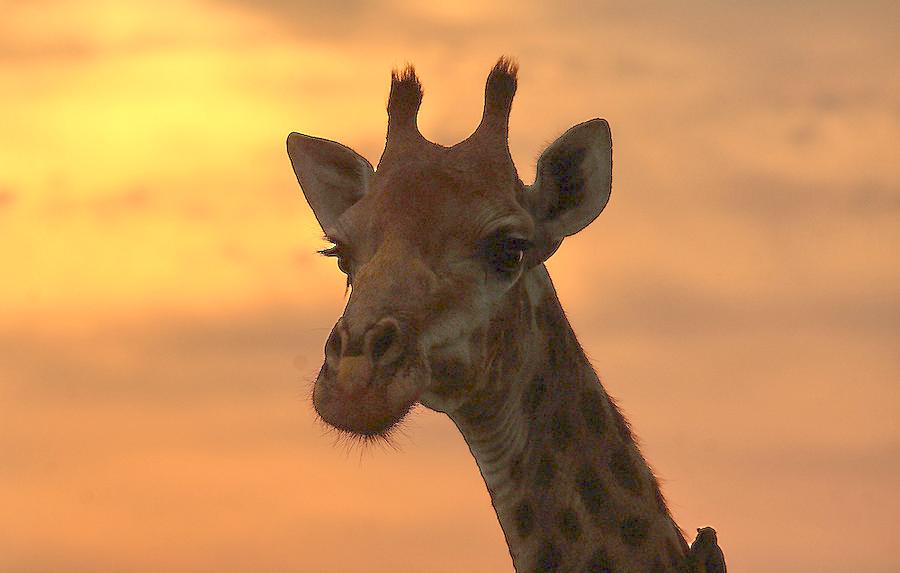}}}
\centering{\footnotesize (b) CRM~\cite{Ying_2017_ICCV} }
\subfigure{\raisebox{-0.15cm}{\includegraphics[width=1\textwidth]{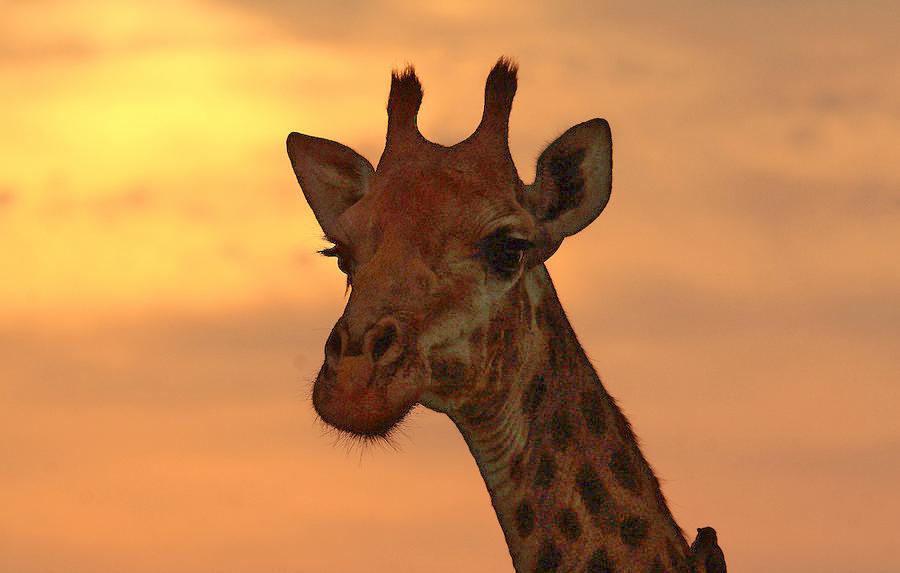}}}
\centering{\footnotesize (g) NPE~\cite{npe_tip2013} }
\end{minipage}
\begin{minipage}[t]{0.156\textwidth}
\subfigure{\raisebox{-0.15cm}{\includegraphics[width=1\textwidth]{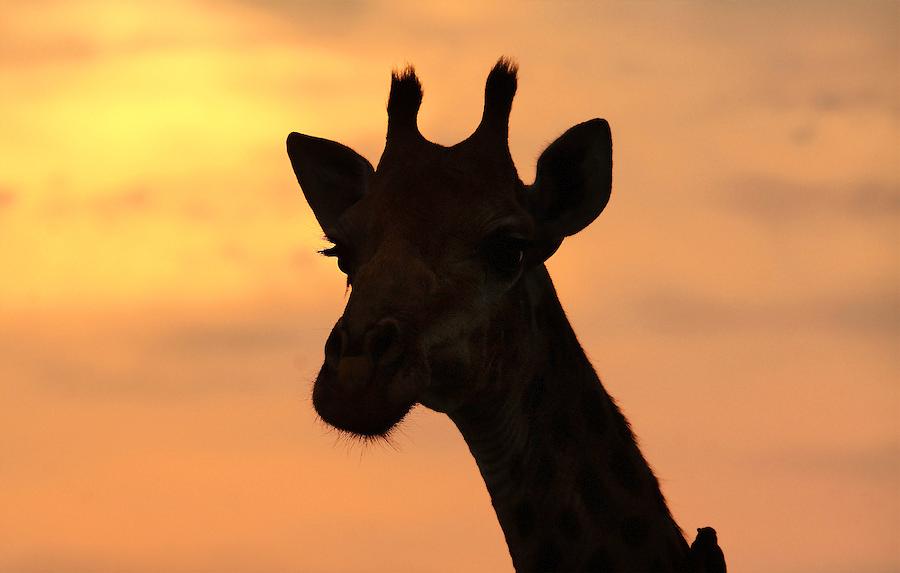}}}
\centering{\footnotesize (c) SIRE~\cite{pmiesire_tip2015} }
\subfigure{\raisebox{-0.15cm}{\includegraphics[width=1\textwidth]{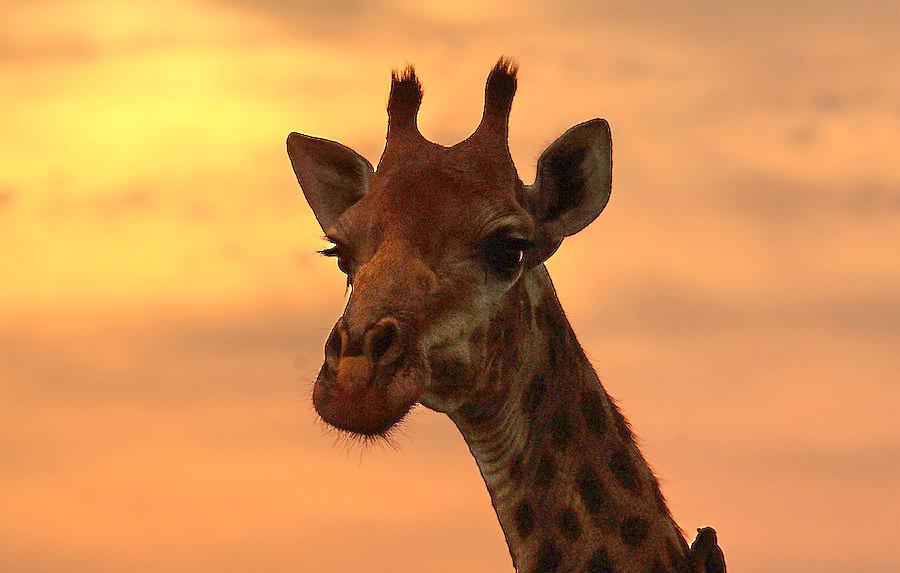}}}
\centering{\footnotesize (h) LIME~\cite{lime_tip2017} }
\end{minipage}
\begin{minipage}[t]{0.156\textwidth}
\subfigure{\raisebox{-0.15cm}{\includegraphics[width=1\textwidth]{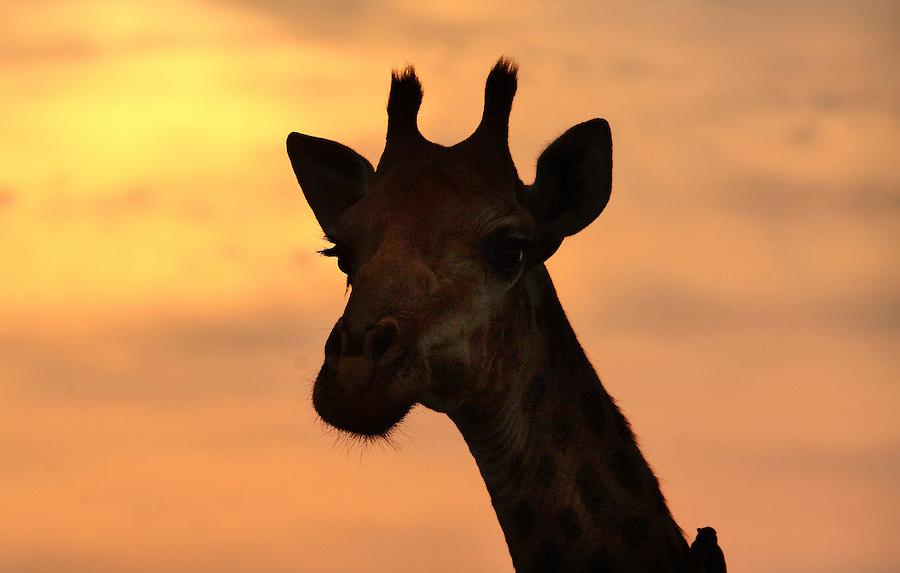}}}
\centering{\footnotesize (d) WVM~\cite{wvm2016} }
\subfigure{\raisebox{-0.15cm}{\includegraphics[width=1\textwidth]{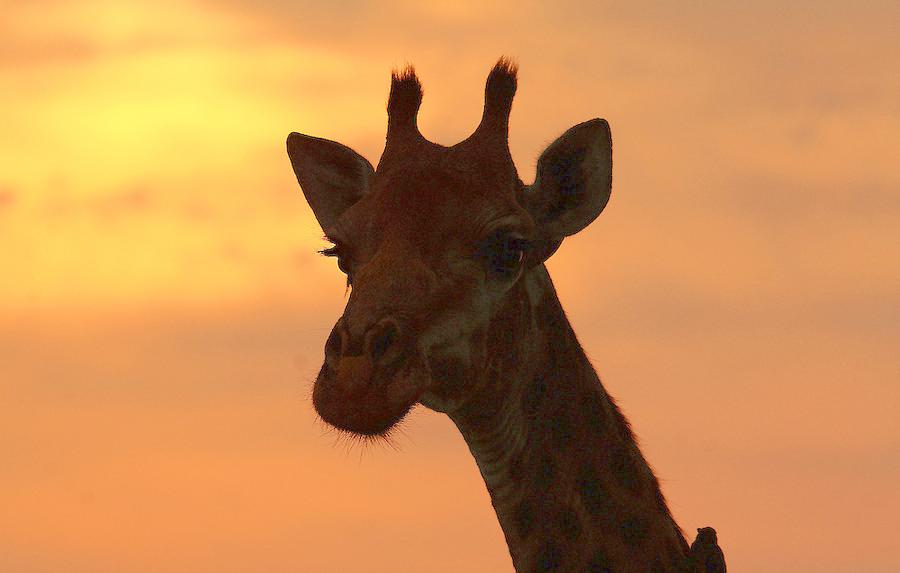}}}
\centering{\footnotesize (i) STAR~\cite{star_tip2020} }
\end{minipage}
\begin{minipage}[t]{0.156\textwidth}
\subfigure{\raisebox{-0.15cm}{\includegraphics[width=1\textwidth]{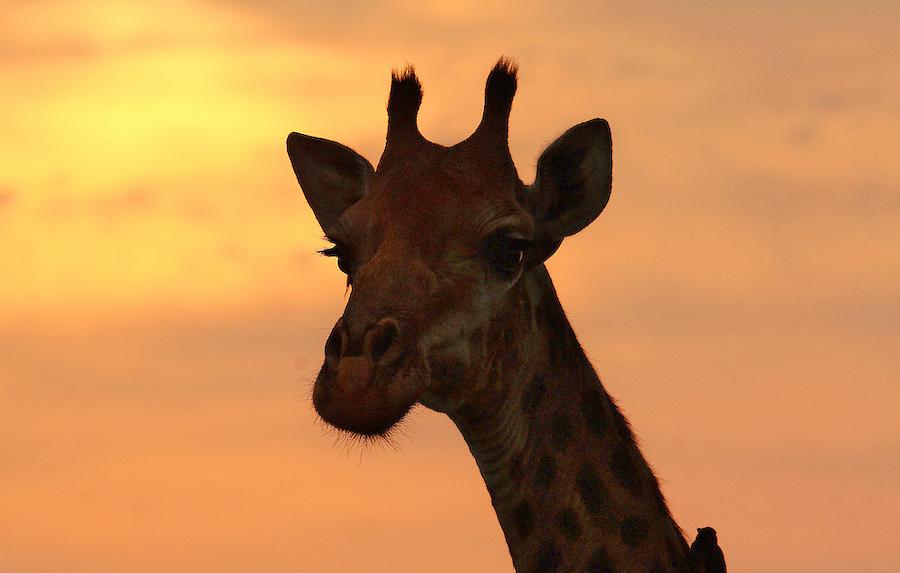}}}
\centering{\footnotesize (e) JieP~\cite{jiep2017} }
\subfigure{\raisebox{-0.15cm}{\includegraphics[width=1\textwidth]{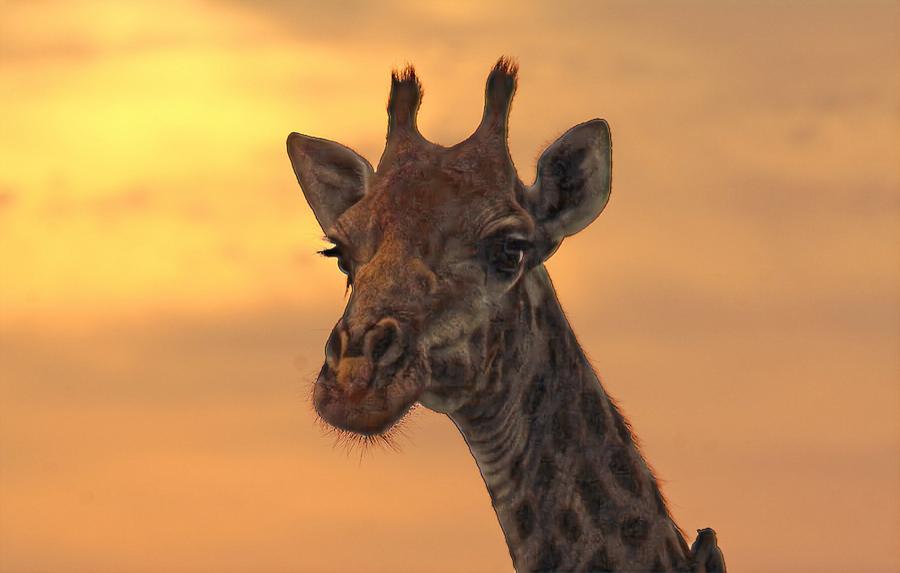}}}
\centering{\footnotesize (j) KinD++~\cite{2021KinD++} }
\end{minipage}
\begin{minipage}[t]{0.156\textwidth}
\subfigure{\raisebox{-0.15cm}{\includegraphics[width=1\textwidth]{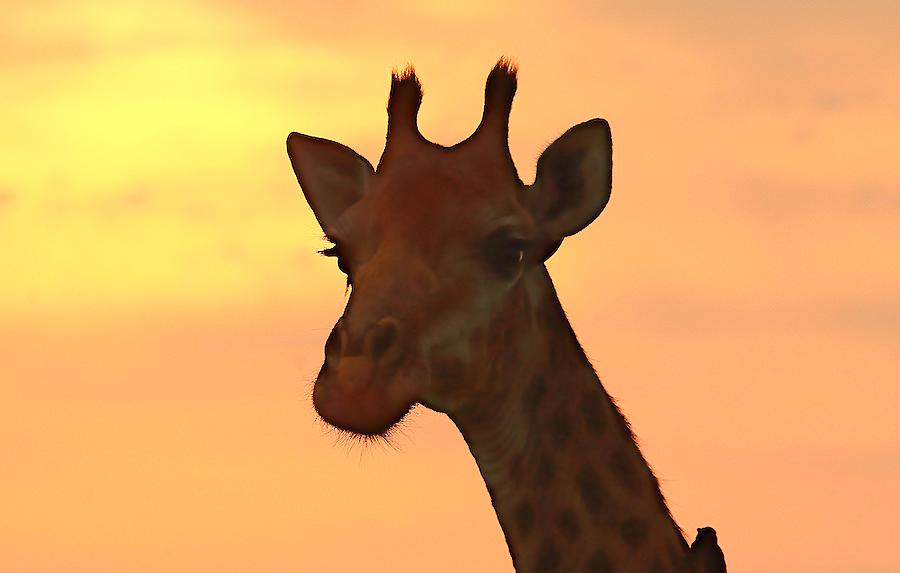}}}
\centering{\footnotesize (f) RRM~\cite{li2018structure} }
\subfigure{\raisebox{-0.15cm}{\includegraphics[width=1\textwidth]{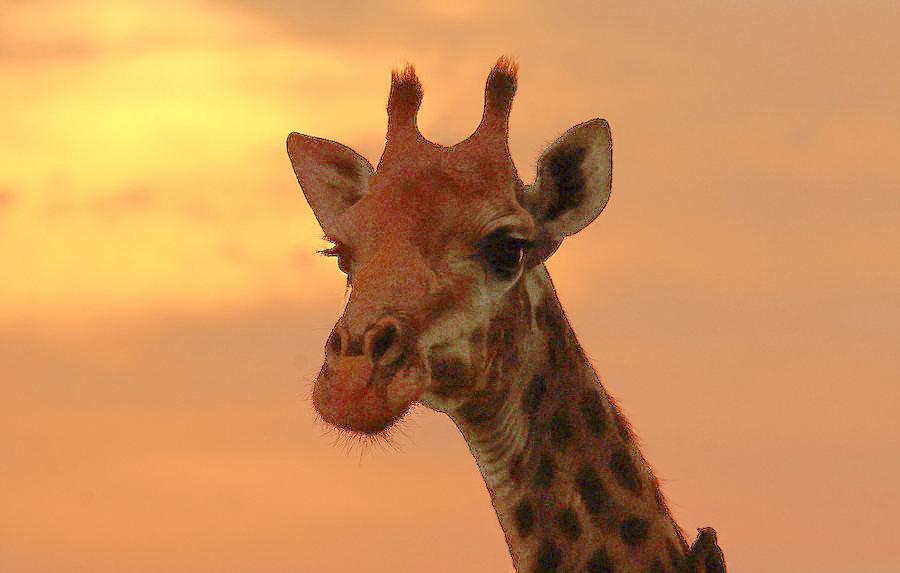}}}
\centering{\footnotesize (k) NLHD (ours) }
\end{minipage}
\vspace{-2mm}
\caption{Comparison of the result images on the image ``32'' in the 35images dataset enhanced by various low-light image enhancement methods.}
\label{giraffe}
\end{figure*}

\begin{figure*}[t]
 \vspace{-4mm}
\centering
\begin{minipage}[t]{0.156\textwidth}
 \vspace{20mm}
\subfigure{\raisebox{-0.15cm}{\includegraphics[width=1\textwidth]{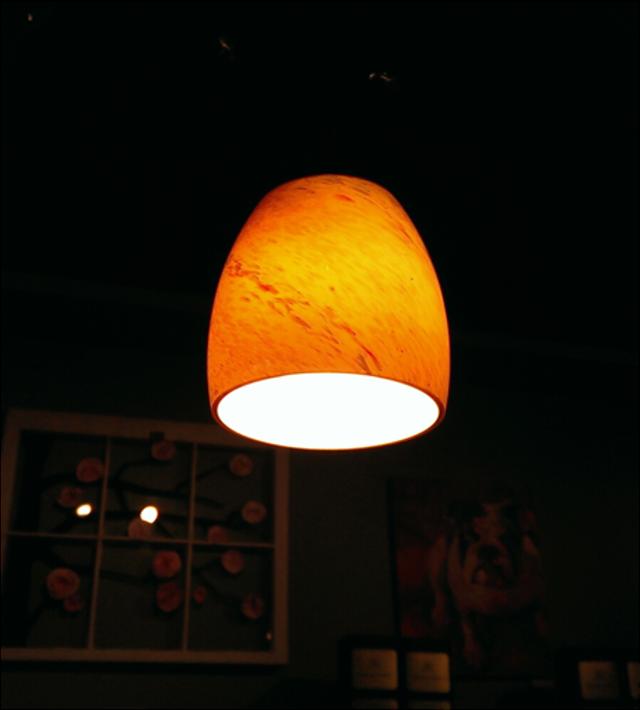}}}
\centering{\footnotesize (a) Original }
\end{minipage}
\begin{minipage}[t]{0.156\textwidth}
\subfigure{\raisebox{-0.15cm}{\includegraphics[width=1\textwidth]{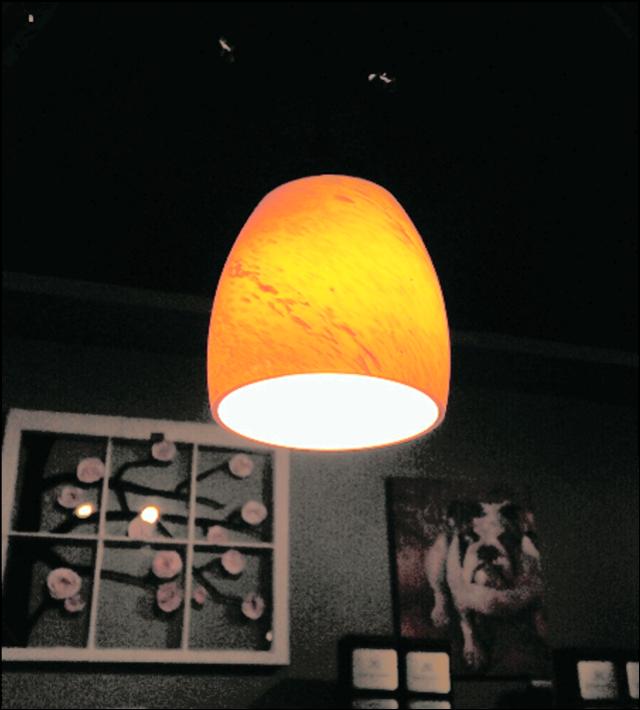}}}
\centering{\footnotesize (b) CRM~\cite{Ying_2017_ICCV} }
\subfigure{\raisebox{-0.15cm}{\includegraphics[width=1\textwidth]{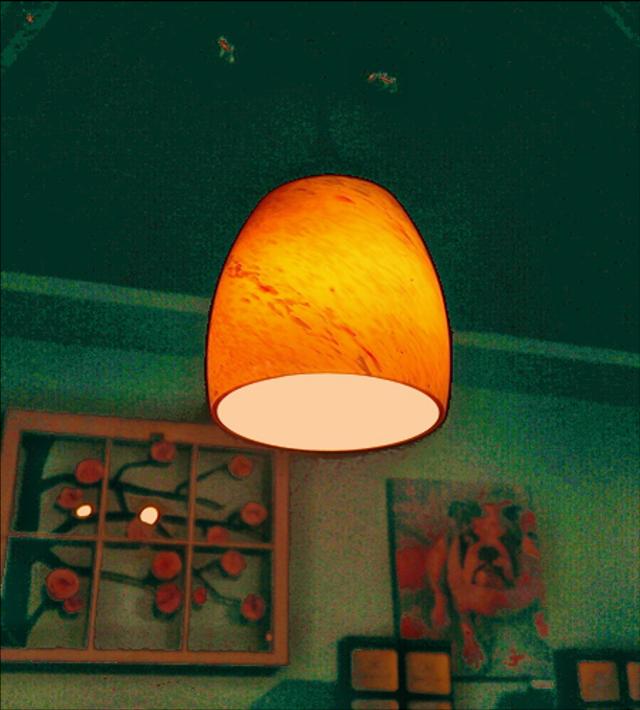}}}
\centering{\footnotesize (g) NPE~\cite{npe_tip2013} }
\end{minipage}
\begin{minipage}[t]{0.156\textwidth}
\subfigure{\raisebox{-0.15cm}{\includegraphics[width=1\textwidth]{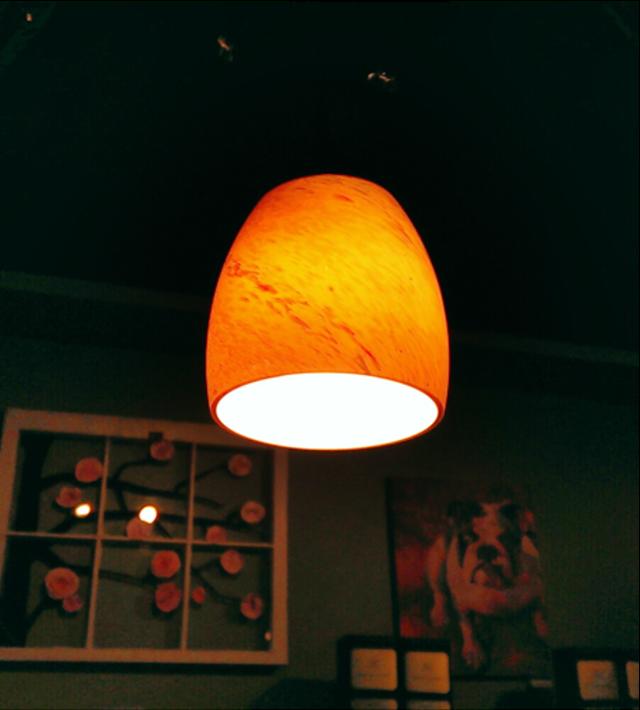}}}
\centering{\footnotesize (c) SIRE~\cite{pmiesire_tip2015}}
\subfigure{\raisebox{-0.15cm}{\includegraphics[width=1\textwidth]{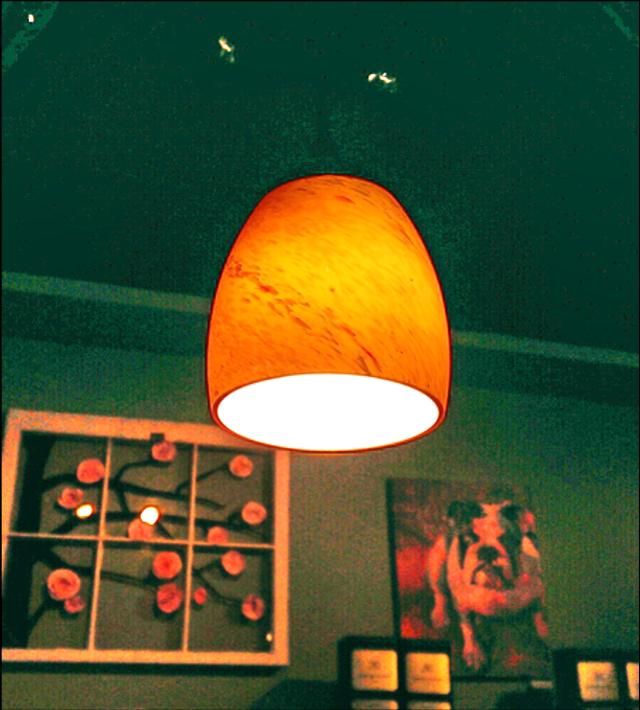}}}
\centering{\footnotesize (h) LIME~\cite{lime_tip2017} }
\end{minipage}
\begin{minipage}[t]{0.156\textwidth}
\subfigure{\raisebox{-0.15cm}{\includegraphics[width=1\textwidth]{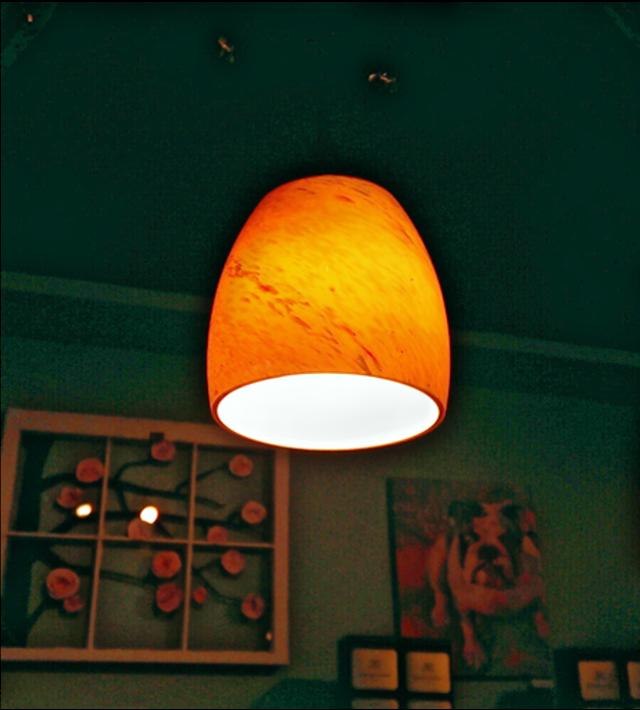}}}
\centering{\footnotesize (d) WVM~\cite{wvm2016} }
\subfigure{\raisebox{-0.15cm}{\includegraphics[width=1\textwidth]{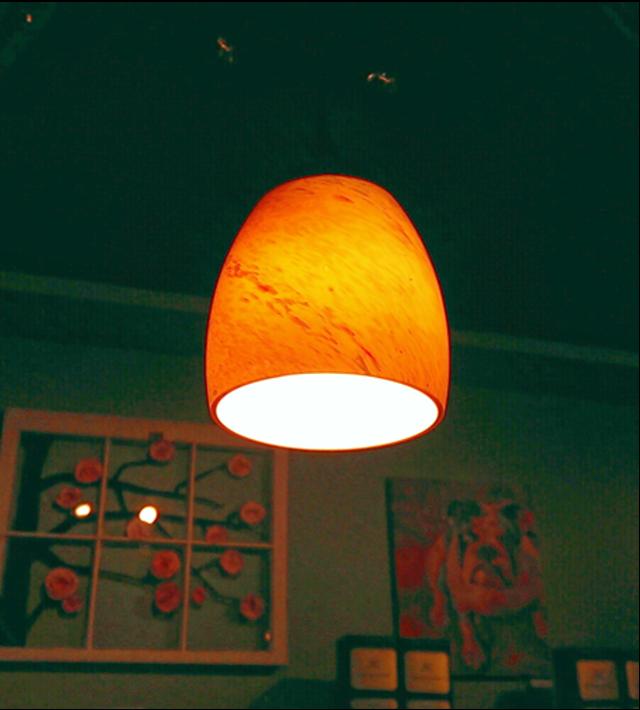}}}
\centering{\footnotesize (i) STAR~\cite{star_tip2020} }
\end{minipage}
\begin{minipage}[t]{0.156\textwidth}
\subfigure{\raisebox{-0.15cm}{\includegraphics[width=1\textwidth]{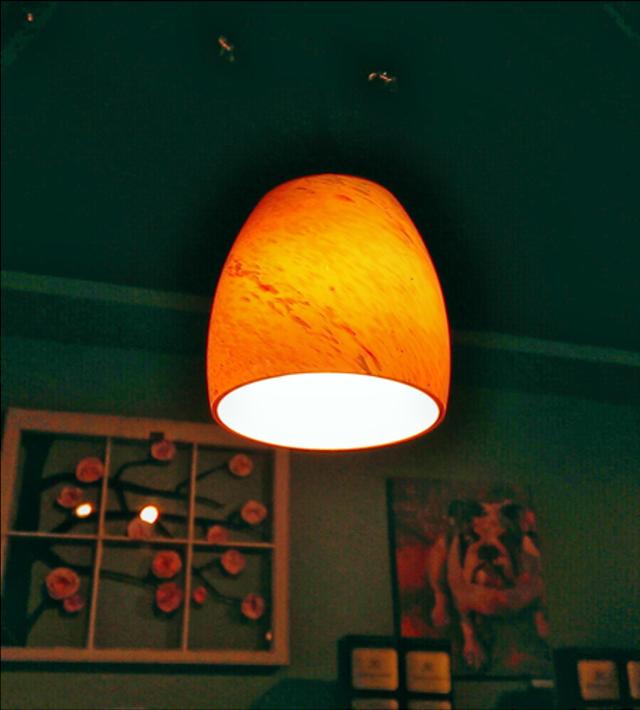}}}
\centering{\footnotesize (e) JieP~\cite{jiep2017} }
\subfigure{\raisebox{-0.15cm}{\includegraphics[width=1\textwidth]{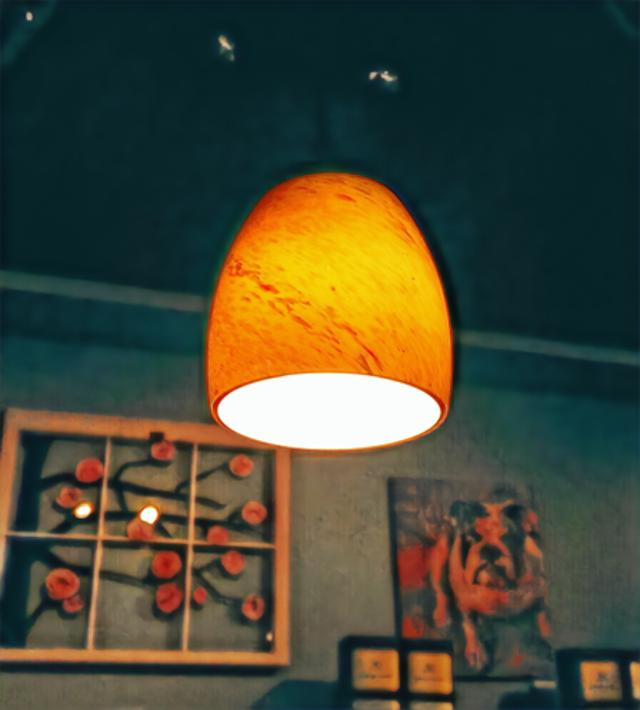}}}
\centering{\footnotesize (j) KinD++~\cite{2021KinD++}}
\end{minipage}
\begin{minipage}[t]{0.156\textwidth}
\subfigure{\raisebox{-0.15cm}{\includegraphics[width=1\textwidth]{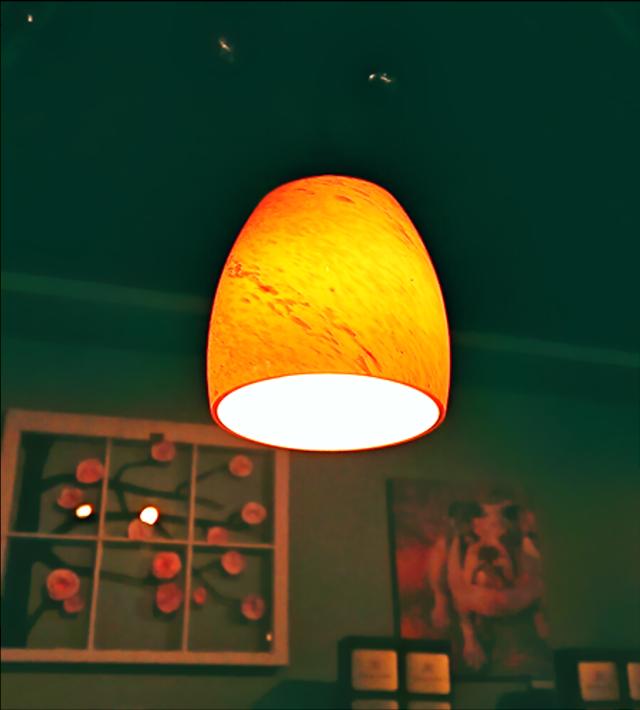}}}
\centering{\footnotesize (f) RRM~\cite{li2018structure} }
\subfigure{\raisebox{-0.15cm}{\includegraphics[width=1\textwidth]{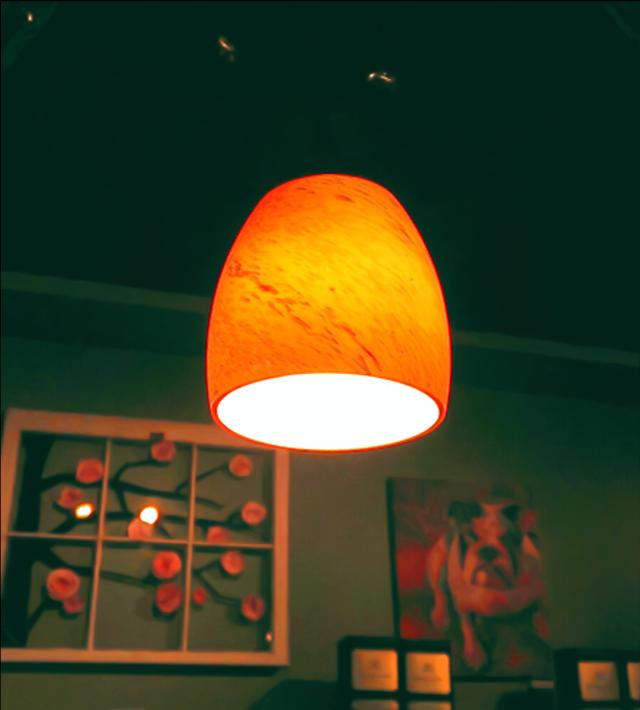}}}
\centering{\footnotesize (k) NLHD (ours) }
\end{minipage}
\vspace{-1mm}
\caption{\linespread{1}\selectfont{{Comparison of the result images on the image ``30'' in the 35images dataset enhanced by various low-light image enhancement methods.}}}
\label{light}
\vspace{-3mm}
\end{figure*}

\begin{figure*}[t]
\centering
\begin{minipage}[t]{0.156\textwidth}
\vspace{12.5mm}
\subfigure{\raisebox{-0.15cm}{\includegraphics[width=1\textwidth]{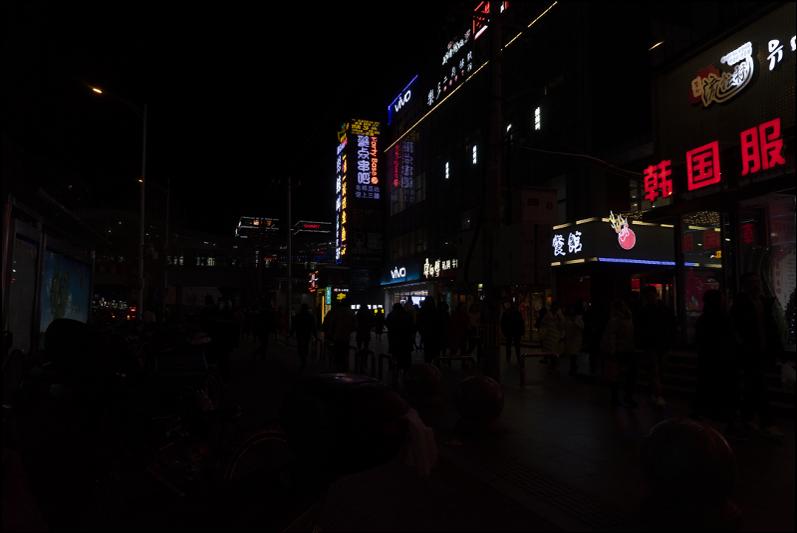}}}
\centering{\footnotesize (a) Original }
\end{minipage}
\begin{minipage}[t]{0.156\textwidth}
\subfigure{\raisebox{-0.15cm}{\includegraphics[width=1\textwidth]{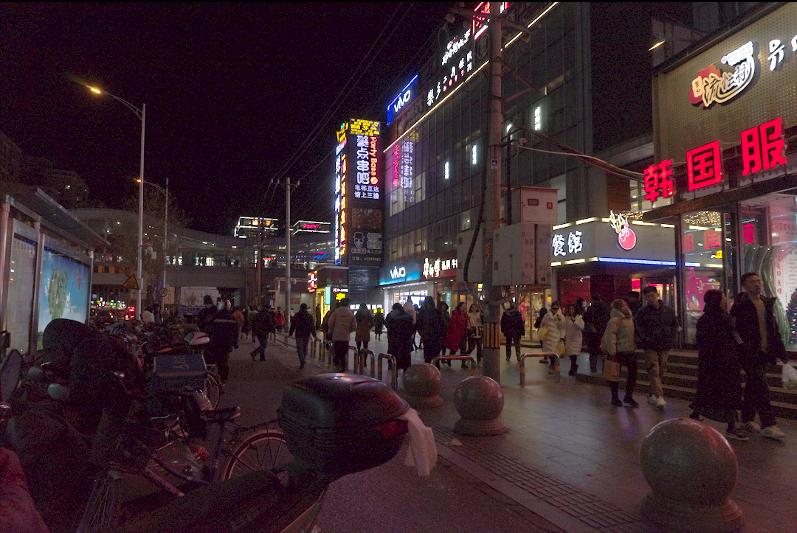}}}
\centering{\footnotesize (b) CRM~\cite{Ying_2017_ICCV} }
\subfigure{\raisebox{-0.15cm}{\includegraphics[width=1\textwidth]{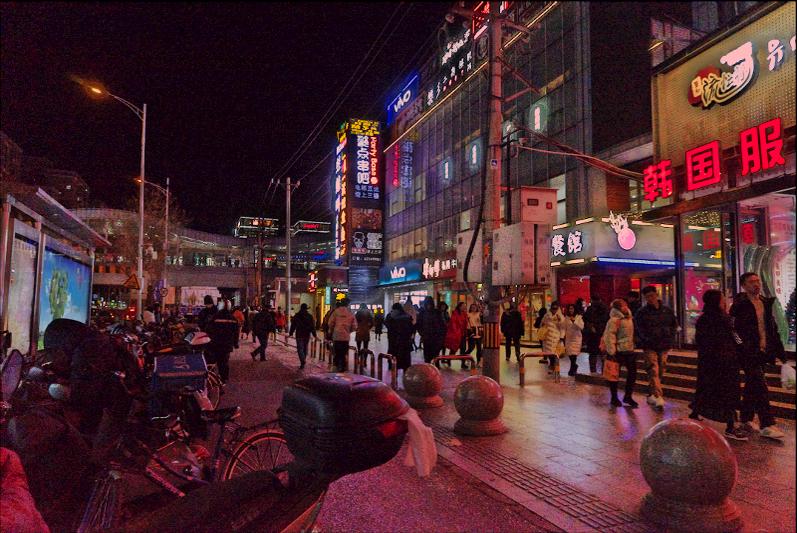}}}
\centering{\footnotesize (g) NPE~\cite{npe_tip2013} }
\end{minipage}
\begin{minipage}[t]{0.156\textwidth}
\subfigure{\raisebox{-0.15cm}{\includegraphics[width=1\textwidth]{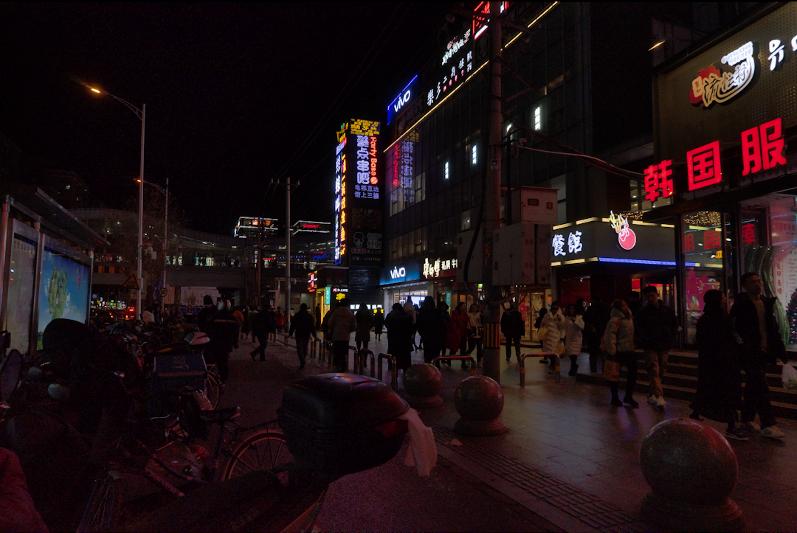}}}
\centering{\footnotesize (c) SIRE~\cite{pmiesire_tip2015} }
\subfigure{\raisebox{-0.15cm}{\includegraphics[width=1\textwidth]{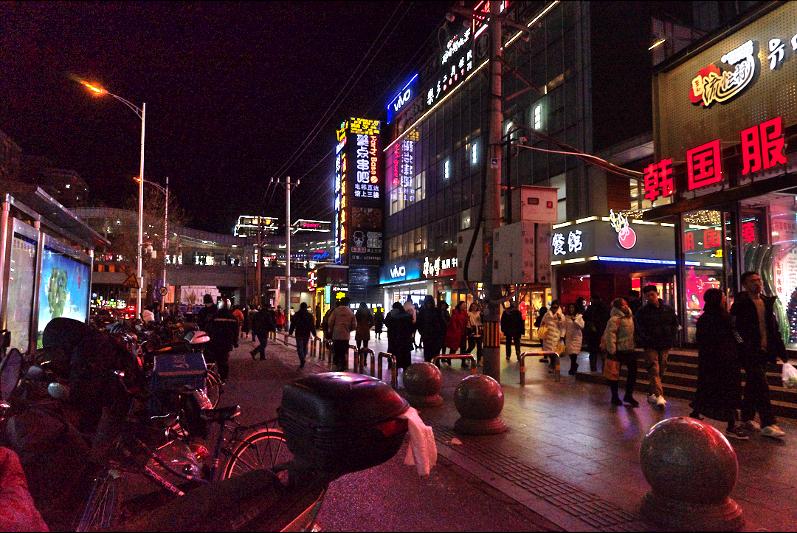}}}
\centering{\footnotesize (h) LIME~\cite{lime_tip2017} }
\end{minipage}
\begin{minipage}[t]{0.156\textwidth}
\subfigure{\raisebox{-0.15cm}{\includegraphics[width=1\textwidth]{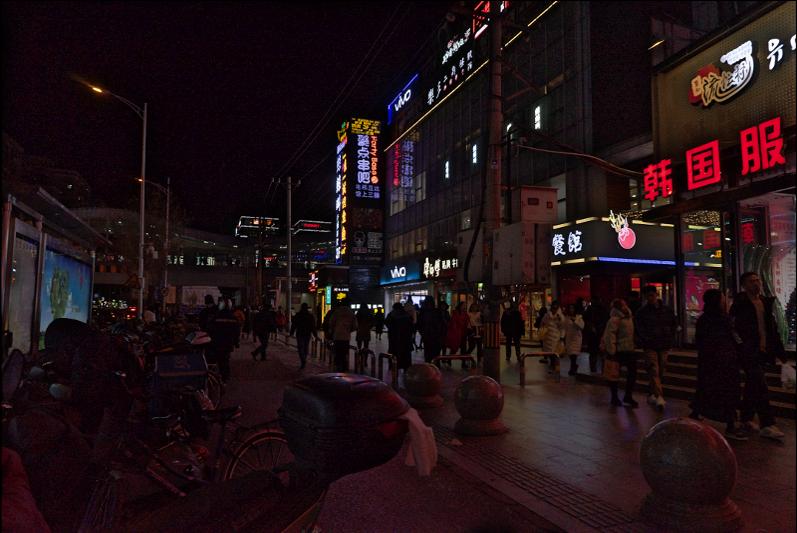}}}
\centering{\footnotesize (d) WVM~\cite{wvm2016} }
\subfigure{\raisebox{-0.15cm}{\includegraphics[width=1\textwidth]{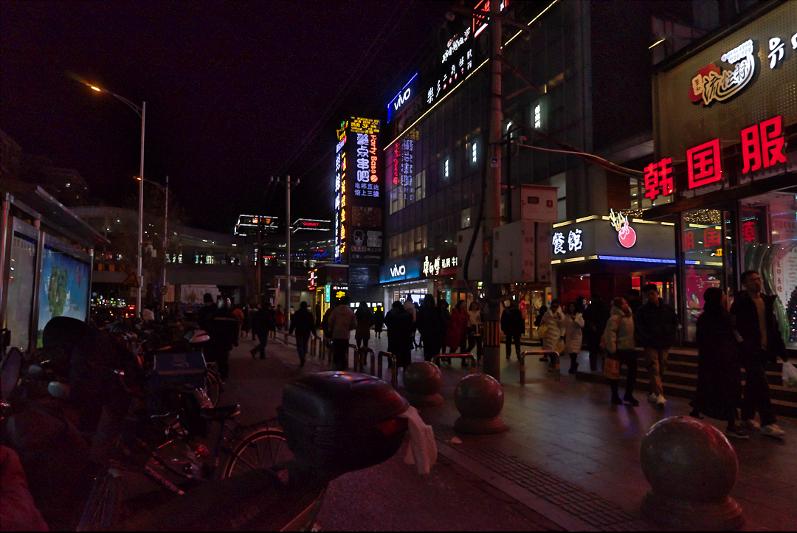}}}
\centering{\footnotesize (i) STAR~\cite{star_tip2020} }
\end{minipage}
\begin{minipage}[t]{0.156\textwidth}
\subfigure{\raisebox{-0.15cm}{\includegraphics[width=1\textwidth]{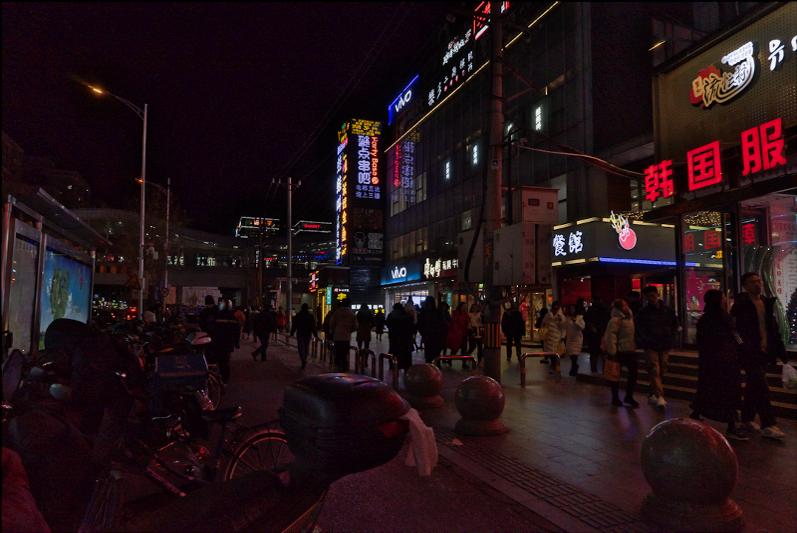}}}
\centering{\footnotesize (e) JieP~\cite{jiep2017} }
\subfigure{\raisebox{-0.15cm}{\includegraphics[width=1\textwidth]{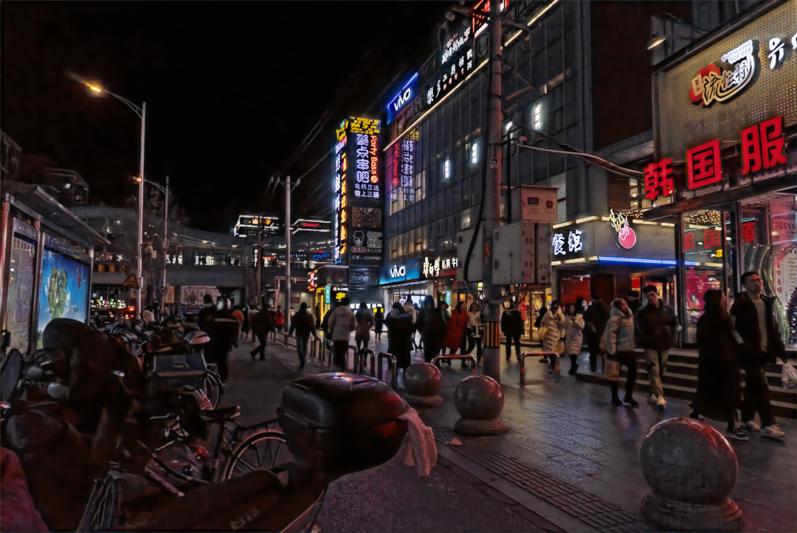}}}
\centering{\footnotesize (j) KinD++~\cite{2021KinD++} }
\end{minipage}
\begin{minipage}[t]{0.156\textwidth}
\subfigure{\raisebox{-0.15cm}{\includegraphics[width=1\textwidth]{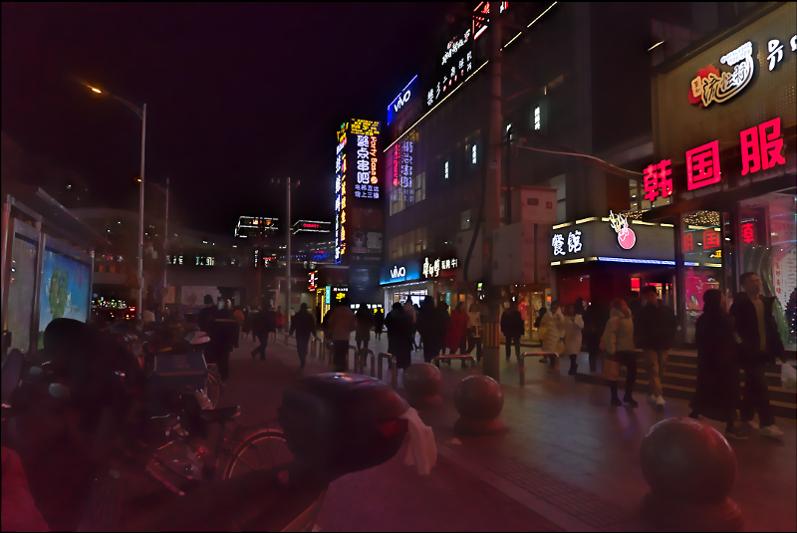}}}
\centering{\footnotesize (f) RRM~\cite{li2018structure} }
\subfigure{\raisebox{-0.15cm}{\includegraphics[width=1\textwidth]{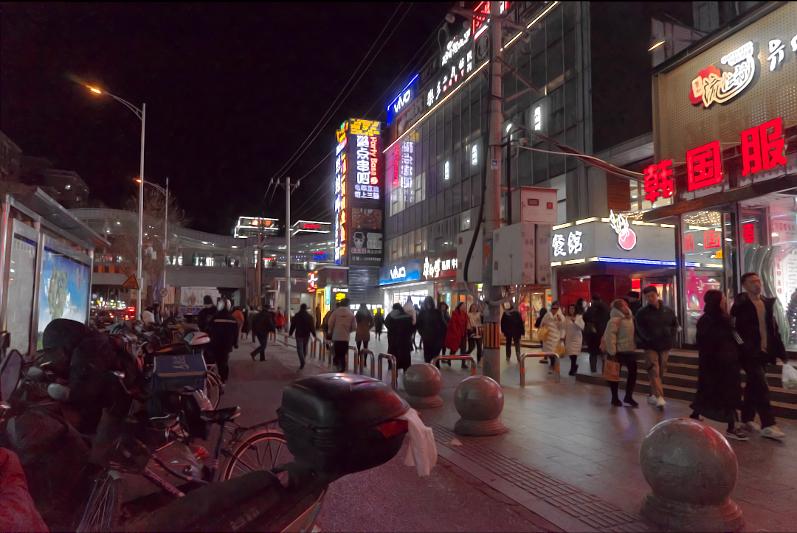}}}
\centering{\footnotesize (k) NLHD (ours) }
\end{minipage}
\vspace{-1mm}
\caption{\linespread{1}\selectfont{{Comparison of the result images on the image ``10'' in the 200darkface dataset enhanced by various low-light image enhancement methods.}}}
\label{darkface}
\vspace{-3mm}
\end{figure*}

\begin{figure*} [htb]
\vspace{-2mm}
\centering
\subfigure{
\begin{minipage}[t]{0.24\textwidth}
\centering
\raisebox{-0.15cm}{\includegraphics[width=1\textwidth]{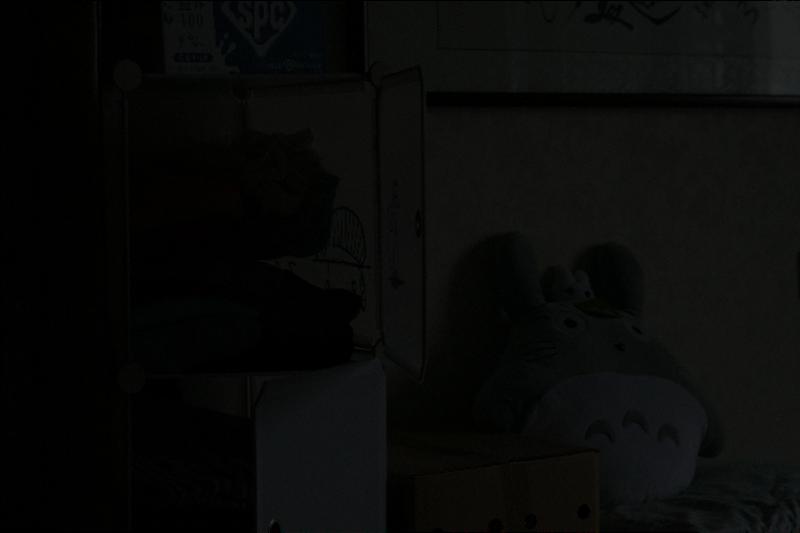}}
{\footnotesize (a) Low-light}
\end{minipage}
\begin{minipage}[t]{0.24\textwidth}
\centering
\raisebox{-0.15cm}{\includegraphics[width=1\textwidth]{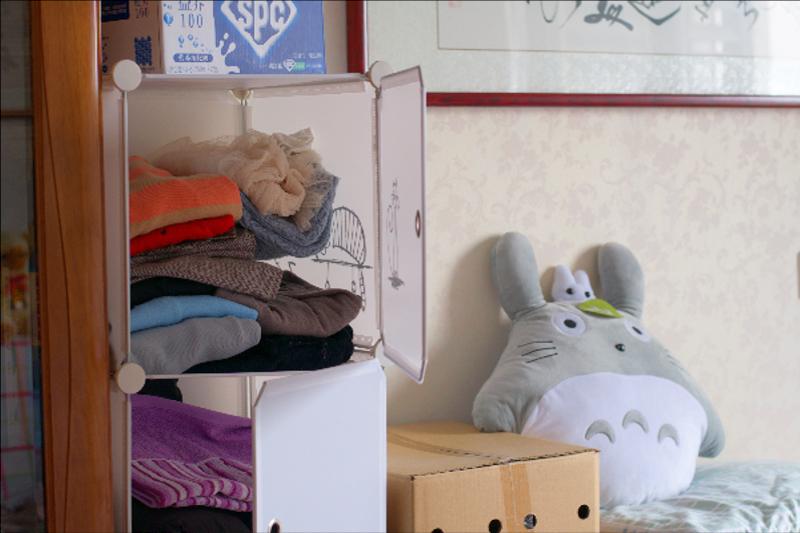}}
{\footnotesize (b) High-light}
\end{minipage}
\begin{minipage}[t]{0.24\textwidth}
\centering
\raisebox{-0.15cm}{\includegraphics[width=1\textwidth]{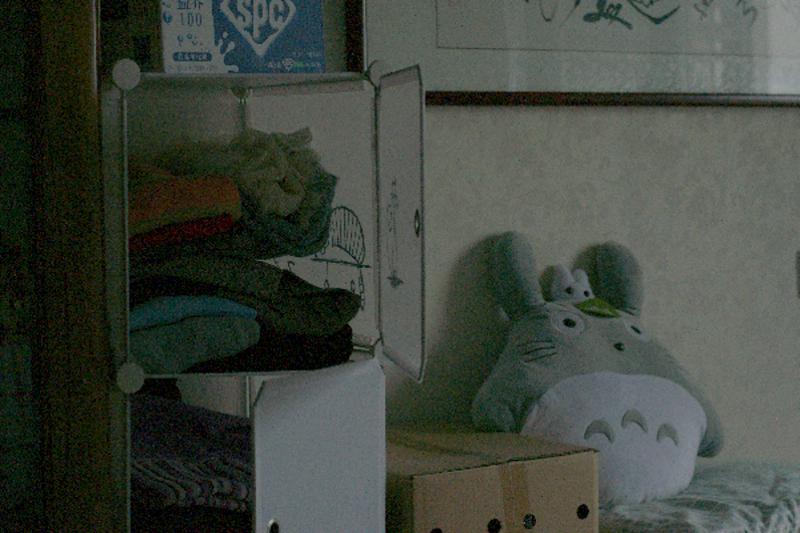}}
{\footnotesize (c) CRM~\cite{Ying_2017_ICCV} }
\end{minipage}
\begin{minipage}[t]{0.24\textwidth}
\centering
\raisebox{-0.15cm}{\includegraphics[width=1\textwidth]{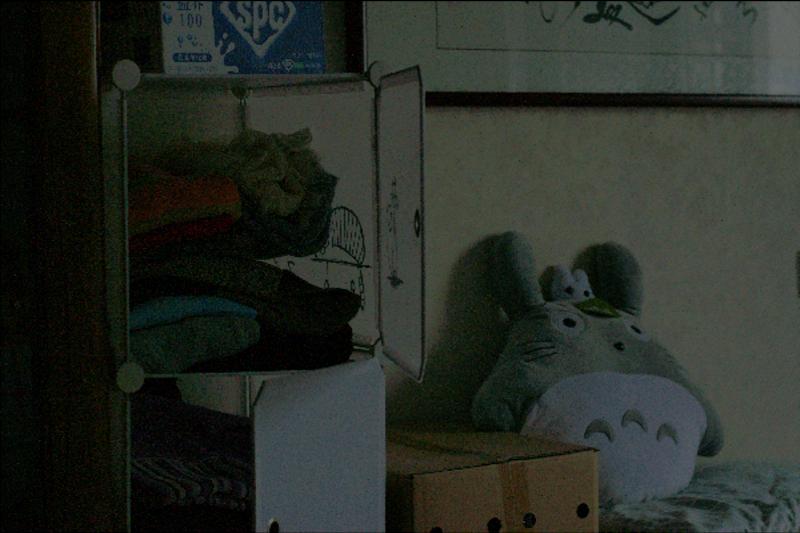}}
{\footnotesize (d) SIRE~\cite{pmiesire_tip2015} }
\end{minipage}
}\vspace{-3mm}
\subfigure{
\begin{minipage}[t]{0.24\textwidth}
\centering
\raisebox{-0.15cm}{\includegraphics[width=1\textwidth]{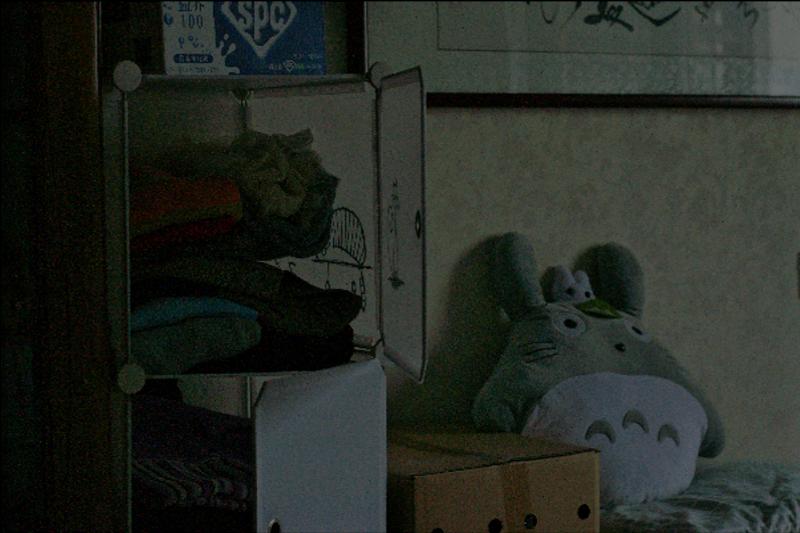}}
{\footnotesize (e) WVM~\cite{wvm2016} }
\end{minipage}
\begin{minipage}[t]{0.24\textwidth}
\centering
\raisebox{-0.15cm}{\includegraphics[width=1\textwidth]{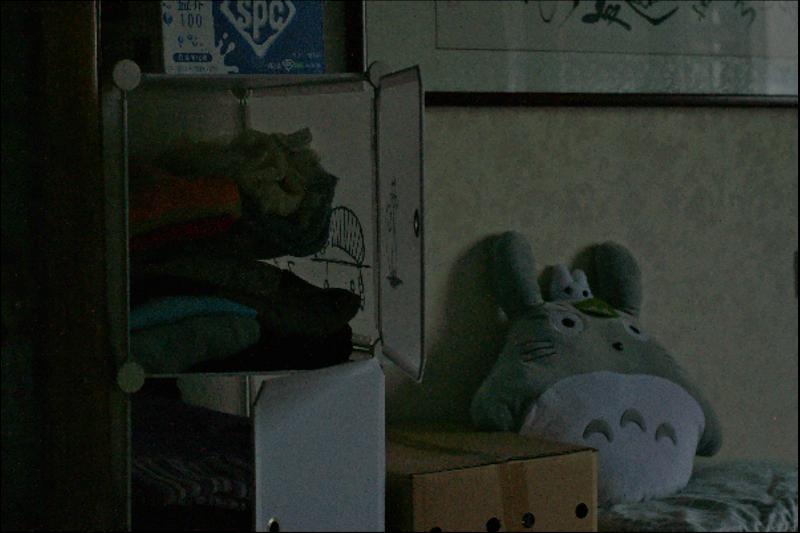}}
{\footnotesize (f) JieP~\cite{jiep2017} }
\end{minipage}
\begin{minipage}[t]{0.24\textwidth}
\centering
\raisebox{-0.15cm}{\includegraphics[width=1\textwidth]{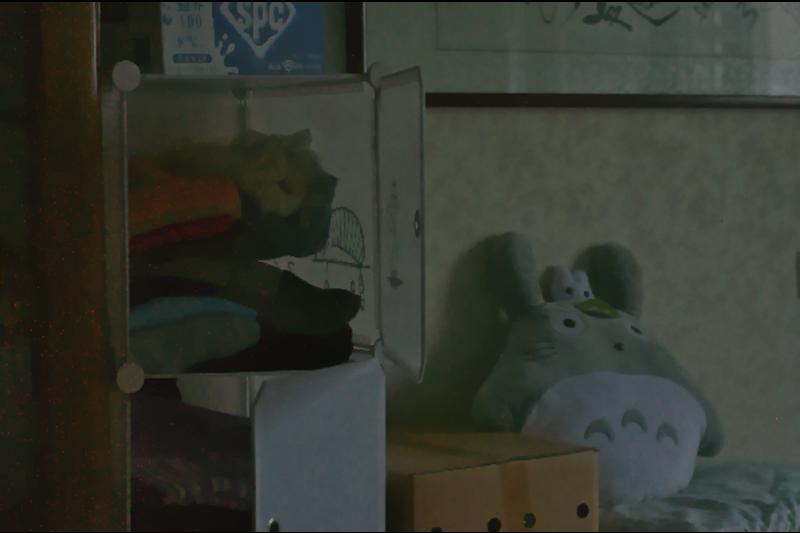}}
{\footnotesize (g) RRM~\cite{li2018structure} }
\end{minipage}
\begin{minipage}[t]{0.24\textwidth}
\centering
\raisebox{-0.15cm}{\includegraphics[width=1\textwidth]{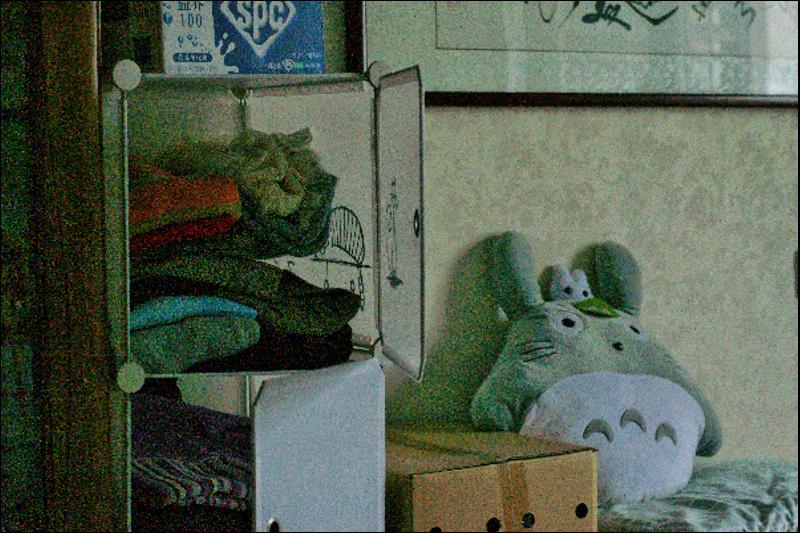}}
{\footnotesize (h) NPE~\cite{npe_tip2013} }
\end{minipage}
}\vspace{-3mm}
\subfigure{
\begin{minipage}[t]{0.24\textwidth}
\centering
\raisebox{-0.15cm}{\includegraphics[width=1\textwidth]{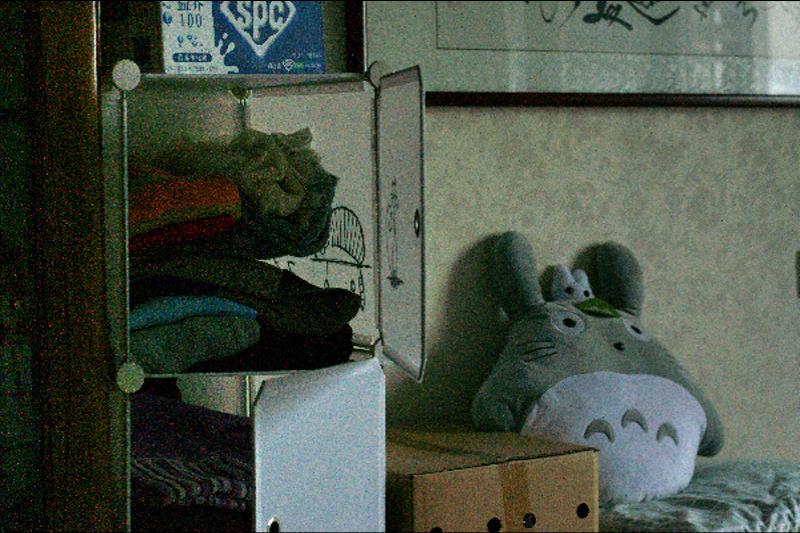}}
{\footnotesize (i) LIME~\cite{lime_tip2017} }
\end{minipage}
\begin{minipage}[t]{0.24\textwidth}
\centering
\raisebox{-0.15cm}{\includegraphics[width=1\textwidth]{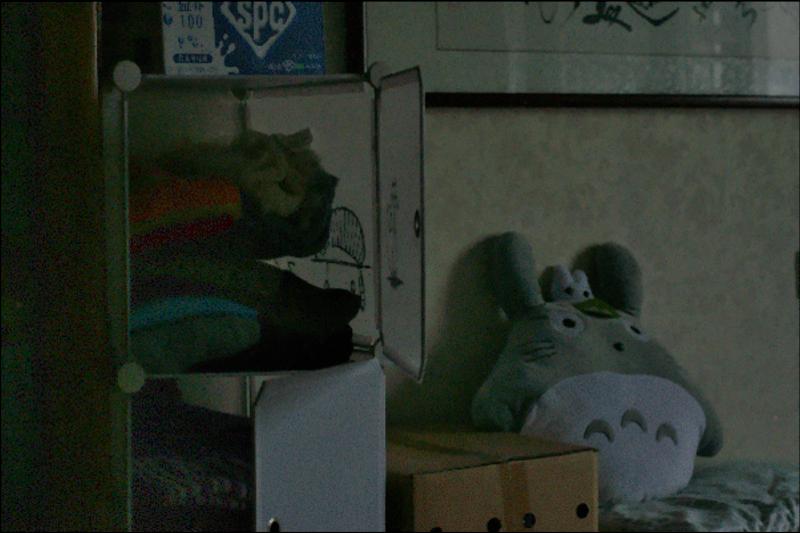}}
{\footnotesize (j) STAR~\cite{star_tip2020} }
\end{minipage}
\begin{minipage}[t]{0.24\textwidth}
\centering
\raisebox{-0.15cm}{\includegraphics[width=1\textwidth]{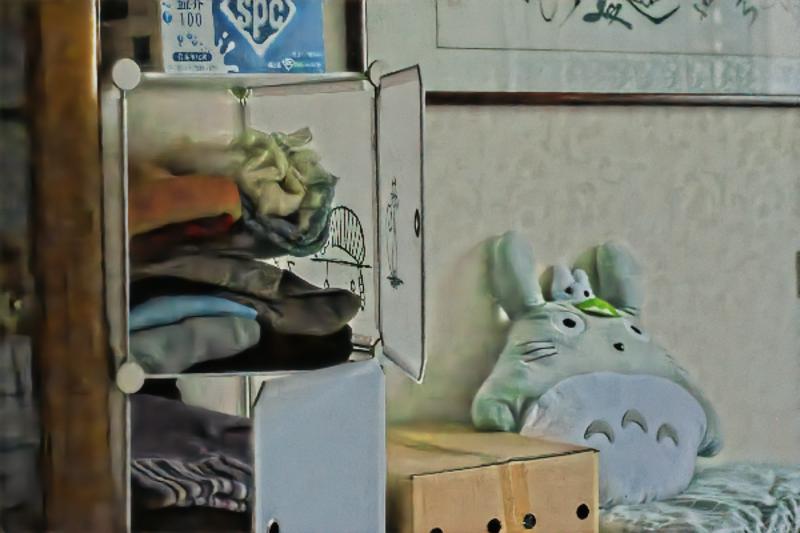}}
{\footnotesize (k) KinD++~\cite{2021KinD++} }
\end{minipage}
\begin{minipage}[t]{0.24\textwidth}
\centering
\raisebox{-0.15cm}{\includegraphics[width=1\textwidth]{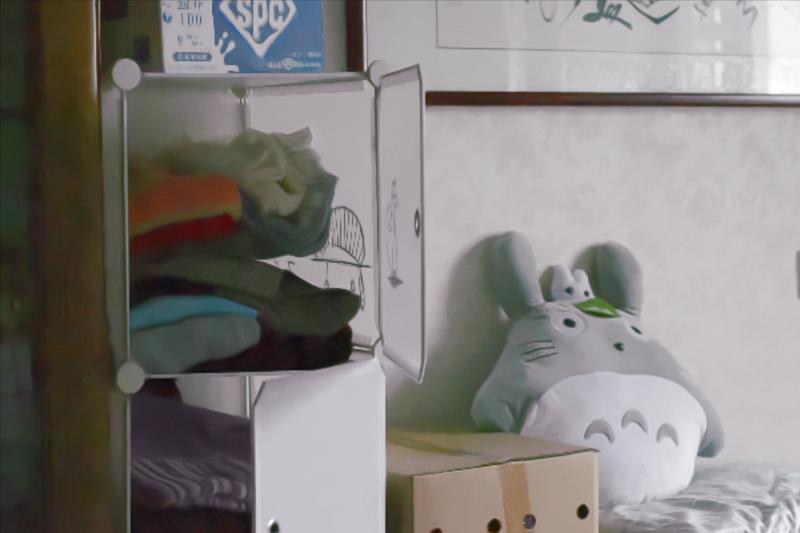}}
{\footnotesize (l) NLHD (ours) }
\end{minipage}
}\vspace{-1mm}
\caption{Comparison of the result images on the image ``24'' in the LOL dataset~\cite{drde_bmvc2018} enhanced by various low-light image enhancement methods.}
\label{LOL}
\vspace{-3mm}
\end{figure*}

We run the codes provided by the authors or collected from the Internet of CRM~\cite{Ying_2017_ICCV}, SIRE~\cite{pmiesire_tip2015}, Weighted Variational Model (WVM)~\cite{wvm2016}, Joint intrinsic-extrinsic Prior (JieP) model~\cite{jiep2017}, Robust Retinex Model (RRM)~\cite{li2018structure}, NPE~\cite{npe_tip2013}, LIME~\cite{lime_tip2017}, STAR~\cite{star_tip2020} and the latest deep learning based method KinD++~\cite{2021KinD++} on 35images and 200darkface datasets respectively. Table~\ref{table1} compares Average NIQE value and average LOE value of these methods in the two datasets. One can see that our results are better than other methods. Fig.~\ref{giraffe} and Fig.~\ref{light} respectively show an image from 35images enhancement result comparison among the proposed method and other existing methods; Fig.~\ref{darkface} shows an image from 200darkface enhancement result comparison among the proposed method and other existing methods. We can see from Fig.~\ref{giraffe}, Fig.~\ref{light} and Fig.~\ref{darkface} that the enhanced result images by the proposed method have better subjective visual quality than other methods. We also implement the proposed method on LOL dataset~\cite{drde_bmvc2018} to verify the effectiveness of extremely dark image enhancement; the objective evaluation results can be seen in Table~\ref{table2}, the proposed method achieved much better objective evaluation results on SSIM, PSNR, and $\Delta$E. Fig.~\ref{LOL} shows an extreme dark low-light image, the corresponding ground truth image, and all the enhanced images by different methods; we can see from this figure that the proposed NLHD method achieved a much better enhancement result than all other methods; the proposed method effectively suppressed the noise and achieved the color deviation correction.

\begin{table} [htb]
\caption{Average NIQE~\cite{niqe} and LOE~\cite{npe_tip2013} results of different methods on 35images  and 200darkface~\cite{ug2_dark_face}. The best results are highlighted in bold. Because the codes of RRM and KinD++ fail to implement on some bigger images in 35images dataset or for other reasons, the evaluation results can not be presented for this dataset.}
\vspace{-3mm}
\begin{center}
\footnotesize
\begin{tabular}{r|c|c|c|c}
\hline
\hline
\rowcolor[rgb]{ .851,  .851,  .851}
Dataset 
& \multicolumn{2}{c|}
{35images}
& \multicolumn{2}{c}
{200darkface} 
\\
\hline
\rowcolor[rgb]{ .851,  .851,  .851}
Metric 
& NIQE $\downarrow$ & LOE $\downarrow$ & NIQE $\downarrow$ & LOE $\downarrow$ 
\\
\hline
\textbf{CRM}~\cite{Ying_2017_ICCV}    & 3.13 & 744.61 & 3.66  & 531.16      
\\
\textbf{SIRE}~\cite{pmiesire_tip2015}        & 3.01 & 637.70  & 3.14 & 482.42 
\\
\textbf{WVM}~\cite{wvm2016}       & 2.99 & 633.40  & 3.41 &265.12
\\
\textbf{JieP}~\cite{jiep2017}     & 2.99 & 724.52 & 3.44 & 362.79
\\
\textbf{RRM}~\cite{li2018structure}  & ---   & ---   & 3.58 & 387.17 
\\
\textbf{NPE}~\cite{npe_tip2013}       & 3.22 & 710.21  & 4.29 & 920.02
\\
\textbf{LIME}~\cite{lime_tip2017}     & 3.39 & 779.73 & 3.98 & 896.30
\\
\textbf{STAR}~\cite{star_tip2020}     & 2.93  & 677.43 & 3.84 &390.42   
\\
\textbf{KinD++}~\cite{2021KinD++}     & ---   & ---  & 3.04  & 664.21   
\\
\hline
\hline
\multicolumn{1}{r|}{\textbf{NLHD}}
& \textbf{2.76} & \textbf{546.63} & \textbf{2.80} & \textbf{250.79}
\\
\hline
\hline
\end{tabular}
\end{center}
\vspace{-3mm}
\label{table1}
\end{table}

\begin{table}[htb]
\caption{Quantitative comparison on LOL dataset in terms of PSNR, SSIM, and $\Delta$E. The best results are highlighted in bold.}
 \vspace{-1mm}
\centering
\setlength{\tabcolsep}{1.1mm}{ 
\begin{tabular}{r|ccccc}
\hline
\hline
\rowcolor[rgb]{ .851,  .851,  .851}
Metrics & CRM~\cite{Ying_2017_ICCV}  & SIRE~\cite{pmiesire_tip2015} & WVM~\cite{wvm2016}  
& JieP~\cite{jiep2017}  &RRM~\cite{li2018structure}  \\
\hline
 PSNR    $\uparrow$   & 16.3425    & 11.8140  &11.4164  &11.5782      & 13.0741  \\

 SSIM   $\uparrow$    & 0.6055     & 0.4771   &0.4634  &0.4789      & 0.6044 \\

 $\Delta$E $\downarrow$  & 16.8154    & 24.8859  &25.7354  &25.3589     & 21.7515 \\
\hline
\hline
\rowcolor[rgb]{ .851,  .851,  .851}
Metrics &NPE~\cite{npe_tip2013} & LIME~\cite{lime_tip2017}  & STAR~\cite{star_tip2020} &  KinD++~\cite{2021KinD++} & NLHD  \\
 \hline
 PSNR  $\uparrow$    &15.9962  & 15.8115  &12.3084   & 16.9706  & \textbf{21.1121} \\

 SSIM  $\uparrow$    &0.4830   & 0.5000   &0.5026   & 0.6883    & \textbf{0.8101}  \\

 $\Delta$E $\downarrow$ & 17.1756 & 17.2276  &24.1980   & 13.6101   & \textbf{9.4792} \\
\hline
\hline
\end{tabular}}\label{table2}
\vspace{-1mm}
\end{table}

\textbf{Comparison on Speed.}
Table~\ref{table3} compares the running time of different Retinex decomposition methods on a low-light $1200 \times 900$ RGB image and the image enhancement procedure. We notice that the speed of our method is not faster than all other methods; we will further refine our algorithms and programs to improve the running speed in our future work.

\begin{table}[htb] 
\caption{Comparison of computational times (in seconds) of different Retinex image decomposition methods. The time is computed by averaging the 10 running times for each method on an RGB image with a size of $1200\times900$.}
\vspace{-3mm}
\begin{center}
\setlength{\tabcolsep}{1.2mm}{
\begin{tabular}{c|ccccc}
\hline
\hline
\rowcolor[rgb]{ .9,  .9,  .9}
Method & CRM~\cite{Ying_2017_ICCV} & SIRE~\cite{pmiesire_tip2015} & WVM~\cite{wvm2016} &  JieP~\cite{jiep2017} & RRM~\cite{li2018structure}
\\
\hline
Time &  0.68 &  2.91 & 34.77 & 10.81 &  85.09
\\
\hline
\hline
\rowcolor[rgb]{ .9,  .9,  .9}
Method  &  NPE~\cite{npe_tip2013} &  LIME~\cite{lime_tip2017} & STAR~\cite{star_tip2020}  &  KinD++~\cite{2021KinD++} & NLHD
\\
\hline

Time  & 30.85 & 1.15 &  3.58  & 20.85  & 13.93
\\
\hline
\hline
\end{tabular}} 
\vspace{-1mm}
\label{table3}
\vspace{-5mm}
\end{center}
\end{table}

\subsection{Application to Object Detection}\label{face detection}
\begin{figure}[htb]
\includegraphics[width=0.48\textwidth]{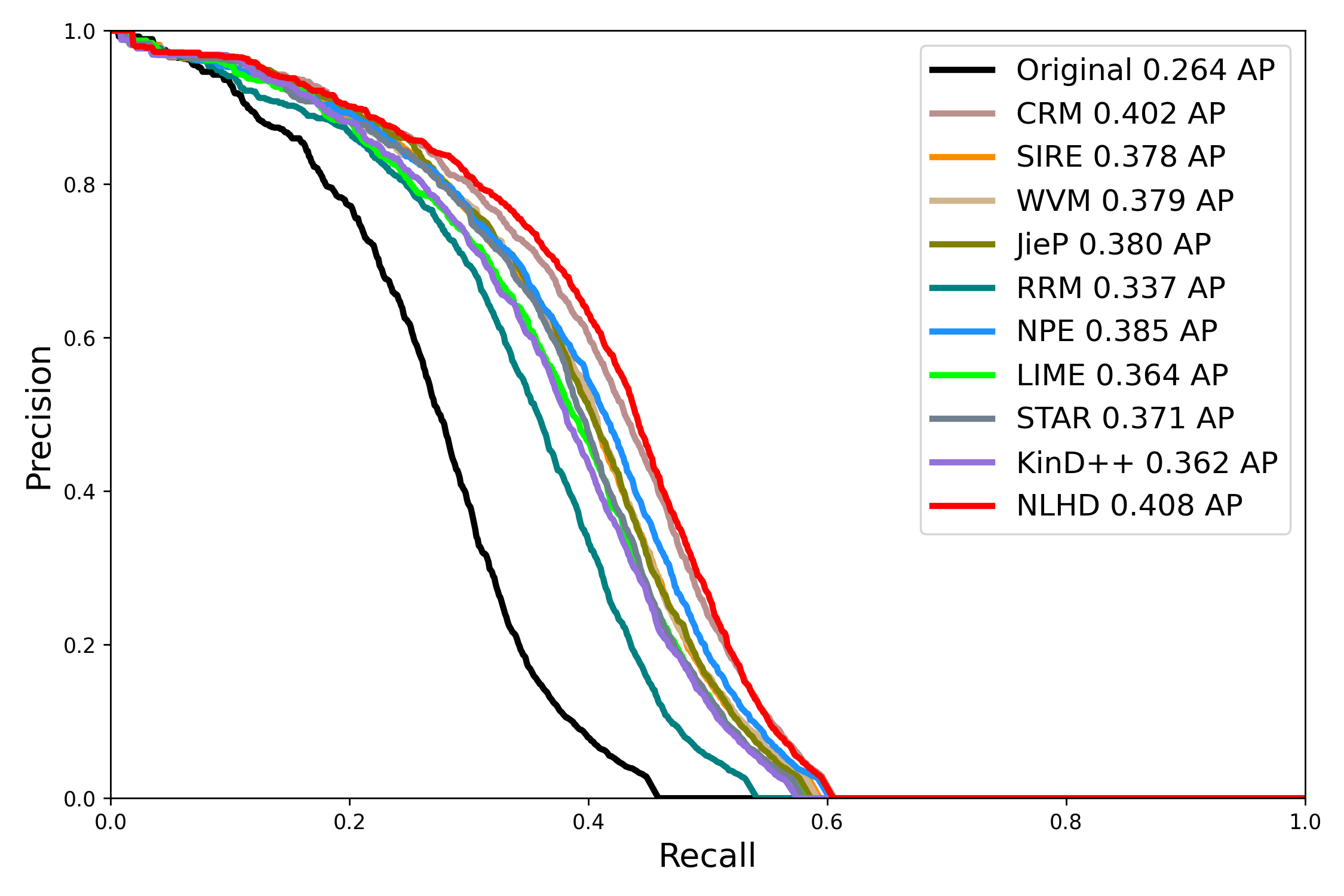}
\vspace{-3mm}
\caption{Comparison of Precision-Recall curves for face detection results on the original low-light images and enhanced images by various low-light enhancement methods. The test images are from the Dark Face dataset~\cite{ug2_dark_face}.}
\label{PRcurve}
\vspace{-3mm}
\end{figure}

\begin{figure*}[ht]
\par\setlength\parindent{1em}(a)
\begin{minipage}[c]{0.95\linewidth}
\includegraphics[width=4.1cm,height=3cm]{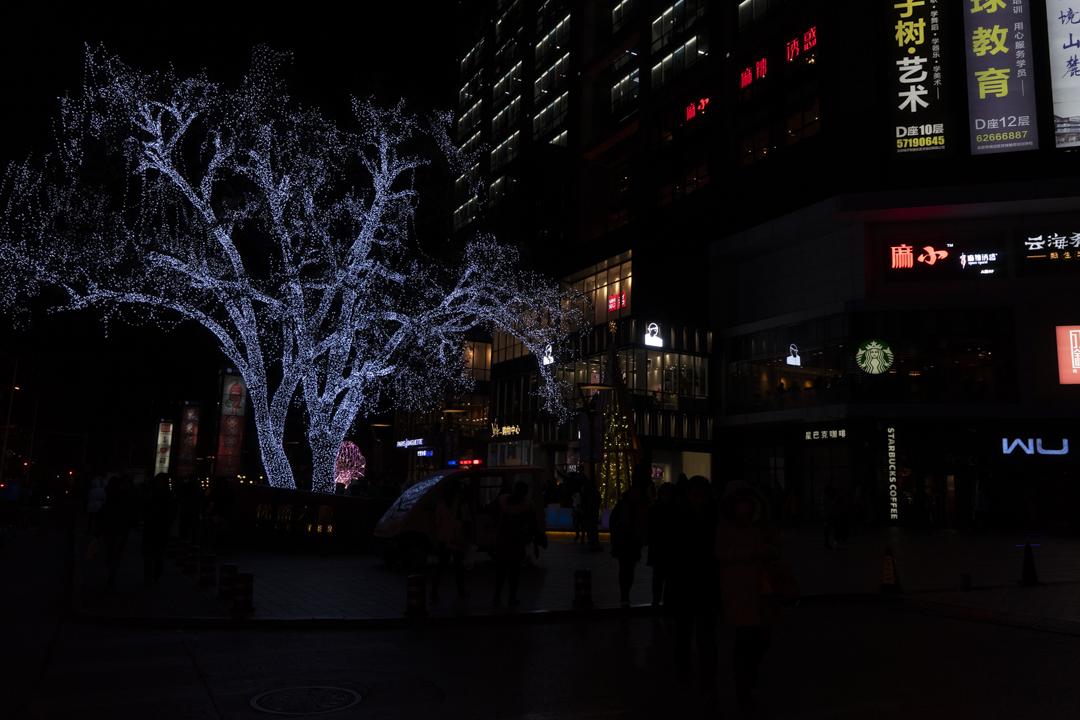} 
\includegraphics[width=4.1cm,height=3cm]{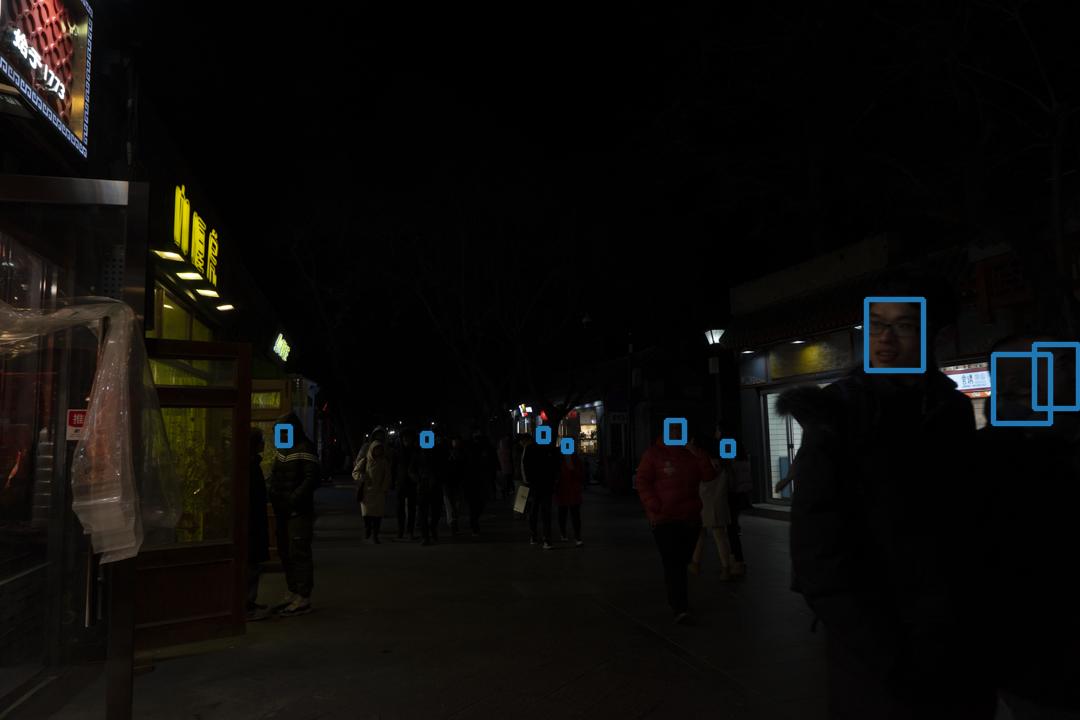} 
\includegraphics[width=4.1cm,height=3cm]{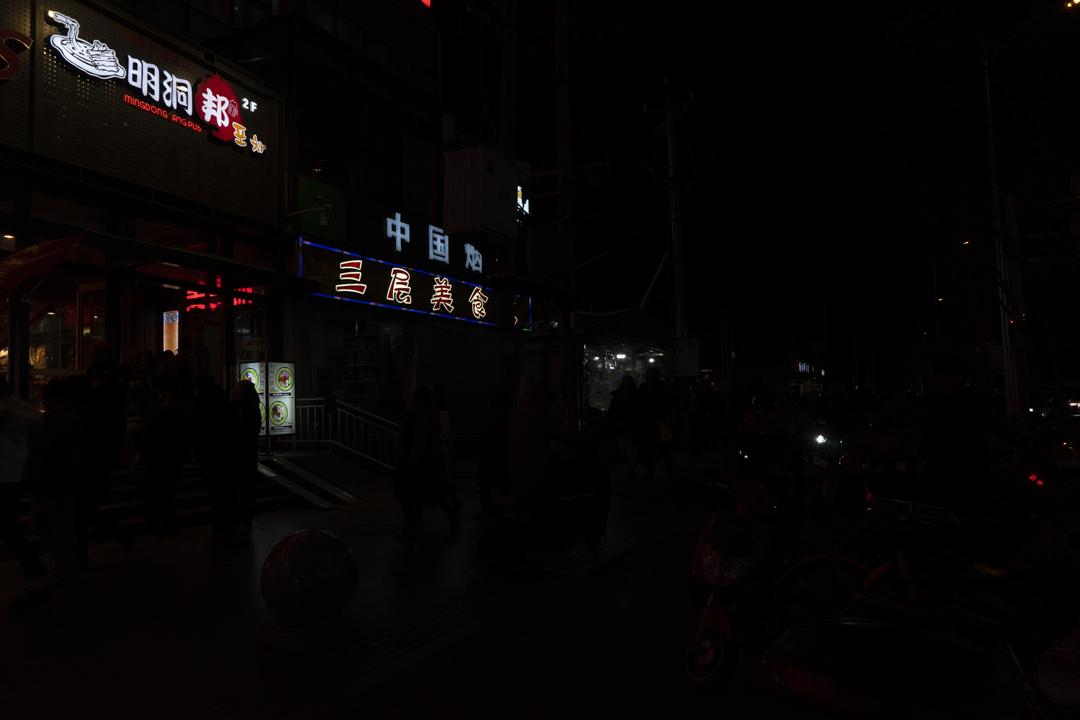}
\includegraphics[width=4.1cm,height=3cm]{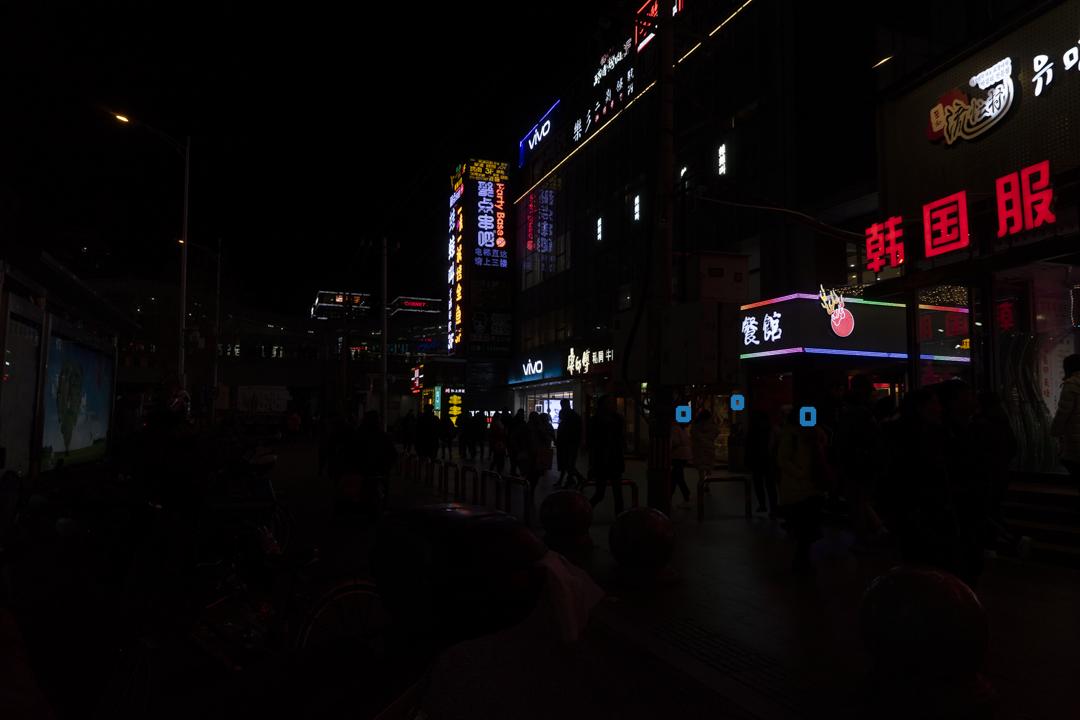}
\end{minipage}
\vspace{1mm}
\par\setlength\parindent{0.95em}(b)
\begin{minipage}[c]{0.95\linewidth}
\includegraphics[width=4.1cm,height=3cm]{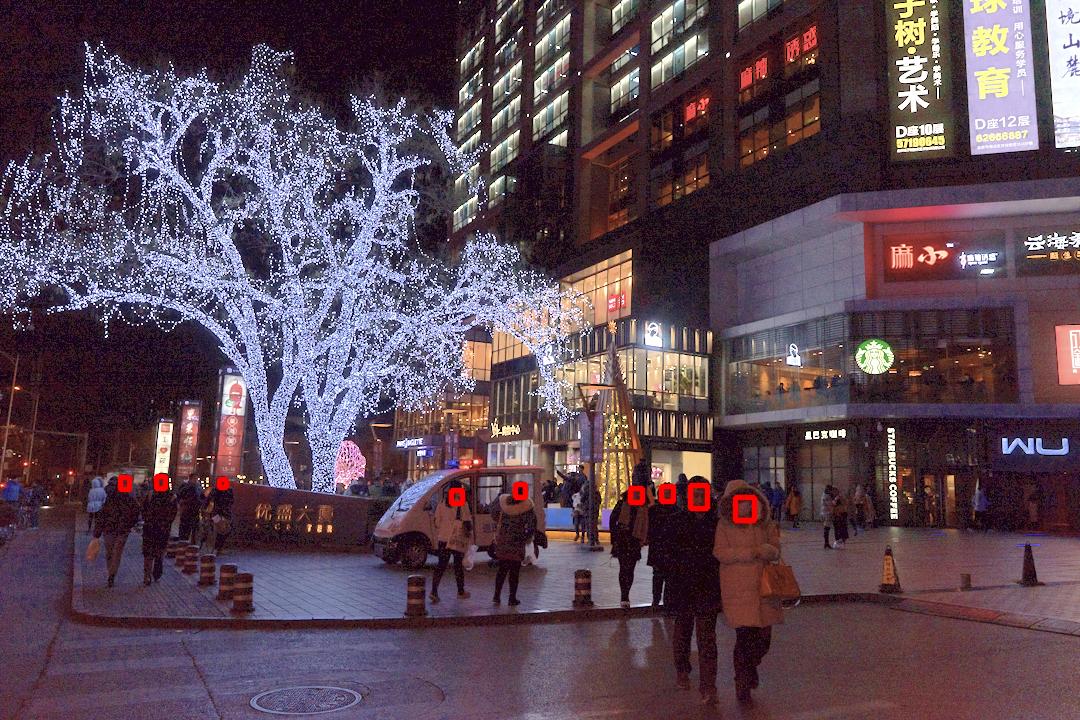} 
\includegraphics[width=4.1cm,height=3cm]{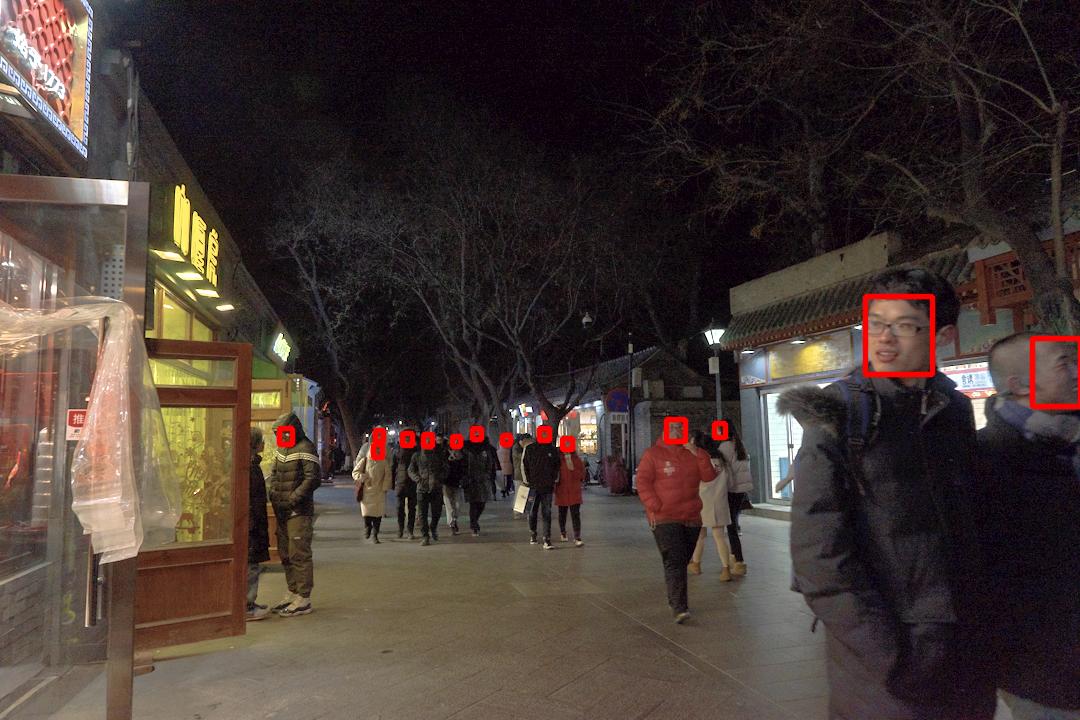} 
\includegraphics[width=4.1cm,height=3cm]{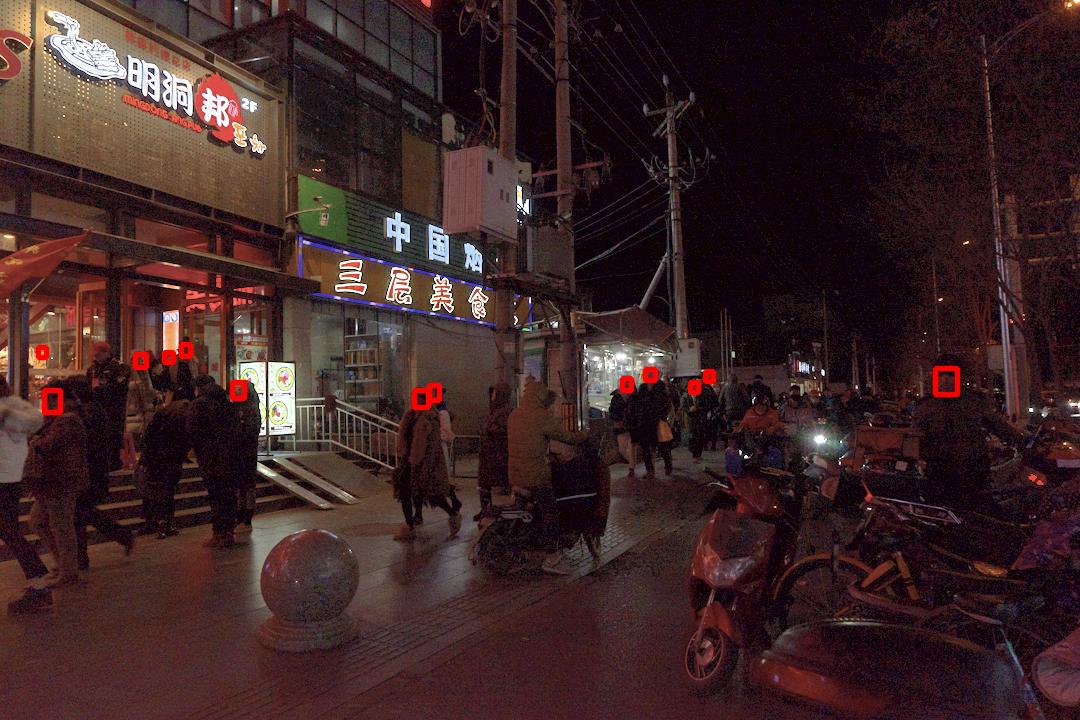} 
\includegraphics[width=4.1cm,height=3cm]{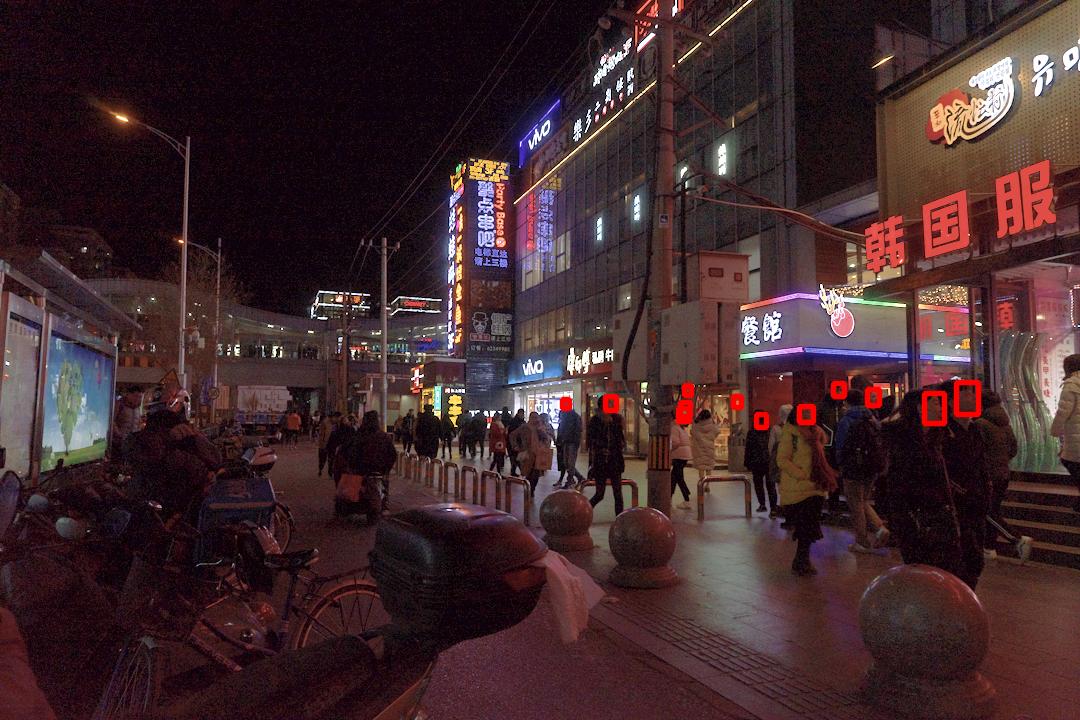}
\end{minipage}
\vspace{1mm}
\par\setlength\parindent{1.03em}(c)
\begin{minipage}[c]{0.95\linewidth}
\includegraphics[width=4.1cm,height=3cm]{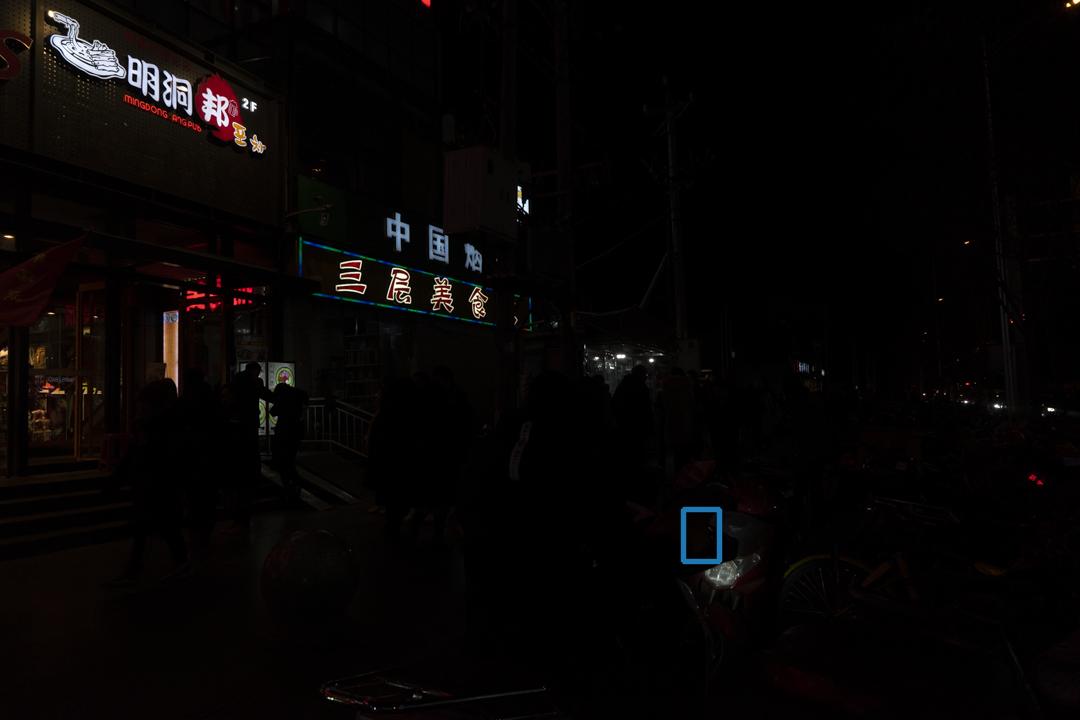} 
\includegraphics[width=4.1cm,height=3cm]{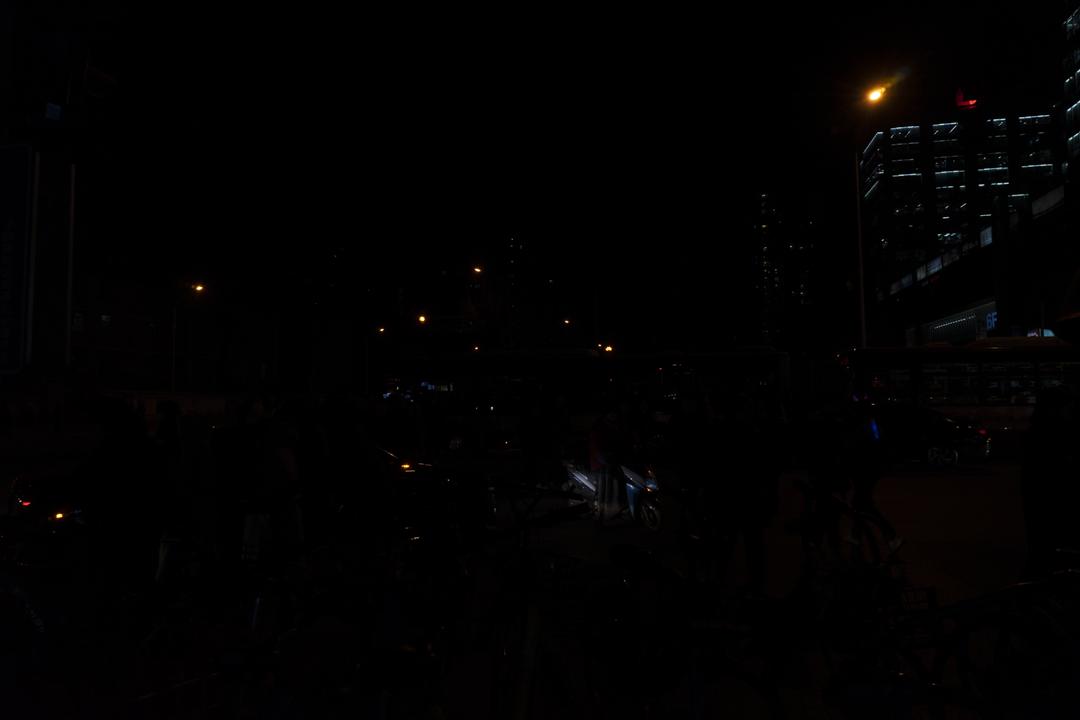} 
\includegraphics[width=4.1cm,height=3cm]{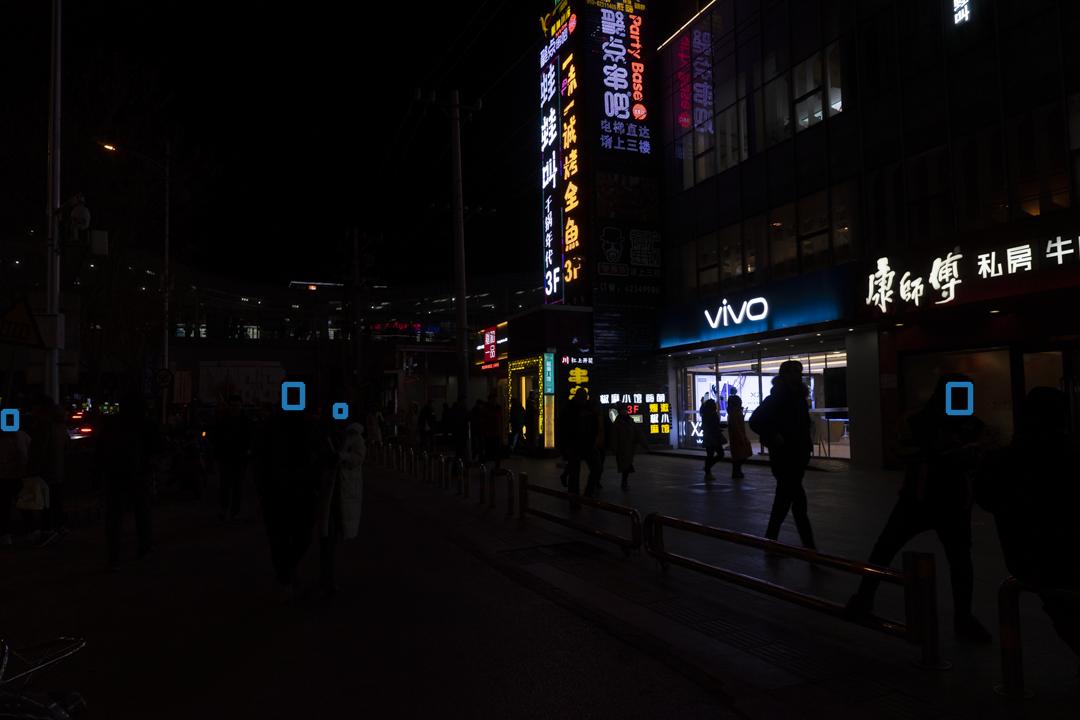} 
\includegraphics[width=4.1cm,height=3cm]{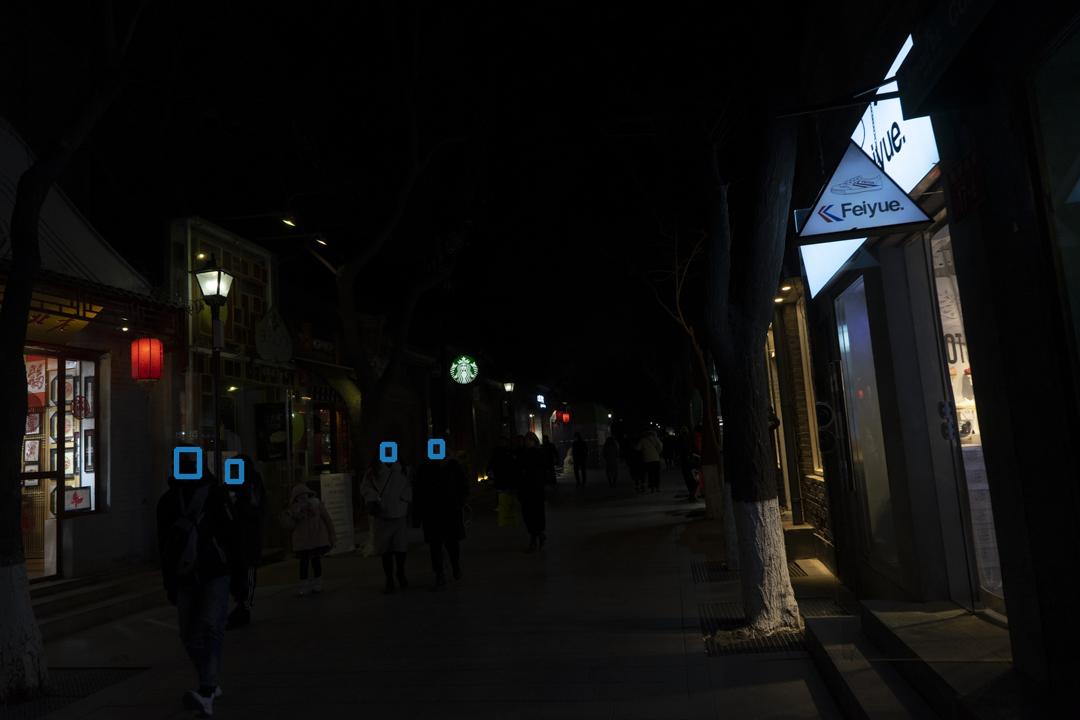} 
\end{minipage}
\vspace{1mm}
\par\setlength\parindent{1.05em}(d)
\begin{minipage}[c]{0.95\linewidth}
\includegraphics[width=4.1cm,height=3cm]{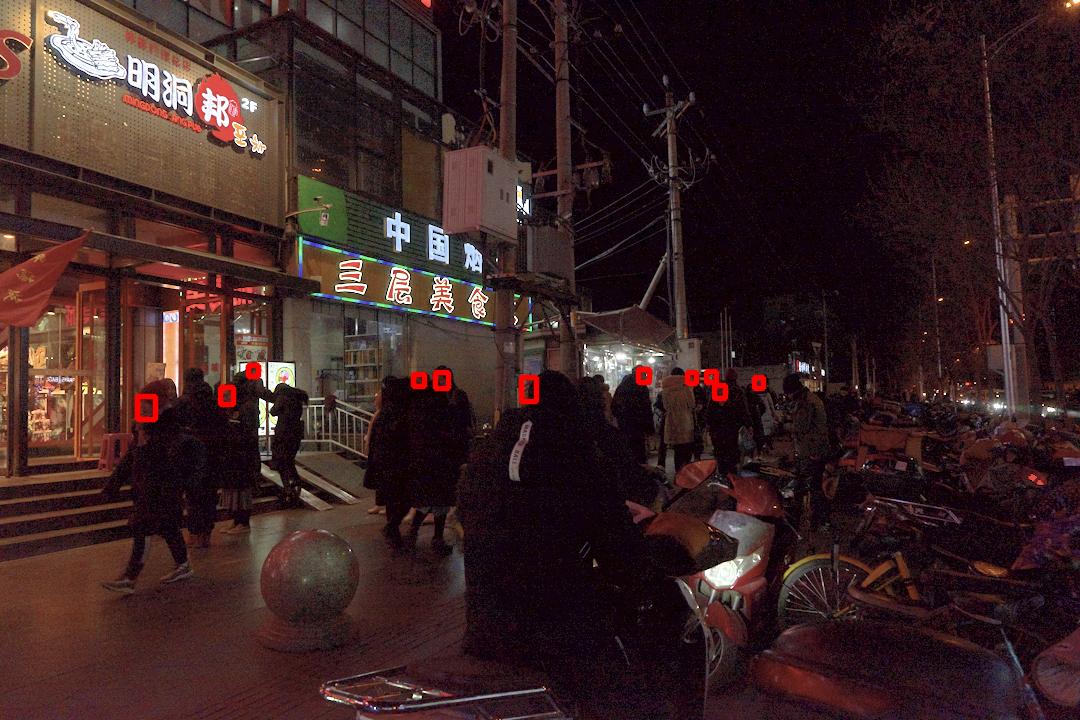} 
\includegraphics[width=4.1cm,height=3cm]{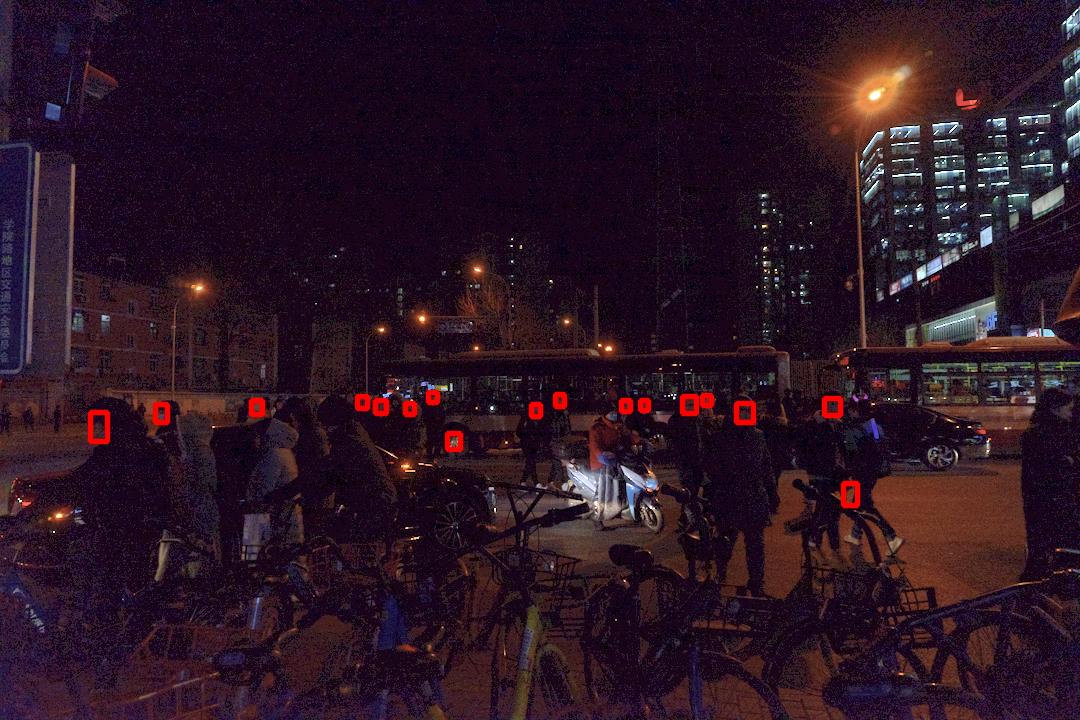}
\includegraphics[width=4.1cm,height=3cm]{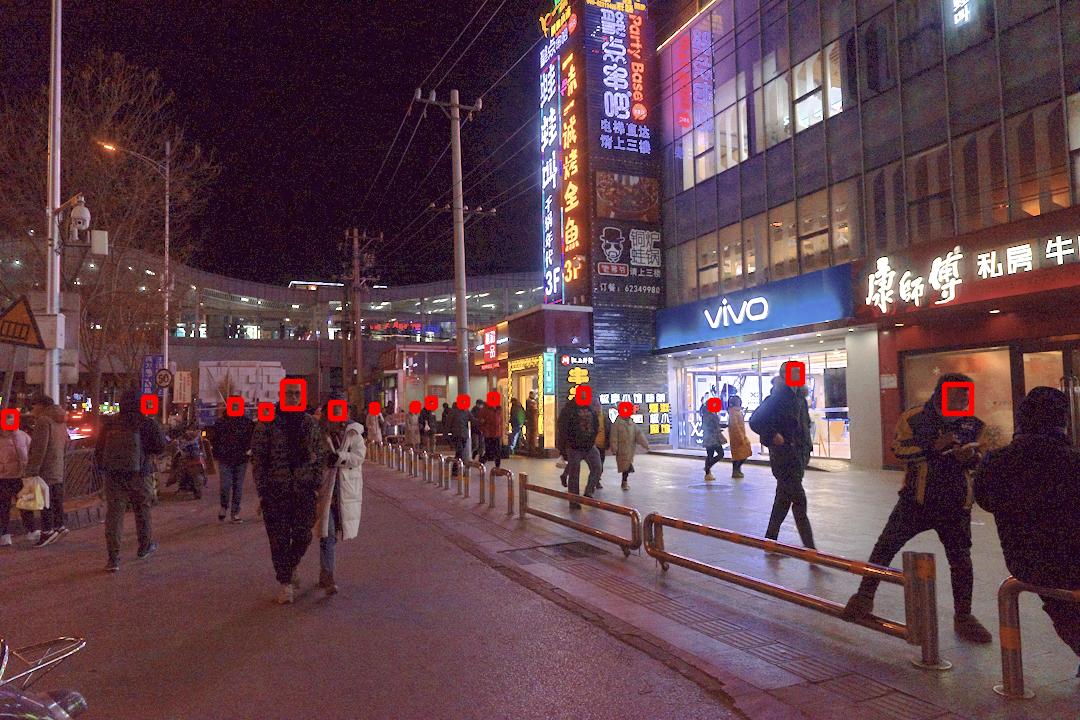} 
\includegraphics[width=4.1cm,height=3cm]{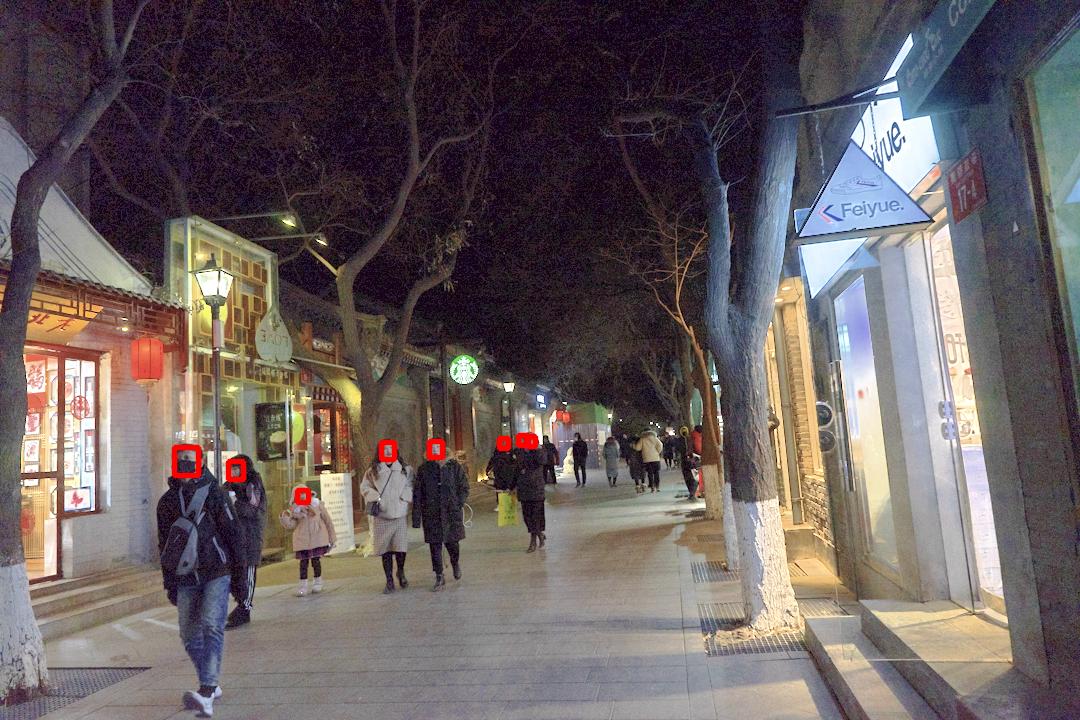}
\end{minipage}
\vspace{-1mm}
\caption{Comparisons of face detection results between the original image and its correspondingly enhanced image by our NLHD method, the original low-light images are from Dark Face dataset~\cite{ug2_dark_face}. The images in row \textbf{(a)} and \textbf{(c)} present the detection results on the original images, the images in row \textbf{(b)} and \textbf{(d)} present the detection results on the enhanced image by our NLHD method.}
\vspace{-3mm}
\label{f10}
\end{figure*}

To further verify the effectiveness of the NLHD method, we compare the face detection results on the enhanced Dark Face dataset~\cite{ug2_dark_face} with various low-light image enhancement methods. We choose the current excellent object detection method YOLOv5~\cite{glenn_jocher_2021_4679653} to achieve face detection on the original low-light images and enhanced images by various low-light enhancement methods, respectively. We firstly train the YOLOv5 model on the Wider Face dataset~\cite{widerface} and then use the Dark Face dataset to test. Since the label of the test set is not public, we randomly select 600 images from the training set of the Dark Face dataset for evaluation. These images are enhanced by various methods, including CRM~\cite{Ying_2017_ICCV}, SIRE~\cite{pmiesire_tip2015}, Weighted Variational Model (WVM)~\cite{wvm2016}, Joint intrinsic-extrinsic Prior (JieP) model~\cite{jiep2017}, Robust Retinex Model (RRM)~\cite{li2018structure}, NPE~\cite{npe_tip2013}, LIME~\cite{lime_tip2017}, STAR~\cite{star_tip2020} and the latest deep learning based method KinD++~\cite{2021KinD++}. Fig.~\ref{f10} compares the visual results of face detection between the original images and the enhanced images by the proposed NLHD method. It can be seen that the face detection results on the enhanced images can be improved to a certain extent; even many faces hidden in the darker regions can be detected. 
In order to more objectively compare the test results, we also depict the precision-recall (P-R) curves under 0.6 IoU threshold in Fig.~\ref{PRcurve}. As shown in Fig.~\ref{PRcurve}, the proposed NLHD based low-light image enhancement method effectively improved the low-light environment face detection performance by the YOLOv5 model. It can be seen that the face detection result of the proposed NLHD method achieved the best performance among all the compared methods listed in the figure; it can obtain the highest Average Precision (AP) score. So the proposed NLHD method can be used for the low-light environment object detection task.

\vspace{-2mm}
\subsection{Ablation Study}\label{Ablation Study}
a) Is the proposed NLHD necessary? To show the necessity of NLHD, we implement low-light image enhancement experiments by the NLHD method (w/ NLHD) and directly image enhancement by the bright channel of RGB space (wo NLHD) on the 35images dataset and the 200darkface dataset, respectively. We first get the bright channel of RGB space, the maximum pixel values in R, G, and B channels are fused to construct a bright channel (Actually, channel V in HSV space is the bright channel of RGB space),  then we implement the exponential transform and the logarithmic transform respectively on the bright channel with the same parameters as the NLHD based enhancement method. We use the minimum fusion result of the enhanced bright channel to replace the V channel in HSV space to achieve the final enhanced image. We compare NLHD with the above method; the objective evaluation results in Table~\ref{table4} show that the performance of the NLHD is much better than that of the above method. The results show that the NLHD method can use pixel-level non-local self-similarity to achieve better Retinex decomposition, thus achieving better low-light image enhancement results.

\begin{table} [htb]
\caption{Average NIQE~\cite{niqe} and LOE~\cite{npe_tip2013} results of wo NLHD and w/ NLHD on 35images and 200darkface datasets.}
\vspace{-3mm}
\begin{center}
\footnotesize
\begin{tabular}{r|c|c|c|c}
\hline
\hline
\rowcolor[rgb]{ .851,  .851,  .851}
Dataset 
& \multicolumn{2}{c|}
{35Images}
& \multicolumn{2}{c}
{200darkface} 
\\
\hline
\rowcolor[rgb]{ .851,  .851,  .851}
Metric 
& NIQE $\downarrow$ & LOE $\downarrow$ & NIQE $\downarrow$ & LOE $\downarrow$ 
\\
\hline
wo NLHD            & 3.31 & 636.86  & 3.58 & 262.99 
\\
w/ NLHD            & \textbf{2.76} & \textbf{546.63}  & \textbf{2.80} & \textbf{250.79} 
\\
\hline
\hline

\end{tabular}
\end{center}
\vspace{-2mm}
\label{table4}
\end{table}

\begin{figure*}[ht]
\par\setlength\parindent{0.9em}(a)
\begin{minipage}[c]{0.95\linewidth}
\includegraphics[width=4.1cm,height=3cm]{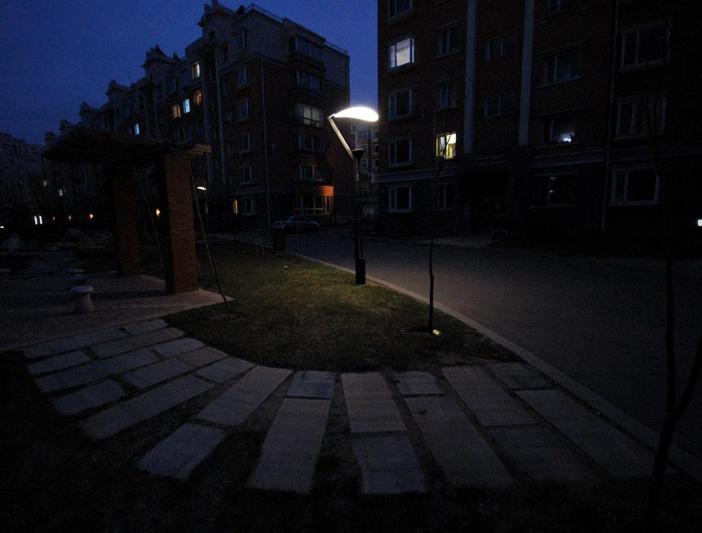} \includegraphics[width=2.7cm,height=3cm]{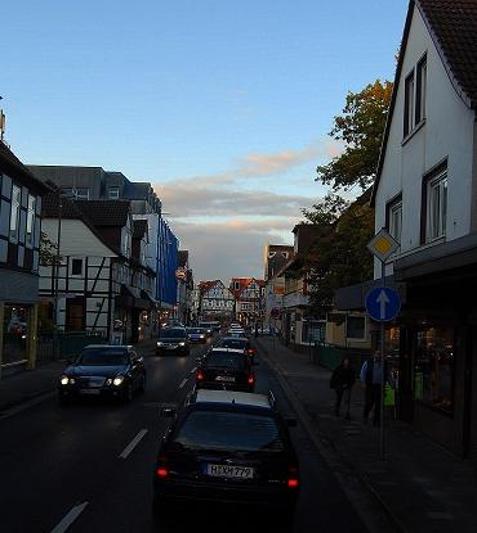} \includegraphics[width=3cm,height=3cm]{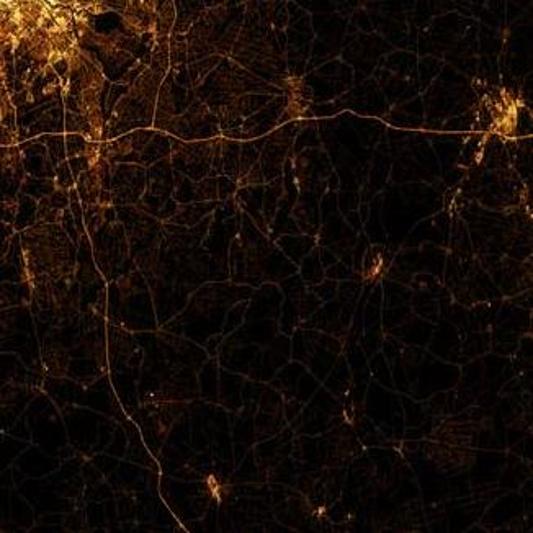} \includegraphics[width=4.1cm,height=3cm]{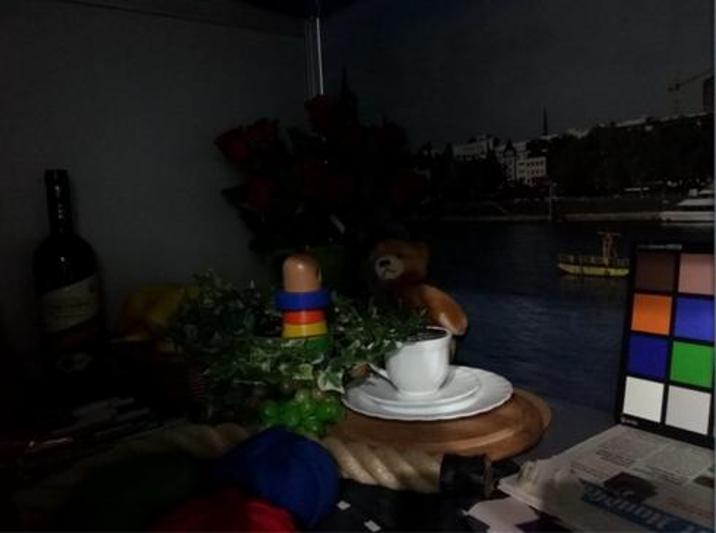}
\includegraphics[width=2.7cm,height=3cm]{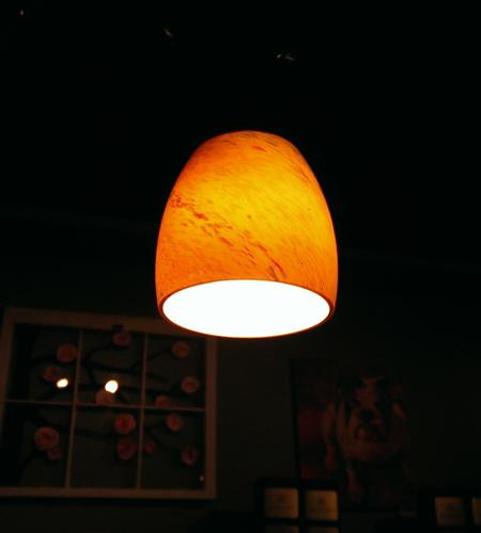}
\end{minipage}
\vspace{1mm}
\par\setlength\parindent{0.85em}(b)
\begin{minipage}[c]{0.95\linewidth}
\includegraphics[width=4.1cm,height=3cm]{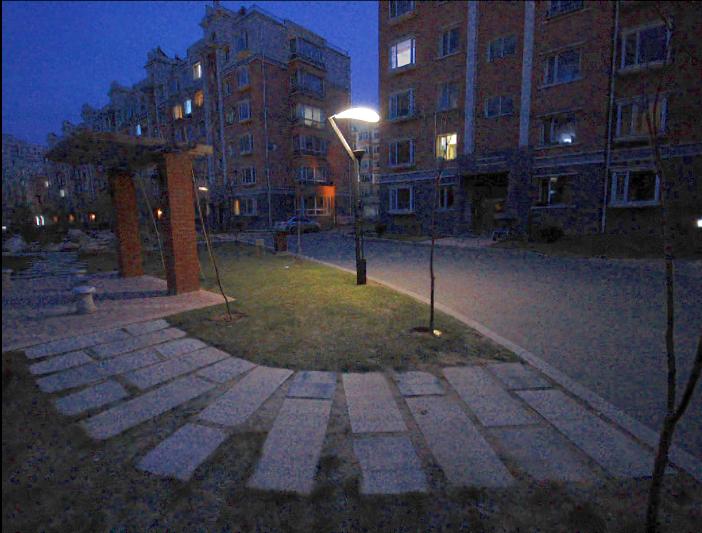} \includegraphics[width=2.7cm,height=3cm]{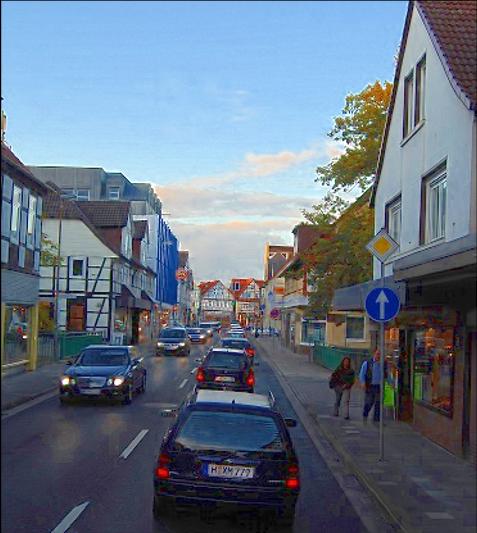} \includegraphics[width=3cm,height=3cm]{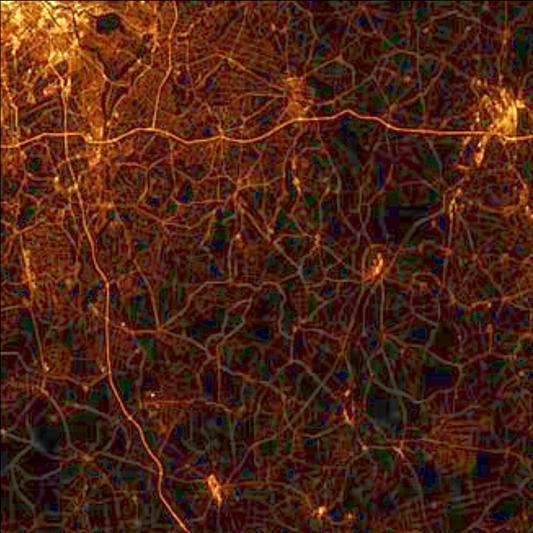} \includegraphics[width=4.1cm,height=3cm]{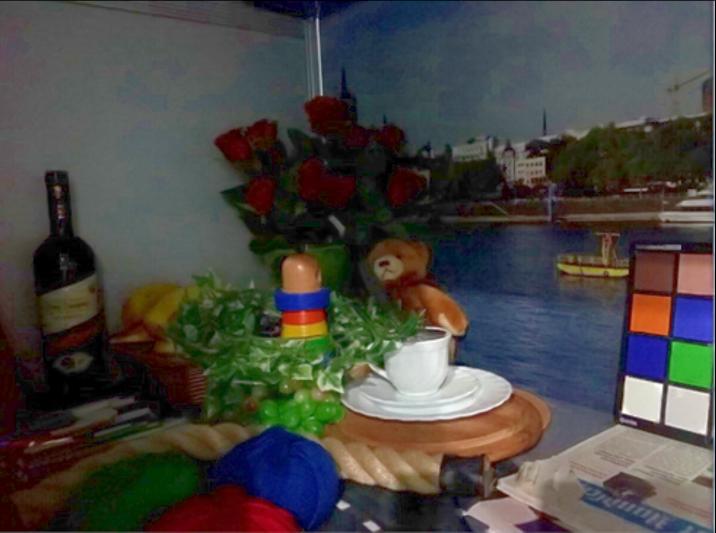}
\includegraphics[width=2.7cm,height=3cm]{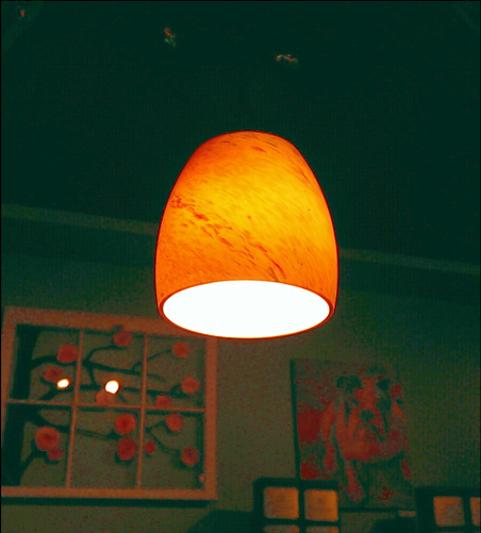}
\end{minipage}
\vspace{1mm}
\par\setlength\parindent{0.9em}(c)
\begin{minipage}[c]{0.95\linewidth}
\includegraphics[width=4.1cm,height=3cm]{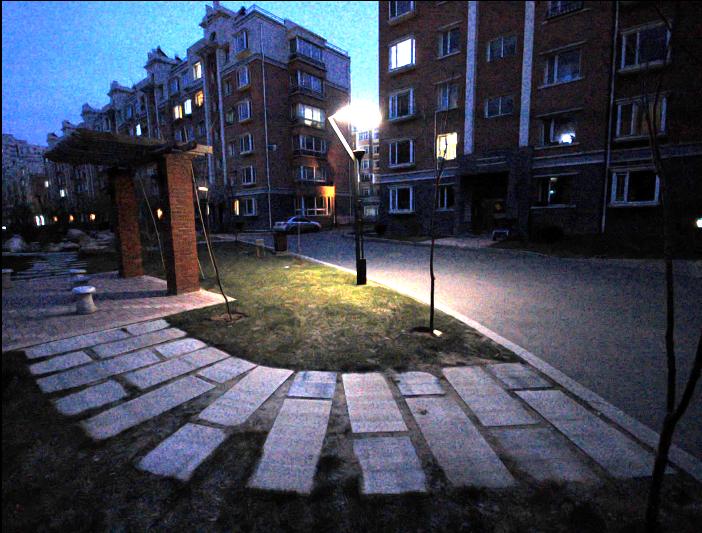} \includegraphics[width=2.7cm,height=3cm]{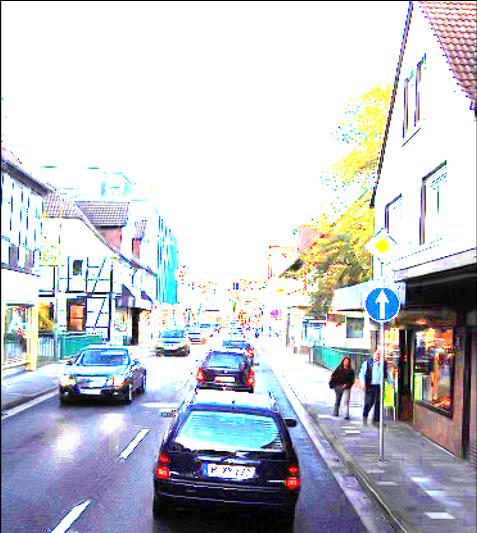} \includegraphics[width=3cm,height=3cm]{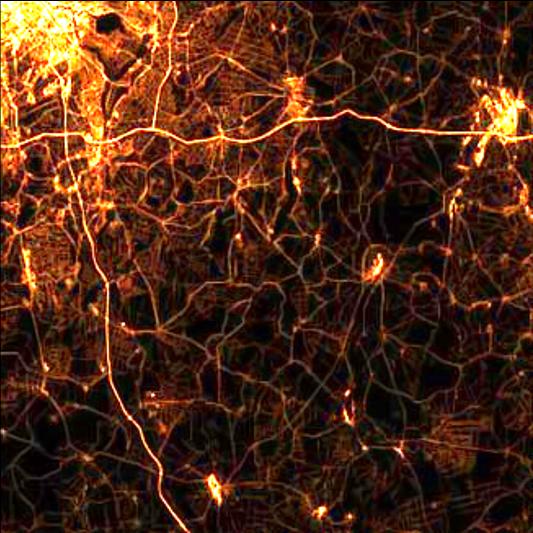} \includegraphics[width=4.1cm,height=3cm]{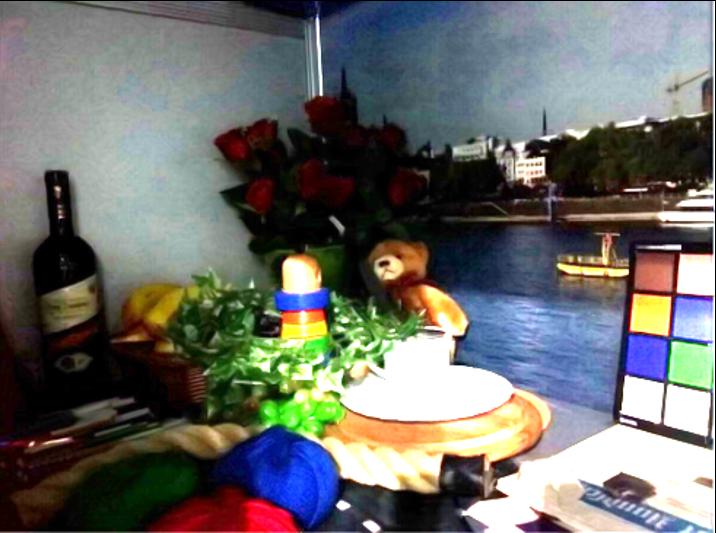}
\includegraphics[width=2.7cm,height=3cm]{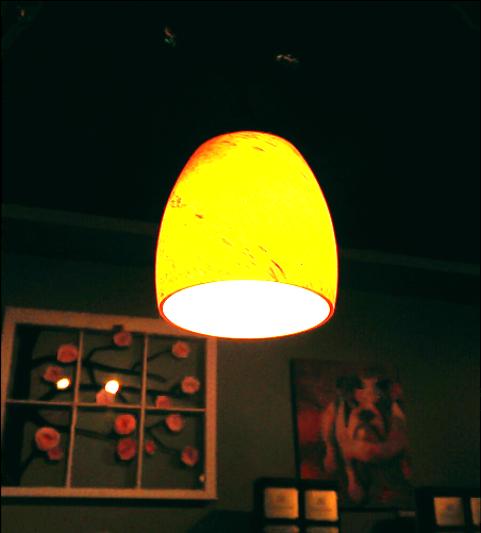}
\end{minipage}
\vspace{1mm}
\par\setlength\parindent{0.8em}(d)
\begin{minipage}[c]{0.95\linewidth}
\includegraphics[width=4.1cm,height=3cm]{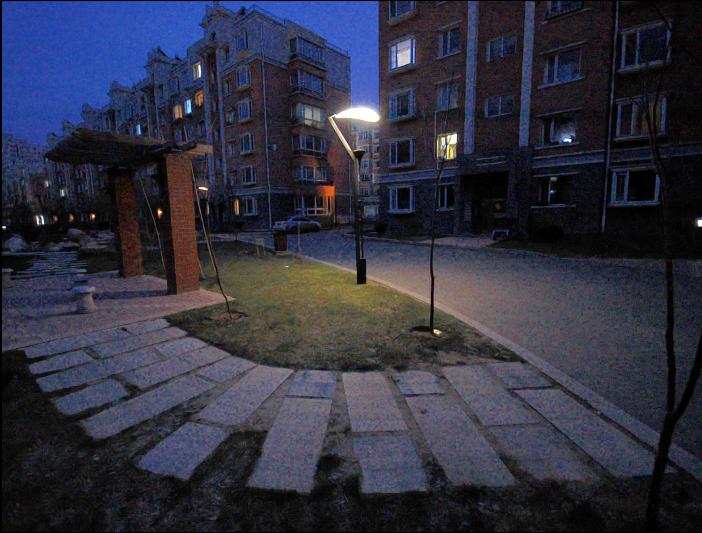} \includegraphics[width=2.7cm,height=3cm]{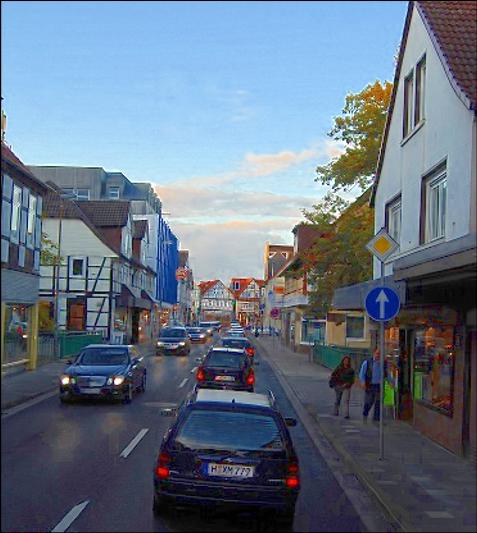}
\includegraphics[width=3cm,height=3cm]{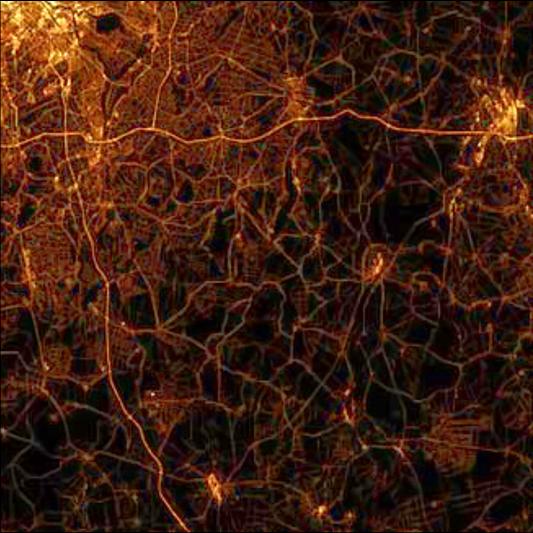} \includegraphics[width=4.1cm,height=3cm]{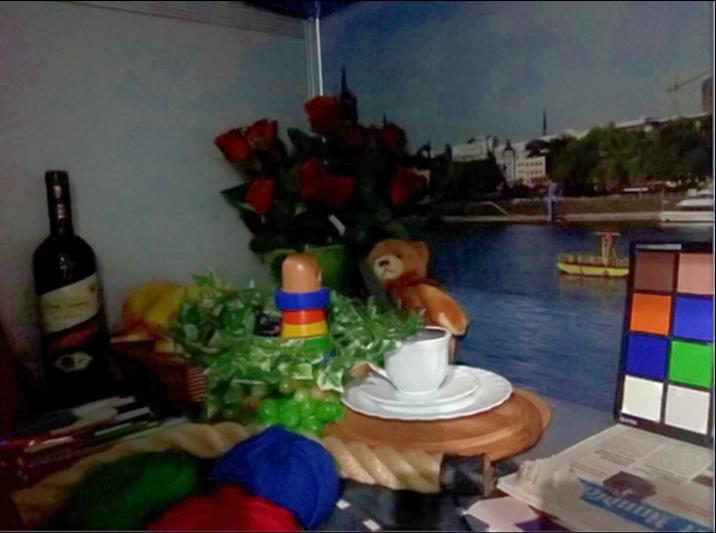}
\includegraphics[width=2.7cm,height=3cm]{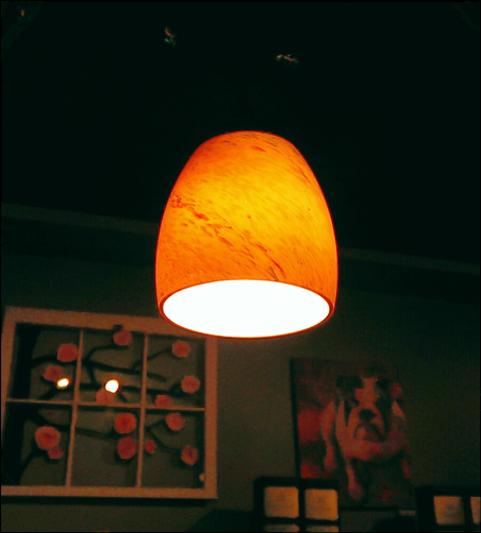}
\end{minipage}
\vspace{-1mm}
\caption{Comparisons of the result images enhanced by separate exponential transform, separate logarithmic transform, and the proposed minimum fusion strategy. The images in the row \textbf{(a)} are original low-light images, the images in the row \textbf{(b)} are obtained by separate exponential transform, the images in the row \textbf{(c)} are obtained by separate logarithmic transform, and the proposed minimum fusion strategy obtains the images in the row \textbf{(d)}.}
\label{f11}
\vspace{-3mm}
\end{figure*}

\par b) Minimum fusion strategy \emph{v.s.} separate exponential transform or separate logarithmic transform.
The proposed exponential and logarithmic transform minimum fusion strategy can achieve better objective and subjective image enhancement results. In order to further verify the effectiveness of this strategy, we compare the three enhancement results, separate exponential transform (NLHD w/o exp), separate logarithmic transform (NLHD w/o log), and the minimum fusion (NLHD), as shown in Table~\ref{table5}. We just use NIQE to evaluate the objective metric and use the 35images dataset here. Fig.~\ref{f11} shows the comparison of the results with these three situations. We can see that the minimum fusion strategy achieved the best subjective visual image enhancement results.

\begin{table}[htb]
\caption{Average NIQE value of separate exponential transform (NLHD w/o exp), separate logarithmic transform (NLHD w/o log), and the minimum fusion (NLHD) in the 35images dataset were compared.}
\vspace{-3mm}
\begin{center}
\footnotesize
\begin{tabular}{r|c|c|c}
\hline
\hline
\rowcolor[rgb]{ .851,  .851,  .851}
Metric & NLHD w/o exp & NLHD w/o log  & NLHD  
\\
\hline
NIQE $\downarrow$  & 2.80 & 3.78  & \textbf{2.76} 
\\
\hline
\hline
\end{tabular}
\end{center}
\vspace{-3mm}
\label{table5}
\end{table}

c) Retinex model \emph{v.s.} separate illumination or separate reflectance.
We reconstruct the enhanced image by separate illumination component (NLHD w/o illumination), separate reflectance component (NLHD w/o reflectance), and the whole Retinex model based NLHD method (NLHD). We use NIQE to evaluate the objective metric and use the 35images dataset here. We can see from Table~\ref{table6} that the NLHD method presents the best performance, NLHD w/o illumination can achieve a close result to the NLHD method, NLHD w/o reflectance result is the worst one. The objective evaluation results correspond to the subjective visual quality, partly referring to Fig.~\ref{landr}. 

\begin{table}[htb]
\caption{Average NIQE values of separate illumination reconstruction (NLHD w/o illumination), separate reflectance reconstruction (NLHD w/o reflectance), and the Retinex model based NLHD reconstruction on the 35images dataset.}
\vspace{-2mm}
\begin{center}
\footnotesize
\setlength{\tabcolsep}{1.6mm}{
\begin{tabular}{r|c|c|c}
\hline
\hline
\rowcolor[rgb]{ .851,  .851,  .851}
Metric &  NLHD w/o illumination & NLHD w/o reflectance  & NLHD  
\\
\hline
NIQE $\downarrow$           & 2.98 & 3.94  & \textbf{2.76}  
\\
\hline
\hline
\end{tabular}}
\end{center}
\label{table6}
\end{table}

\section{Conclusion}\label{Conclusion}
This paper proposes a low-light image enhancement method based on pixel-level non-local Haar transform decomposition (NLHD) and the Retinex model. The proposed image decomposition method makes full use of the pixel-level self-similarity in images, thus achieving better illumination and reflectance decomposition results, further achieving better low-light image enhancement results by using the decomposition results to the Retinex model. The proposed exponential and logarithm transform enhancement results minimum fusion strategy achieved more natural enhancement results than existing most low-light image enhancement methods. The extensive experiments of Retinex decomposition and low-light image enhancement show the effectiveness of the proposed method. The experimental results of the application under low-light environment face detection show that our low-light image enhancement method is competitive with other computer vision based applications, such as object detection; the performance of our method is a competitive one.  

 {
\small
\bibliographystyle{ieee}
\bibliography{nlhd}
 }

\end{document}